\documentclass[11pt,a4paper]{article}
\usepackage{jheppub,amsmath,amssymb,slashed,url,bm,textgreek,upgreek}
\usepackage{jheppub}  
\usepackage{tikz,lipsum,lmodern}
\usepackage[most]{tcolorbox}
\usepackage{amssymb} 
\usepackage{amsmath}
\usepackage{mathtools}
\usepackage{amsfonts}    
\usepackage{dsfont}
\usepackage{pdfpages}
\usepackage{verbatim}
\usepackage{tensor}
\usepackage{mathrsfs}
\usepackage{textgreek} 
\usepackage[mathscr]{euscript}
\usepackage[normalem]{ulem}
\usepackage{tikz}
\usepackage{makecell}
\usepackage{caption}
\usepackage{subcaption}
\usetikzlibrary{3d, arrows.meta, decorations.pathreplacing, decorations.markings,calc,shapes.misc,decorations.pathmorphing,patterns.meta, math}

\newcommand{\beq}{\begin{equation}}
\newcommand{\eeq}{\end{equation}}
 
\newcommand{\bea}{\begin{eqnarray}}
\newcommand{\ea}{\end{eqnarray}}
\newcommand{\barr}{\begin{array}}
\newcommand{\earr}{\end{array}}

\def\ie{\begin{equation}\begin{aligned}}
\def\fe{\end{aligned}\end{equation}}

\def\d{{\rm d}}
\def\i{{\rm i}}

\newcommand\Sp{\text{Sp}}
\newcommand\PSU{\text{PSU}}
\newcommand\U{\text{U}}

\newcommand\SL{\text{SL}}
\newcommand\SU{\text{SU}}
\newcommand\SO{\text{SO}}
\newcommand\OSp{\text{OSp}}

\newcommand{\cN}{{\cal N}}

\newcommand{\ket}[1]{{\left| {#1} \right>}}
\title{
Can black holes preserve $\mathcal{N} > 4$ supersymmetry?
\vspace{-0.6cm}
}

 \author{Matthew Heydeman${}^1$, Xiaoyi Shi${}^2$, Gustavo J. Turiaci${}^2$}
 \affiliation{${}^1$ Department of Physics \& The Black Hole Initiative, Harvard University, Cambridge, MA, USA}
 \affiliation{${}^2$ Physics Department, University of Washington, Seattle, WA, USA}

\vspace{1cm}

\abstract{The dynamics of near-BPS black holes are governed by the breaking of the conformal symmetry that emerges near their horizons. Using the classification of superconformal symmetries, we systematically classify and quantize all effective theories that can arise in the near-BPS limit of black holes. Using these results, we argue—under certain physical assumptions—that BPS black holes cannot preserve more than four supercharges. This conclusion is consistent with existing constructions in string theory. }
\begin{document}\maketitle

\thispagestyle{empty} 

\newpage
\setcounter{page}{1}

\section{Introduction and motivation} 

Near-extremal charged and/or rotating black holes are characterized by the emergence of a conformal symmetry. They universally present a long $AdS_2 \times X$ throat near their horizon, as shown by Kunduri, Lucietti and Reall \cite{Kunduri:2007vf}\footnote{This work typically assumes $D=4,5$ total spacetime dimension but the logic as well as known examples seem to cover a wider set of situations. Their proof accounts for the possibility of additional matter fields and higher derivative corrections beyond Einstein gravity.}, with the isometry $\SL(2,\mathbb{R})$ acting in the $AdS_2$ region and extended to the 1d conformal group. This emergent symmetry can be used to organize several aspects of the low-temperature dynamics of these black holes, through the Schwarzian mode which is the gravitational effective field theory when the symmetry is softly broken~\cite{Nayak:2018qej,Moitra:2018jqs,Castro:2018ffi,Sachdev:2019bjn,Ghosh:2019rcj,Iliesiu:2020qvm,Heydeman:2020hhw,Iliesiu:2022onk,Choi:2023syx,Rakic:2023vhv,Kapec:2023ruw,Kolanowski:2024zrq}.  From the gravity perspective, this mode arises from a specific metric fluctuation. From the perspective of the quantum system describing the black hole (which might have a microscopic description via string theory or AdS/CFT), its origin  is less transparent, but nevertheless dictated by the pattern of symmetry breaking.


In this paper, we want to raise the following somewhat ambitious sounding question-- What are all the possible effective theories describing the near-extremal dynamics and spectrum of black holes; i.e., what patterns of near horizon conformal symmetry breaking are possible at low temperatures? This question has a surprisingly simple answer in the case of non-supersymmetric theories. Because the extremal solution always has a decoupled $\SL(2,\mathbb{R})$ factor together with potentially other global symmetries, the only possibility for the soft mode is the Schwarzian mode together with decoupled gauge modes. These gauge modes arise from the soft breaking of those other global isometries besides the conformal one (they may be isometries of the internal space $X$, or boundary modes of gauge fields in addition to the metric). An example is the  $\SU(2)$ mode arising from rotations in the case of near-extremal Reissner-Nordstr\"om black holes, which is broken to a $\U(1)$ mode in the Kerr solution.


This question becomes richer in the context of supergravity, and in particular there are more nontrivial possibilities when the extremal black hole in question is BPS; namely the extremal solution has unbroken supercharges before turning on finite temperature. Intuitively from the point of view of symmetry, supersymmetry is one of the few ways we may extend $\SL(2,\mathbb{R})$ beyond product groups while maintaining a compact horizon. In effective field theories with supersymmetry, fermionic degrees of freedom in the effective theory emerge from the gravitino. These fermions can introduce nontrivial couplings between the Schwarzian and some of the gauge modes, more precisely those that are part of the $R$-symmetries. Thus supersymmetry gives ways to extend the Schwarzian theory while still preserving locality and consistency with the bosonic symmetries of the extremal solution. The problem is constrained enough that the possibilities are finite and few, and we will proceed to enumerate them.


There is a loose analogy between the effective field theory of near extremal black holes as dictated by (super)symmetry principles and the possible symmetries of the interacting S-matrix. The Coleman-Mandula theorem~\cite{Coleman:1967ad} famously shows that the possible symmetries of the S-matrix must factorize as $\textbf{Poincare} \times \textbf{Internal}$, which is analogous to the $\SL(2,\mathbb{R}) \times G$ symmetry implied by the work of Kunduri, Lucietti and Reall. Of course, this theorem has a famous ``loophole'' in the form of the Haag–Łopuszański–Sohnius theorem~\cite{Haag:1974qh}, which allows for nontrivial mixing of $\textbf{Poincare} \times \textbf{Internal}$ by the addition of fermionic generators. For near-BPS black holes, a similar mixing of near-horizon symmetries is possible via fermionic isometries, and can be related to the classification of superconformal groups. We will show how this leads to new interesting interacting theories and classify the possibilities.


What are the effective theories of near-BPS black hole dynamics? This classification is directly related to the classification of superconformal groups \cite{Nahm:1977tg,Knizhnik:1986wc}. To each superconformal group, there is an associated super Schwarzian theory. When one tries to think about this Schwarzian theory as a black hole EFT, we will show that, surprisingly, not all of these theories are unitary, allowing us to discard a large class of them. Finally, physics input is needed to further constrain the possibilities. Motivated by general features of $D\geq 4$ dimensional flat or AdS space, we propose that the algebra of conserved charges should be linear, which is the usual case for Lie algebras. This means that the (anti)commutator of two generators should be a linear combination of the generators (and not some more general nonlinear function). Beyond this condition on the linearity of the algebra, we can also ask about the nature of the BPS bound relating the black hole's mass to its charges. First, we may impose that the BPS bound is also linear in the $R$-charges. With these two constraints, plus unitarity, we find there are only two possible near-BPS black hole effective theories, where we indicate the amount of supersymmetry as well as the superconformal algebra the theory is based on\footnote{It is important to note in general that the number $\mathcal{N}$ of supersymmetries present in the Schwarzian description has nothing to do with the number of supersymmetries found in a gravitational vacuum solution; the relevant question is which symmetries are present for the extremal black hole with given boundary conditions.}:
\begin{itemize}
    \item 
    \textbf{$\mathcal{N}=2$ Schwarzian -} [$\SU(1,1|1)$], with $\U(1)$ $R$-symmetry and a complex fermion. This theory typically appears for near-BPS black holes in AdS \cite{Boruch:2022tno,Castro:2021wzn,Maulik:2024dwq, Heydeman:2024ezi,Heydeman:2024fgk,Cabo-Bizet:2024gny} and makes predictions about the BPS and near-BPS spectrum of the dual superconformal field theories.
    \item 
    \textbf{``Small'' $\mathcal{N}=4$ Schwarzian -}  $[\PSU(1,1|2)]$, with an $\SU(2)$ $R$-symmetry and a complex fermion in the fundamental representation. As an example, this theory appears for near-BPS black holes in flat space  as well as black holes in the $AdS_3$ limit of the D1-D5 system \cite{Heydeman:2020hhw}. This is relevant for justifying Strominger-Vafa style microstate counting~\cite{Strominger:1996sh} by showing the longstanding issue raised in \cite{Preskill:1991tb} is circumvented by a supersymmetric mass gap.
\end{itemize}
We indicate in brackets the superconformal group that is used to construct these theories. The physics of these two theories has been extensively studied. A quantization reveals that there are BPS states in the first (second) case preserving 2 (4) supercharges. A surprising conclusion of our work is that these two are the only possibilities given the assumptions listed earlier. Besides these two, there is also a $\mathcal{N}=1$ Schwarzian theory \cite{Fu:2016vas, Stanford:2019vob,Johnson:2021owr}, although it does not appear in any higher-dimensional black hole scenario we are aware of outside of $AdS_3$ examples. 


We now relax the assumption that the BPS relation is linear, but still insist on a linear algebra of conserved charges. In this work, we find only two new possibilities:
\begin{itemize}
    \item 
    \textbf{$\mathcal{N}=3$ Schwarzian - }$[\OSp(3|2)]$, with $\SO(3)$ $R$-symmetry and a fermion in the vector representation. Depending on global considerations, the quantum $R$-symmetry could be $\SU(2)$, $\SO(3)$ or an anomalous $\SO(3)$. 
    \item 
    \textbf{``Large'' $\mathcal{N}=4$ Schwarzian - } [D$(2,1|\upalpha)$], with $\SU(2)_+ \times \SU(2)_-$ $R$-symmetry and a complex fermion in the bi-fundamental.  The lower scripts are simply to distinguish the two groups. This theory additionally comes with a  continuous parameter $\upalpha$ and the model depends sensitively on this.
\end{itemize}
We will study the quantization of these theories in this paper. We find that all the theories considered so far present BPS states that preserve four or fewer supercharges at extremality.
 The ``large'' $\mathcal{N}=4$ theory is quite special, and is relevant for black holes in asymptotically $AdS_3 \times S^3 \times S^3 \times S^1$ string theory backgrounds. The fact that the BPS relation is non-linear might explain ultimately why it seems hard to embed this geometry as a near-horizon approximation of an asymptotically 10d flat solution, see for example \cite{Tong:2014yna,Gukov:2004ym}. This can be done if we replace the last factor $S^1 \to \mathbb{R}$ in which case the central charge is infinite and there is no dynamical gravity. The $\mathcal{N}=3$ theory appears in orbifolds of the same solution \cite{Giveon:2003ku,Eberhardt:2018sce}. Finally, another example where the large $\mathcal{N}=4$ Schwarzian is relevant are 6d black holes with $AdS_2 \times S^2 \times S^2$ throats \cite{Boonstra:1998yu}, without any $AdS_3$ region.  In another publication \cite{WOP_US},  we will point out interesting predictions regarding the spectrum of BPS black holes in these string backgrounds.


It is interesting to note which theories do not appear in the list of near-BPS black hole dynamics. In particular, there is no theory that preserves more than four supercharges if we insist on the algebra of conserved charges being linear. This is interesting in the context of the ``small black holes'' in string theory. These are solutions with a horizon that becomes singular in the BPS limit\footnote{The  argument in string theory is that one can generate BPS black hole solutions using dualities. Their area may be written in terms of an invariant on the lattice of charges. Three or four nonzero charges leave $\frac18$-th or $\frac{1}{16}$-th supersymmetry remaining, while setting some charges to zero preserves more supercharges but causes the invariant to vanish. Our techniques are more general and will not make any reference to this argument.}. One proposal has been that stringy corrections resolve the singularity and one ends up with a string-size classical smooth black hole \cite{Sen:1995in}. Nevertheless, there is no effective theory that describes this regime since, in all cases, these small black holes preserve a larger group of supercharges. An example is the heterotic string itself and the Dabholkar-Harvey states, preserving 8 supercharges. As pointed out in \cite{Chen:2024gmc} this conclusion is consistent with a different proposal claiming that there is a Horowitz-Polchinski transition before the BPS regime is reached \cite{Chen:2021dsw}. The methods in this paper allow one to make the argument in \cite{Chen:2024gmc} more precise.


Finally, we can consider what happens if we relax our assumption of linear algebra of conserved charges. We take this situation as more exotic since all examples we have involve asymptotically $AdS_2$ or $AdS_3$ geometries \cite{Henneaux:1999ib}; see also \cite{Dibitetto:2018ftj,Legramandi:2020txf, Macpherson:2023cbl, Hong:2019wyi,Conti:2025djz,Lozano:2025ief}. If we expect the black hole to be embedded in a $D>3$ asymptotically flat or AdS spacetime, it is reasonable to exclude these possibilities. In any event, we find new cases:
\begin{itemize}
    \item 
    \textbf{$\mathcal{N}=n$ Schwarzian - }[$\OSp(n|2)$ for $n\geq 4$], with several choices. At the quantum level there are three versions of the theory, depending on global considerations. They are distinguished by the $R$-symmetry being $\SO(n)$ (anomalous or non-anomalous) and $\text{Spin}(n)$. In the latter case, the fermion fields in the Schwarzian are in vector representations, but this theory has black hole states transforming in spinor representations.\footnote{In the 1d Schwarzian or the 2d CFT context, a theory based on $\OSp(2|4)$ is not the $\upalpha \to 1$ limit of D$(1,2|\upalpha)$. This is only true when the $\OSp(2|4)$ theory is quantized such that the $R$-symmetry group is $\text{Spin}(4)$ and not $\SO(4)$. In 2d CFT this appears as a choice in how to sum over integer spectral flow.  }
    \item 
    \textbf{$\mathcal{N}=7$ Schwarzian - }[G$(3)$], with $R$-symmetry group G$_2$ and fermions transforming in the 7-dimensional fundamental of G$_2$.
    \item 
    \textbf{$\mathcal{N}=8$ Schwarzian -} [F$(4)$], which has the $R$-symmetry group $\text{Spin}(7)$ (note that F$(4)$ is the superconformal algebra and not the $R$-symmetry). The fermions transforms in the spinor representation of $\text{Spin}(7)$.
\end{itemize}
We will quantize these theories for completeness, although we believe they cannot arise from near-extremal black holes in higher-dimensional flat or AdS spaces since the commutation relations of conserved charges of these theories are nonlinear. We find that for a theory with $\mathcal{N}$ supercharges, there are full (partial) BPS states preserving $\mathcal{N}$ (or less) supercharges. Theories based on other superconformal groups, such as $\text{OSp}(4^*|2n)$ and $\SU(1,1|n)$ with $n>2$, are shown to be non-unitary and do not appear. This exhausts the possibilities for theories considered in \cite{Maloney:1999dv,Britto-Pacumio:1999dnb} in the context of black holes and superconformal quantum mechanics, but note that the modern interpretation is very different. In our case, the quantum mechanical system is the interacting super-soft modes arising from broken conformal symmetry, rather than a quantum system with an exact conformal symmetry \cite{Maldacena:2016upp}.

In a similar vein to the above discussion of small black holes, one can also consider black holes appearing in $AdS_3$ theories with \emph{more} than $\mathcal{N} = (4,4)$ supersymmetry, again using the fundamental heterotic string. Besides the one mentioned above, a series of these proposals appeared in \cite{Dabholkar:2007gp,Lapan:2007jx,Johnson:2007du,Kraus:2007vu} in which it was conjectured that the heterotic string on $AdS_3 \times S^2 \times T^5$ is dual to the worldvolume theory of $N$ coincident heterotic strings. It was argued that the symmetry of the vacuum of this theory should be $\OSp(4^*|4)$, with the $\SU(2) \times \text{USp}(4)$ acting geometrically on the horizon from the first factor, while the second factor acts non-geometrically as spin frame rotations of the $T^5$. This proposal has a number of puzzling features, including the fact that the affine enhancement of $\OSp(4^*|4)$ leads, as we discuss later, to a nonlinear and nonunitary algebra. We consider this to be a strong indication that instead there is a transition to a different phase similar to the proposal in \cite{Chen:2024gmc}. There is one more context where $\OSp(4^*|4)$ appears, which will be shown to be nonunitary. This is in the context of half-BPS Wilson loops in $\mathcal{N}=4$ Super Yang-Mills. The dynamics of this system is not described by the Schwarzian theory because the conformal symmetry is not broken. This is not inconsistent with our analysis since it is outside the framework of symmetry breaking relevant for black holes. This can be explicitly verified using the construction of the effective theory describing the dynamics of Wilson loops in \cite{Giombi:2022pas}.

Another interesting scenario is the case of more general black holes in $AdS_5$. Large $\frac{1}{16}$-BPS black holes dual to $\frac{1}{16}$-BPS states in $SU(N)$ $\mathcal{N}=4$ SYM are known to have an order $N^2$ entropy and are a case where the Schwarzian analysis is successful~\cite{Boruch:2022tno}. Non-singular black hole solutions preserving $\frac18$-th supersymmetry do not seem to exist in the literature, despite the $N^2$ growth of the Macdonald Index~\cite{Choi:2018hmj}. The analysis of our paper does not strictly rule out these black holes, but it does put constraints on their non-BPS fluctuation spectrum; even more supersymmetry than this is essentially ruled out. Evidence supporting the non-existence of these black holes comes from recent field theory discussion~\cite{Chang:2023ywj}, which suggests that the $\frac18$-th Schur sector does not have non-graviton cohomologies, in the sense of \cite{Chang:2022mjp,Choi:2022caq,Choi:2023znd,Chang:2023zqk,Budzik:2023vtr,Choi:2023vdm,Chang:2024zqi,Chang:2024lxt,Chang:2025rqy}. The lack of an extended super-Schwarzian description in the near-BPS limit for $\mathcal{N}>4$ may explain why the $\frac12$ and $\frac14$-BPS sectors in field theory do not exhibit strong chaos~\cite{Chen:2024oqv}. All this being said, the results of our paper rely on the assumption of Einstein supergravity in the near horizon region. One might speculate that black holes preserving more supersymmetry may exist if they are dressed with other fields or branes beyond the minimal supergravity approximation \cite{Choi:2025lck}.

As a final point before turning to the setup for our argument, we can comment on other kinds of consistency conditions for quantum gravity which are similar in spirit, but different in practice. In particular, certain Swampland Conjectures (see \cite{Arkani-Hamed:2006emk,Ooguri:2006in,Palti:2019pca,Harlow:2022ich} and references therein) form a set of criterion for which seemingly consistent classical black holes and gravity might be inconsistent with UV principles. This is in contrast to our calculation, which only uses the low energy, near horizon degrees of freedom at one loop and the constraints of extended superconformal symmetry. Nevertheless, our findings seem consistent with known results in string theory; it would be interesting to further strengthen this connection, but we will not attempt to do so here.

The paper is organized as follows. In \textbf{section \ref{sec:GPF}} we classify all possible effective theories of near-BPS black holes and quantize them. In \textbf{section \ref{sec:N3SCH}} we extract the near-BPS spectrum of theories with three supercharges. In \textbf{section \ref{sec:LN4SCH}} we extend it to theories with four supercharges described by the large $\mathcal{N}=4$ Schwarzian. In \textbf{section \ref{sec:N>4}}, we show that theories with more than four supercharges are either non-unitary or display nonlinear commutation relations between supercharges.

\section{Effective theories for near-BPS horizons}\label{sec:GPF}

Our goal is to classify the possible effective theories that describe quantum effects in the near-BPS limit of black holes. This section will describe our general strategy, as well as the definitions of the models we will study. We will also quantize these theories using exact path integral techniques and then discuss their implications in the rest of the paper. Some special cases have received considerable attention in recent years because they are relevant for well studied black holes in string theory. Rather than go into those extensive details (which are contained in the references in the Introduction), we will be brief and emphasize mostly the new features that appear in the theories we uncover in this work. We stress that we view these models as gravitational effective field theories which in principle apply universally to black holes preserving the desired symmetries, independent of whether there exists a specific embedding in string theory. 

\subsection{The general approach}

Characterizing effective theories for near-BPS states in generic supersymmetric quantum systems is too broad of a problem to address. In this paper, we focus on theories that are holographic, and BPS states that are described by black holes. BPS black hole states are necessarily extremal geometries in Lorentzian signature\footnote{We mean here Lorentzian geometries that can be interpreted as states in the dual quantum system. Of course, there are also complex non-extremal BPS geometries that contribute to the gravitational path integral of boundary protected quantities \cite{Cabo-Bizet:2018ehj,Iliesiu:2021are}. That is not the meaning intended in this context. }. This implies that our black holes are also near-extremal, which has important implications.

It was proven by Kunduri, Lucietti, and Reall in \cite{Kunduri:2007vf} that extremal black holes necessarily have near-horizon geometries with an $AdS_2$ factor, under some assumptions. This implies that there is an emergent $\SL(2,\mathbb{R})$ group of isometries that appears near the horizon of such black holes. Depending on the case, there can also be other compact global isometries
\beq
\SL(2,\mathbb{R}) \times G,~~~G=G_1 \times G_2 \times \ldots 
\eeq
with $G_1,G_2,\ldots$ some set of compact groups. Some of these groups can arise from the metric, while others can arise from gauge fields in higher-dimensional spacetime. For example, in the case of Reissner-Nordstrom the geometric isometries near the horizon are $\SL(2,\mathbb{R}) \times \SU(2)$ and there is a $\U(1)$ symmetry arising from the Maxwell field. In the case of the Kerr black hole the isometries are $\SL(2,\mathbb{R}) \times \U(1)$ with $\U(1)$ corresponding to the unbroken group of rotations \cite{Bardeen:1999px}. The near-horizon isometries of the $1/16$-BPS black hole in $AdS_5$ are $\SL(2,\mathbb{R}) \times \U(1)\times \U(1)$ with another $\U(1)^3$ arising from gauge fields, which themselves are geometric symmetries in ten dimensions. This being said, depending on our boundary conditions, in general not all symmetries of a generic black hole metric will give rise to factors in $G$; in some cases we must work at fixed charges in order to remain near extremality. This was dubbed a `mixed' or `partial' canonical ensemble\cite{Boruch:2022tno}, in which some charges are fixed to maintain extremality while others are allowed to fluctuate, giving rise to the factors in $G$.

In general, the $\SL(2,\mathbb{R})$ symmetry of $AdS_2$ is enhanced to the full one dimensional conformal group of time reparameterizations $t\to f(t)$, by large diffeomorphisms that act at the conformal boundary of $AdS_2$. Other than the original $\SL(2,\mathbb{R})$ transformations which leave the metric intact, the rest of the reparametrization modes are physical and have to be integrated into the gravitational path integral. Similarly, compact isometries $G$ are enhanced to large diffeomorphisms (or large gauge transformations) which are parametrized by their action on the conformal boundary $g_m(t) \in G_m$, with $m=1,2,\ldots$, modulo global transformations. These also have to be integrated in the path integral. 

The gravitational action on $AdS_2$ is independent of these large diffeomorphism (or gauge) transformations, leading to potentially problematic divergences. For this reason, one is forced to work in nearly $AdS_2$, including the leading-order correction to the throat approximation arising at finite temperature. They break the conformal symmetry and the corresponding modes now receive a non-vanishing action. Symmetries constrain the leading order action for these modes to take the form
\ie\label{eq:actionbosonicint}
I = S_0 - \int \d\tau \,\left\{\Phi_r\, \text{Sch}(f,\tau)  + \sum_{m} K_m \, {\rm Tr} [ (g_m^{-1} \partial_\tau g_m)^2]  \right\} + \ldots
\fe
where $\Phi_r$ is the reparametrization mode coupling and $K_m$ the $G_m$ mode coupling. $S_0$ is a parameter that can be interpreted as zero-temperature entropy. The Schwarzian derivative is the leading order local action that can appear, since it has to depend on the reparametrization mode and at the same time be invariant under $\SL(2,\mathbb{R})$ reparametrizations acting on $f(t)$. This action breaks both the reparametrization symmetry and the $\SL(2,\mathbb{R})$ symmetry acting on $t$. A similar argument applies to the rest of the action.

The form of the action above can be derived from effective theory arguments \cite{Choi:2023syx,Choi:2021nnq}, but can also be directly derived from the Einstein action via a dimensional reduction of the higher-dimensional theory to $AdS_2$. This does not rule out the possibility of having $\Phi_r=0$ or $K=0$ for specific theories, although we do not have any examples. Another possibility is the presence of a non-local term that is more relevant in the IR than the Schwarzian action. This can happen in some cases in the presence of matter sources \cite{Maldacena:2016upp,Castro:2021wzn,Milekhin:2021sqd,Milekhin:2021cou}. We assume here that this is not the case. Our methods otherwise apply to any theory leading to near-BPS black holes with $AdS_2$ throats.

The Schwarzian theory was exactly quantized in \cite{Stanford:2017thb,Mertens:2017mtv} and also \cite{Kitaev:2018wpr,Yang:2018gdb}. The extension with the gauge fields was also quantized in \cite{Mertens:2019tcm, Iliesiu:2019lfc,Kapec:2019ecr}. The result is quite simple, in an ensemble of fixed $G$ charges, the effective theory reduces to the Schwarzian theory. Therefore, when we work in this ensemble, there is essentially one universal behavior. The supersymmetric generalization is instead far richer.

When considering the near-BPS limit of black holes in supergravity, the gravitini can generate new fermionic symmetries that interact with the 1d conformal modes and the factors of the $G$ modes that correspond to $R$-symmetries (those bosonic symmetries that do not commute with the supercharge). To classify all the possibilities, it is useful to take a step back and analyze the global isometries first. What are all the possible supergroups ${\sf G}$ that include $\SL(2,\mathbb{R})$ as a bosonic subgroup? The answer, worked out by Nahm \cite{Nahm:1977tg}, is given in Table \ref{tab:scalgebras}. In all cases, the bosonic subgroup with $N_b$ generators are ${\sf G} \supset \SL(2,\mathbb{R})\times G_R$ with the $R$-symmetry group $G_R$ specified in the second column. The $N_f$ fermionic generators transform in representation $\rho$ of the group $G_R$ and are in a doublet of $\SL(2,\mathbb{R})$. For each of these global superconformal groups, one can associate local versions. These were classified by Knizhnik \cite{Knizhnik:1986wc}. Besides $f(t)$ and $g(t)$, these theories include local extensions of the supersymmetry transformation that we collectively denote $\eta(t)$ and transforms in the representation $\rho$ of $G_R$. More details on these supergroups can be found in the original references, and here will be given as needed.
\begin{figure}
\begin{tabular}{|c|c|c|c|}
  \hline
  \textbf{Superconformal Isometries} & \textbf{$R$-symmetry $G_R$} & \textbf{$\rho$} & \textbf{Dimensions $(N_b,N_f)$} \\
  \hline
  $\OSp(n|2)$                      & $\SO(n)$                              & $n$                & $(\frac{1}{2}n^2 - \frac{1}{2}n + 3, 2n)$ \\
  $\SU(1,1|n),~~n\ne 2$            & $\SU(n)\times \U(1)$                 & $n+\bar{n}$        & $(n^2 + 3, 4n)$ \\
  $\OSp(4^*|2n)$                   & $\SU(2) \times \text{USp}(2n)$       & $(2,2n)$           & $(2n^2 + n + 6, 8n)$ \\
  $\text{PSU}(1,1|2)$             & $\SU(2)$                             & $2+\bar{2}$        & $(6, 8)$ \\
  $\text{D}(2,1|\upalpha)$        & $\SU(2)_- \times \SU(2)_+$           & $(2,2)$            & $(9, 8)$ \\
  $\text{G}(3)$                   & $G_2$                                & $7$                & $(17, 14)$ \\
  $\text{F}(4)$                   & $\text{Spin}(7)$                             & $8_s$              & $(24, 16)$ \\
  \hline
\end{tabular}
\caption{Global superconformal groups, together with the $R$-symmetry $G_R$, the representation $\rho$ of the fermionic generators, and the number of generators. To each of these supergroups there is an associated local superconformal group \cite{Knizhnik:1986wc}.}
\label{tab:scalgebras}
\end{figure}

To leading order away from the extremal $AdS_2$ limit, these modes receive an action that has to be the appropriate supersymmetric completion of the Schwarzian theory and the $G_R$ mode action \eqref{eq:actionbosonicint}. In the next section, we describe an efficient way to derive the action and present the general result for the thermal partition function valid for all theories. Just like the bosonic case, we could also have gauge modes associated to other flavor symmetries. We will assume from now on we work on fixed-charge sectors of those flavor symmetries.

Of all the theories that are derived from Table \ref{tab:scalgebras}, only a few have been considered so far. The first example is $\OSp(2|2) \sim \SU(1,1|1)$, which leads to the so-called $\mathcal{N}=2$ Schwarzian theory and describes the near-BPS limit of the $1/16$-BPS black hole in $AdS_5$ \cite{Boruch:2022tno,Castro:2021wzn,Maulik:2024dwq} as well as $AdS_4$ \cite{Heydeman:2024ezi,Heydeman:2024fgk}. The other example is $\PSU(1,1|2)$ leading to $\mathcal{N}=4$ Schwarzian theory and describes near-BPS black holes in flat space as well as black strings in 6d such as the D1/D5 system \cite{Heydeman:2020hhw}. String backgrounds leading to near-BPS black holes are known to involve the superconformal group $\text{D}(2,1| \upalpha)$, with $\OSp(4|2)$ as a special case, such as the $AdS_3 \times S^3 \times S^3 \times S^1$ background \cite{deBoer:1999gea,Gukov:2004fh} as well as $\OSp(3|2)$ that is relevant for orbifolds of $AdS_3 \times S^3 \times S^3 \times S^1$ \cite{Giveon:2003ku,Eberhardt:2018sce}. The near-BPS effective theories for the last two examples have not been analyzed so far. Independently of whether there exists any known black hole in supergravity or string theory presenting the symmetries in Table \ref{tab:scalgebras}, we will attempt to evaluate the partition function and spectrum uniformly for all the groups.

\subsection{The action and the partition function}

We are interested in the theory describing the breaking of the local superconformal symmetry associated with the global supergroups in table \ref{tab:scalgebras}. In order to do so, we need an efficient way to generate supersymmetric actions with the correct degrees of freedom. This can be achieved via the $BF$ formulation of JT gravity, which leads to the Schwarzian theory on the boundary. 

For any supergroup ${\sf G} \supset \SL(2,\mathbb{R})\times G_R$ we can write down the $BF$ action 
\ie
I_{BF}=-\i\int \text{Str}\,\phi F,\quad F=\d A-A\wedge A,
\label{eq:BFJT}
\fe
where $\phi$ is a scalar and $A$ a connection one-form, both in the adjoint of ${\sf G}$. The supertrace $\text{Str}$ is defined through the quadratic Casimir of the superalgebra in a standard fashion. We denote by $L_{-1},L_0, L_1$ the generators of the $\SL(2,\mathbb{R})$ subgroup, $T^i$ the generators of the $G_R$ algebra, and $G^a_{\pm 1/2}$ the fermionic generators where $a$ runs over the states in the $\rho$ representation of $G_R$. We consider this theory on a disk such that its boundary is the thermal circle $\tau$, with mixed boundary conditions $\delta(\phi+2 \i \Phi_r A_\tau)|_{\text{bdy}}=0$. The integral over $\phi$ restricts to flat connections. Crucially, the variational problem requires a boundary term $\frac{\i}{2} \oint \text{Str} \phi A$, which is also responsible for the model having any dynamics. The total action, after integrating out the adjoint scalar, is 
\beq\label{eq:IBFAA}
I_{BF} =\Phi_r \oint \text{Str}\, A_\tau^2.
\eeq
The next step is to write $A$ consistent with the mixed boundary condition and evaluate the boundary term. Since the connection is flat, these configurations are parameterized by local transformations which make up the 1d superconformal group. The components along $T^i$ are determined by a local function $g(\tau)\in G_R$ and the components along $G^a_{\pm 1/2}$ are parametrized by a fermionic field $\eta^a(\tau)$. This is easier said than done, but provides a concrete algorithm to obtain the action. The only subtlety, explained in \cite{Saad:2019lba} and \cite{Cardenas:2018krd}, is that we need to impose the extra constraint that the $L_0$ component of $A$ vanishes and the $L_1$ component is constant. This is the first-order version of the asymptotically $AdS_2$ condition\footnote{Up to correctly implementing this condition, this analysis is the dimensional reduction of \cite{Henneaux:1999ib}.}. This exercise has been done explicitly for $\OSp(n|2)$ in \cite{Cardenas:2018krd}, the $\mathcal{N}=4$ Schwarzian action associated to $\PSU(1,1|2)$ was written in \cite{Heydeman:2020hhw}, and the action of the theory based on $\text{D}(2,1|\upalpha)$ was written in \cite{Kozyrev:2021icm},  although most of these theories have not been quantized until now. By this we mean-- there is a significant distinction between finding a classical (super) Schwarzian derivative whose saddle point captures the low temperature expansion of the black hole entropy, as opposed to computing the exact partition function of this theory which in general gives significant logarithmic corrections.

For our calculations, we will only need the explicit form of the bosonic part of the action, which can be written as
\beq
A_\tau \supset L_1 + \text{Sch}(f,\tau) L_{-1} + \left[g^{-1}(\tau) \partial_\tau g(\tau)\right]_i T^i .
\eeq
Plugging this expansion into \eqref{eq:IBFAA} gives us the bosonic action
\ie
I_{\rm super-Schw} = -\Phi_r \int \d\tau \,\left\{\text{Sch}(f,\tau)  + q \, {\rm Tr}_\rho [ (g^{-1} \partial_\tau g)^2] + \text{fermions} \right\}.
\fe
The trace that appears in the $R$-symmetry term is taken in the $\rho$ representation. The relative normalization between the two bosonic actions, denoted as $q$, is determined by supersymmetry. The $BF$ formulations provides an efficient way of extracting it from the supertrace in \eqref{eq:IBFAA}, using the coefficients in front of the $\text{SL}(2,\mathbb{R})$ and $G_R$ generators. Taking the correct value of $q$ is essential to preserve the supersymmetry of the entire super-Schwarzian action. This should become more clear below when we consider special cases.

We now quantize these theories and evaluate the partition function $Z(\beta,\alpha_i)$. The inverse temperature $\beta$ is the circumference of the boundary circle. Let ${H_R}$ denote the Cartan generators of $G_R$, with their number equal to $\text{rank}\, G_R$. We can assign a chemical potential $\alpha_i$ to each generator ${H_R^i}$. They are normalized such that the twisted boundary conditions around the thermal circle are given by:
\ie
f(\tau+\beta)=f(\tau),\quad g(\tau+\beta)=e^{4\pi \i \alpha\cdot H_R}g(\tau),\quad \eta(\tau+\beta)=-e^{4\pi \i \alpha\cdot H_R}\eta(\tau).
\label{eq:bc}
\fe
where $\alpha\cdot H_R$ is short for $\sum_{i=1}^{\text{rank}\, G_R}\alpha_i H_R^i$. We can quantize the theory either using fermionic localization \cite{Stanford:2017thb} or using the canonical approach of \cite{Mertens:2017mtv}. Even though the second approach connects more with the work of \cite{Knizhnik:1986wc}, we follow the localization method which is simpler for our purposes. The integration space for all these theories is a coadjoint orbit of the associated 1d superconformal group, and therefore is a symplectic space. Moreover, we assume that the supersymmetric completion of the Schwarzian action is unique such that it also acts as a generator of a $\U(1)$ symmetry on the supermanifold. This is true for actions generated by $BF$ theories described above\footnote{The BF approach is used simply to systematically generate supersymmetric completions of the Schwarzian action, but otherwise this perspective is not very important for this paper.}. We can then apply the Duistermaat-Heckman theorem and compute the path integral by localization on fixed points of this $\partial/\partial\tau$.
In other words,
\ie
Z(\beta,\alpha_i)=&\int\frac{\mathcal{D}f\mathcal{D}g\mathcal{D}\eta}{G}e^{-I_{\rm super-Schw}}
\\
=&\sum_{\text{fixed points}}Z_{\rm 1-loop}^{\rm Schw} \,Z_{\rm 1-loop}^{R}\, Z_{\rm 1-loop}^{\rm fermions}\,e^{-I_{\text{fixed points}}}
\fe
To compute the one-loop determinants, we apply the shortcut discussed in \cite{Turiaci:2023jfa} and also \cite{Witten:2020bvl}. Without knowing about the details of the quadratic form of fluctuations or the symplectic measure, one can directly compute the eigenmodes and eigenvalues of the $\U(1)$ generator acting on the tangent space of fixed points. The Duistermaat-Heckman theorem also implies that the one-loop determinant is then the product of these eigenvalues after the removal of zero modes. The final result is obtained by combining all  contributions and can be found below
\eqref{eq:pf}. Ignoring numerical overall prefactors, the contributions are:
\begin{itemize}
\item The fixed point and one-loop corrections of the Schwarzian mode have been well known since \cite{Stanford:2017thb}. We only present the exact partition function up to a normalization factor:
\ie
Z^{\rm Schw}_{\rm 1-loop}e^{-I_{\rm Sch}}=\left(\frac{\Phi_r}{\beta}\right)^{\frac{3}{2}}e^{\frac{2\pi^2\Phi_r}{\beta}}
\fe
\item The partition function of the gauge field admits a geometric interpretation as the partition function of a particle moving on the group manifold of $G_R$, which is solved in \cite{Picken:190160}. The fixed points are the product of closed loops $g_1(\tau)$ and a reference path $g_0(\tau)$. Here, $g_0(\tau)=\exp\left(4\pi \i\, \frac{\tau}{\beta}\,\alpha\cdot H_R\right)$ is the shortest geodesic that connects the end points $g(0)=1$ and $g(\beta)=e^{4\pi \i \alpha\cdot H_R}$.
Denote the set of $\nu$ such that $\exp\left(4\pi \i\, \nu \cdot H_R\right) = 1$ as $\check{T}$. The closed loop can then be expressed as $g_1(\tau) = \exp\left(4\pi \i\, \frac{\tau}{\beta} \nu \cdot H_R\right)$. Including these closed loops relates to the shifts in the chemical potentials $\alpha\to\alpha+\nu$ that keep the boundary conditions \eqref{eq:bc} unchanged.
The fixed points are given by
\ie
g(\tau)=\exp\left(4\pi \i\, \frac{\tau}{\beta}\,(\alpha+\nu)\cdot H_R\right)
\fe
Evaluate the action at these fixed points, we obtain
\ie
I^{R}_{\rm fixed points}=q\frac{16\pi^2\Phi_r}{\beta}{\rm Tr}_{\rho}\left[\left((\alpha+\nu)\cdot H_R\right)^2\right]
\fe
Now we turn to the fluctuations around these fixed points. Using the Cartan-Weyl basis for the Lie algebra $\mathfrak{g}_R$, the fluctuations can be parametrized by dim$(R)$ functions $\epsilon(\tau)$'s as:
\ie
g(\tau)=&\exp\left(4\pi \i\, \frac{\tau}{\beta}\,(\alpha+\nu)\cdot H_R+\sum_i \epsilon_i(\tau)H_R^i+\sum_{r\in R}\epsilon_r(\tau) E^r\right)
\fe
where $R$ is the set of roots. The twisted boundary condition for $g(\tau)$ around the thermal circle becomes
\ie
\epsilon_i(\tau+\beta)=\epsilon_i(\tau),\quad\epsilon_r(\tau+\beta)=e^{-4\pi\i\,r\cdot(\alpha+\nu)}\epsilon_r(\tau)
\fe
Therefore, the eigenvalues of $\epsilon_i$ are $2\pi m/\beta$ with $m\in\mathbb{Z}$ while the eigenvalues of $\epsilon_r$ are $2\pi (m-2r\cdot(\nu+\alpha))/\beta$. Remember we need to remove all the zero-modes since they relate to the $G_R$ isometry in the bulk; two black holes related by $G_R$ are not distinct. The regularized one-loop determinant is given by:
\ie
Z_{\rm 1-loop}^{G_R}=&\left(\prod_{i=1}^{\text{rank}\, G_R}\prod_{m\ge 1}\frac{\beta}{2\pi \Phi_r m}\right)\times \left(\prod_{r\in R}\prod_{m\ge 1}\frac{\beta}{2\pi\Phi_r(m-2r\cdot(\alpha+\nu))}\right)
\\
=&\left(\frac{\Phi_r}{\beta}\right)^{\frac{{\rm dim}G_R}{2}}\prod_{r\in R_+}\frac{2\pi r\cdot(\alpha+\nu)}{\sin 2\pi r\cdot(\alpha+\nu)}
\fe
The final result has been obtained using zeta-function regularization and the identity $\sin x= x\prod_{n\ge 1}(1-\frac{x^2}{n^2\pi^2})$.

\item The fermion fields vanish on-shell, so we only need their 1-loop fluctuations. Each of the $\cN$ fermions corresponds to one weight vector $\mu$ in the weight space $V_\rho$ of the $\rho$-representation.\footnote{$\rho$ doesn't have to be irreducible - in the small $\mathcal{N}=4$ case, for example, the fermions transform in the direct sum of fundamental and anti-fundamental representations.} The twisted anti-periodic boundary condition $\eta_\mu(\tau+\beta)=-e^{4\pi\i\,(\alpha+\nu)\cdot H_R}\eta_\mu(\tau)=-e^{4\pi\i\,(\alpha+\nu)\cdot \mu}\eta_\mu(\tau)$ indicates the fermions have eigenvalues $2\pi (m+2\mu\cdot(\nu+\alpha))/\beta$ with $m\in\mathbb{Z}+\frac{1}{2}$. The $m=\pm 1/2$ modes correspond to $N_f=2\cN$ fermionic generators of supergroup ${\sf G}$ and should be discarded in the path integral, analogous to the global bosonic generators. The fermionic one-loop determinant is given by the product of eigenvalues on the numerator:
\ie
Z_{\rm 1-loop}^{\rm fermions}=&\prod_{\mu\in V_\rho}\prod_{m\neq \pm\frac{1}{2}}\frac{2\pi \Phi_r(m+2\mu\cdot(\nu+\alpha))}{\beta}
\\
=&\left(\frac{\beta}{\Phi_r}\right)^{\frac{N_f}{2}}\prod_{\mu\in V_{\rho+}}\frac{\cos2\pi\mu\cdot(\alpha+\nu)}{1-16(\mu\cdot(\alpha+\nu))^2}
\fe
where $V_{\rho+}$ is the set of positive weights and we have used $\cos x=\prod_{n\ge 1/2}(1-\frac{x^2}{n^2\pi^2})$.
\end{itemize}
We can now put all the ingredients together and obtain the exact partition function for the extended Schwarzian theory describing the effective theory of near-BPS black holes. After a shift of $S_0$ that fixes the prefactor, it can be put in the form
\ie
Z(\beta,\alpha_i)=&\sum_{\nu\in \check{T}}\left(\frac{\beta}{\Phi_r}\right)^{(N_f-N_b)/2}\prod_{r\in R_+}\frac{2\pi r\cdot(\alpha+\nu)}{\sin 2\pi r\cdot(\alpha+\nu)}\prod_{\mu\in V_{\rho+}}\frac{\cos2\pi\mu\cdot(\alpha+\nu)}{1-16(\mu\cdot(\alpha+\nu))^2}
\\
&\times\exp\left[S_0+ \frac{2\pi^2\Phi_r}{\beta}\left(1-8q\,{\rm Tr}_{\rho}\left[\left((\alpha+\nu)\cdot H_R \right)^2\right]\right)\right].
\label{eq:pf}
\fe
We can see that this partition function is entirely determined by the supergroup information provided in Table \ref{tab:scalgebras}. The value of $q$ can be inferred from the relative coefficients between the $G_R$ and $\SL(2,\mathbb{R})$ traces in the supertrace. The overall prefactor in this expression is arbitrary since we can always absorb it in a shift of $S_0$. When possible, we will adjust the overall prefactor to simplify the dependence of the BPS spectrum on $S_0$.

\subsection{Some examples}
\label{ssec:exampleZsection}
Here we apply the results above to specific examples. 

\paragraph{Large $\mathcal{N}=4$} Consider the superconformal group $D(2,1|\upalpha)$, whose $R$-symmetry is $\SU(2)_+\times\SU(2)_-$. The group theory data on table \ref{tab:scalgebras} implies:
\ie
N_f-N_b=8-9=-1,\quad R_+=\{(1,0),(0,1)\}
\\
\rho=(2,2),\quad V_{\rho+}=\{(1/2,1/2),(-1/2,1/2)\}
\fe
This is an example without a supermatrix representation. Therefore, we need to extract $q$ from a formal representation of the supertrace.  The bosonic part of the Casimir operator of the superalgebra\footnote{We follow the conventions for the specific algebra $\text{D}(2,1|\upalpha)$ in \cite{Gukov:2004ym}.} is given by $C_2=L_0^2-\frac{1}{2}\left\{L_{-1}, L_1\right\}-\gamma T^{+i} T^{+i}-(1-\gamma) T^{-i} T^{-i}$, where $L_{0,\pm 1}$ generate $\text{SL}(2,\mathbb{R})$ and $T^{\pm i}$ generate two $\SU(2)$, respectively. It is convenient sometimes to introduce $\gamma$ via  $\upalpha= (1-\gamma)/\gamma$. From its inverse, we can determine the relative normalization of the two $\SU(2)$ gauge field actions as: $q_+=1/\gamma$ and $q_-=1/(1-\gamma)$. Based on the information above, we can write down the partition function for large $\cN=4$ Schwarzian as follows
\bea
Z &=& \sum_{n\in\mathbb{Z}}\sum_{m\in\mathbb{Z}} \frac{\Phi_r^{1/2}}{\beta^{1/2}}\frac{(\alpha_++n)}{\sin 2\pi \alpha_+ }\frac{(\alpha_-+m)}{\sin 2\pi \alpha_- }\, \cos \pi (\alpha_+ + \alpha_-) \cos \pi (\alpha_+ - \alpha_-)\nonumber\\
&&\times \frac{e^{{S_0+ \frac{2\pi^2\Phi_r}{\beta}(1- \frac{4(1+\upalpha)}{\upalpha}(\alpha_+ +n)^2-4(1+\upalpha)(\alpha_- +m)^2)}}}{(1-4(\alpha_+ +\alpha_- + m + n)^2)(1-4(\alpha_+ -\alpha_- - m + n)^2)}.\label{eqn:ptD21}
\ea
We have derived $\check{T}=\mathbb{Z}\times\mathbb{Z}$ from $\exp\left(4\pi\i\nu\cdot H_R \right)=\exp\left(2\pi\i \,\text{Diag}(\nu_1-\nu_2,\nu_1+\nu_2)\right)=1$. We will extract the physical information encoded in this result in Section \ref{sec:LN4SCH}.

\paragraph{Orthosymplectic groups} Another example we discuss in this article is the effective theory describing the breaking of $\OSp(n|2)$ superconformal symmetry. The bosonic part of the Casimir operator of the superalgebra is given by $C_2=L_0^2-\frac{1}{2}\left\{L_{-1}, L_1\right\}- T^{i} T^{i}$, where $L_{0,\pm 1}$ generate $\text{SL}(2,\mathbb{R})$ and $T^{ i}$ generate $\SO(n)$. Therefore $q=1$, which is obvious from a supermatrix representation of the group. The $R$-symmetry group $\SO(n)$ has a different root structure for even and odd $n$, which are listed in Appendix \ref{app:Osp}.
According to Table \ref{tab:scalgebras}, the fermions transform in the fundamental representation of $\SO(n)$. For $\OSp(2l|2)$, the weights of the fermions are given by $V_{\text{fund}}=\{\pm e^j|j=1,\dots,l\}$, while the Cartan generators can be represented by $\left[H_m\right]_{j k}=-i\left(\delta_{j, 2 m-1} \delta_{k, 2 m}-\delta_{k, 2 m-1} \delta_{j, 2 m}\right)$. Therefore, the partition function is given by
\ie
Z^{\OSp(2l|2)}=&\sum_{n_1,\dots,n_l\in\mathbb{Z}}\left(\frac{\Phi_r}{\beta}\right)^{(l-1)(l-\frac{3}{2})}\prod_{i=1}^l\frac{\cos 2\pi \alpha_i}{1-16(\alpha_i+n_i/2)^2}
\\
&\prod_{1\le j<l\le l}\frac{(\alpha_j+n_j/2)^2-(\alpha_k+n_k/2)^2}{\sin 2\pi (\alpha_j-\alpha_k)\sin 2\pi (\alpha_j+\alpha_k)}
\times\, e^{S_0+\frac{2\pi^2\Phi_r}{\beta}(1-\sum_{i=1}^l 16(\alpha_i+n_i/2)^2)}
\label{eqn:ptospeven}
\fe
The superalgebra for $\OSp(4|2)$ is isomorphic to $D(2,1|\upalpha)$ when the parameter is set to be $\upalpha=1$. We can verify this at the level of partition functions \eqref{eqn:ptD21} and \eqref{eqn:ptospeven} using the identification $\alpha_++n=\alpha_1+n_1/2-\alpha_2-n_2/2$ and $\alpha_-=\alpha_1+\alpha_2$. Nevertheless, there is one important difference. In the above, we take the $R$-symmetry group of the $\OSp(4|2)$ theory to be the `non-anomolous' $\SO(4)$, which is different from $\text{Spin}(4) =\SU(2) \times \SU(2)$. This appears explicitly in the fact that, although the partition functions are the same saddle by saddle, the range in the sum over integral shifts is different. Roughly speaking, the sum in $\text{D}(2,1|1)$ is twice the sum in this version of $\OSp(4|2)$. We will comment more on this later in section \ref{sec:anomalyosp}.

For $\OSp(2l+1|2)$, $2l$ fermions have the same weight as in the $n=2l$ case. The additional fermion corresponds to zero weight, giving no contribution to the one-loop correction other than a factor of $(\beta/\Phi_r)^{1/2}$ after regularization. Taking into account the addition roots, we obtain
\ie
Z^{\OSp(2l+1|2)}=&\sum_{n_1,\dots,n_l\in\mathbb{Z}}\left(\frac{\Phi_r}{\beta}\right)^{(l-1)(l-\frac{1}{2})}\prod_{i=1}^l\frac{\cos 2\pi \alpha_i}{1-16(\alpha_i+n_i/2)^2}
\\
&\prod_{1\le j<k\le l}\frac{(\alpha_j+n_j/2)^2-(\alpha_k+n_k/2)^2}{\sin 2\pi (\alpha_j-\alpha_k)\sin 2\pi (\alpha_j+\alpha_k)}\prod_{m=1}^{l}\frac{\alpha_m+n_m/2}{\sin 2\pi\alpha_m}
\\
&\quad\times\, e^{\frac{2\pi^2\Phi_r}{\beta}\left(1-\sum_{i=1}^l 16(\alpha_i+n_i/2)^2\right)}
\label{eqn:ptospodd}
\fe
In the next section, we will analyze $\cN=3$ theories, where the partition function is obtained by setting $l=1$ in the above expression. Again, by making the sum run over even integers $n$ we can construct a theory with $R$ symmetry group $\text{Spin}(n)$. This is physical if there are other matter fields beyond the Schwarzian multiplet transforming in spinorial representation.

\paragraph{Other cases:}  The theories associated with $\SU(1,1|n)$ with $n>2$ as well as $\OSp(4^*|2n)$ will be shown in section \ref{sec:rulingout} to be problematic. The values of $ q$ of some of the factors of the $R$ symmetry are required to take negative values due to supersymmetry. This implies the partition function is either non-unitary or divergent. Therefore, even though they are well-defined superconformal groups, they cannot possibly appear in the near-BPS regime of a black hole described by a unitary quantum system. Besides the cases we already analyzed, this only leaves the exceptional supergroups G(3) and F(4) which we analyze in some detail in section \ref{sec:exceptionalgroup}.

\section{Black holes with $\mathcal{N}=3$}\label{sec:N3SCH}

In this section, we will analyze in some detail the spectrum of theories with $\mathcal{N}=3$ supercharges. This case arises in some string theory backgrounds constructed in \cite{Giveon:2003ku} 
 and also in \cite{Eberhardt:2018sce}. In contrast to the $\mathcal{N}=2$ and small $\mathcal{N}=4$ Schwarzian theories (which have numerous realizations in string theory as explained in the introduction), black holes preserving exactly $\mathcal{N}=3$ supersymmetry may seem somewhat exotic. However, the analysis we perform here serves as an important test case for many of the features we encounter again for more familiar cases with extended supersymmetry, such as the large $\mathcal{N}=4$ theory. This includes the important feature of nonlinear BPS bounds and non-linear algebras, as well as properties of the index.
 
 There are three possible theories with three supercharges and a softly broken conformal symmetry, which we describe here together with their solution. They can be distinguished by the $R$-symmetry being $\SO(3)$, anomalous $\SO(3)$, and $\SU(2)$.

\subsection{Two theories with $\SO(3)$ $R$-symmetry and their solution}

The $\mathcal{N}=3$ theory has a classical $\SO(3)$ $R$-symmetry. We refer to the generators by $J^i$ with $i=1,2,3$. The fields, which we can collectively denote as $\Phi(\tau)$, satisfy the following twisted boundary conditions when subject to a chemical potential $\upalpha$: $\Phi(\tau+\beta) = \pm e^{4\pi \i \alpha J^3}\,\Phi(\tau)$, where the overall sign depends on whether the field is fermionic or bosonic, and $J^3$ in the exponent acts according to the $\SO(3)$ representation of the field $\Phi$.  


Prior to summing over saddles with nontrivial gauge mode winding number $n$, the one-loop partition function for the disc in the trivial winding sector $(n=0)$ is
\beq
Z_{n=0} \sim \frac{8\cos(2\pi \alpha)}{\pi(1-16 \alpha^2)} \frac{\alpha}{\sin(2\pi \alpha)} \, e^{S_0+\frac{2\pi^2\Phi_r}{\beta}(1-16 \alpha^2)}.
\eeq
The prefactors arise from quantum effects of the reparametrization mode, the fermions, and the $\SO(3)$ gauge fields. The exponential term arises from the on-shell action of the theory. To obtain the full partition function, we need to incorporate global considerations. Since the fields are singlet or adjoint, it is reasonable to impose that the chemical potential is only defined $\text{mod}~\mathbb{Z}/2$. (For comparison, for theories with small $\mathcal{N}=4$, the fermions are in the fundamental making $\alpha$ defined mod $\mathbb{Z}$.) The partition function is
\beq\label{eq:ZN3un}
Z =\sum_{n\in\frac{1}{2}\cdot\mathbb{Z}} \frac{8\cos(2\pi \alpha)}{\pi(1-16 (\alpha+n)^2)}  \frac{\alpha+n}{\sin(2\pi \alpha)} \, e^{S_0+\frac{2\pi^2\Phi_r}{\beta}(1-16 (\alpha+n)^2)}.
\eeq
We can think of configurations with $n\in \mathbb{Z}$ as those that uplift to $\SU(2)$ holonomies that are identical, while those with $n\in \mathbb{Z}+1/2$ as being twisted in the $ \SU(2)$ uplift by a sign. 

There are actually two ways of performing this sum over the $\SO(3)$ bundles, corresponding to the possibility of adding a discrete topological term to the action. It is useful to pass to a $d+1$-dimensional holographic description and view the Schwarzian partition function as equivalent to the gravitational path integral on the disk. There are two different ways of summing over bulk $\SO(3)$ gauge fields, since we are free to add to the action
\beq
I \to I+ \i \pi \nu \int_{\mathcal{M}_{\text{bulk}}} w_2(\SO(3)),
\eeq
where $w_2$ is the second Stiefel-Whitney class. The coefficient $\nu$ is only defined mod $2$ and therefore only takes values $\nu=0,1$. For our purposes, the effect of this term is to reverse the sign of the contributions with $2n$ odd. Therefore, there are two $\mathcal{N}=3$ theories, one with partition function \eqref{eq:ZN3un}, and the other with partition function
\bea
Z = \sum_{n\in\frac{1}{2}\cdot\mathbb{Z}} (-1)^{2n \nu}\frac{8\cos(2\pi \alpha)}{\pi(1-16 (\alpha+n)^2)} \frac{\alpha+n}{\sin(2\pi \alpha)} \, e^{S_0+\frac{2\pi^2\Phi_r}{\beta}(1-16 (\alpha+n)^2)}. \label{eq:ZN3t}
\ea
We will now analyze the implications of these results to the low-energy regime of the theory. This is analogous to the possible $\theta$-angle one can incorporate in the $\mathcal{N}=2$ theory.\footnote{The partition function above matches term-by-term the classical $c\to\infty$ limit of the one corresponding to the $\mathcal{N}=3$ Virasoro algebra studied by Miki~\cite{Miki:1989ri}. The possibility of having two theories was not considered in the $\text{AdS}_3$ gravity literature, as far as we know.} One key difference is that the $\theta$-angle is a classically continuous parameter, which means that there is a large landscape of possible $\mathcal{N}=2$ Schwarzian theories that may be engineered in string theory~\cite{Heydeman:2024fgk}, while in contrast here we see there is only a discrete choice. We will comment later which choice seems to have a realization in string theory.


Let us begin with the most basic facts about the theory. Since it has a (classical) global $\SO(3)$ symmetry we expect the Hilbert space to decompose into multiplets $\mathcal{H}=\oplus_j \mathcal{H}_j$ where $\mathcal{H}_j$ are states transforming in the spin-$j$ representation. We will see that the sum includes $j\in \mathbb{Z}$ if $\nu=0$, and $j \in \mathbb{Z}+1/2$ if $\nu=1$, implying the latter has an $\SO(3)$ anomaly at the quantum level. Based on this multiplet structure, the dependence of the partition function on the chemical potential $\alpha$ is captured by the expansion in spin-$j$ characters,
\beq
\chi_j(\alpha) = \frac{\sin(2\pi (2j+1)\alpha)}{\sin(2\pi\alpha)}.
\eeq
To extract the spectrum from the partition function it is useful to consider momentarily the partition function in which we only sum over saddles labeled by integers, defining
\beq
\tilde{Z} = \sum_{n\in\mathbb{Z}} \frac{8\cos(2\pi \alpha)}{\pi(1-16 (\alpha+n)^2)} \frac{\alpha+n}{\sin(2\pi \alpha)} \, e^{S_0+\frac{2\pi^2\Phi_r}{\beta}(1-16 (\alpha+n)^2)}
\eeq
Then the two partition functions are 
\beq
Z_\nu(\beta,\alpha) = \tilde{Z}(\beta,\alpha) +(-1)^\nu \tilde{Z}(\beta,\alpha+1/2).
\eeq
This expression can be naturally interpreted from the boundary point of view as a manifestation that $\SO(3) = \SU(2)/\mathbb{Z}_2$ where we gauge the discrete symmetry generated by $e^{2\pi \i J}$. As we show in Appendix \ref{app:N3}, $\tilde{Z}$ can be expanded in $\SU(2)$ characters. Since $\chi_j(\alpha+1/2)=(-1)^{2j} \chi_j(\alpha)$ we see that the contribution of states with half-integer (integer) spin cancels when $\nu=0$ ($\nu=1$). Using the results derived in Appendix \ref{app:N3} we find
\bea
Z_\nu = e^{S_0}\delta_{\nu,0} + \sum_{j\in\mathbb{Z}+\frac{\nu}{2}, \, j>0} \int_{E_0(j)}^\infty \d E \, e^{-\beta E} \,2(\chi_j(\alpha)+\chi_{j-1}(\alpha))\, \rho_j(E),\label{eq:N311}
\ea
We see that, other than a number $e^{S_0}$ of isolated ground states with spin $j=0$ in the non-anomalous theory, all states come in a supermultiplet $2(j)\oplus 2(j-1)$. We will derive this structure from the supercharges below. Each supermultiplet labeled by its maximal spin $j$ has a spectral gap given by
\beq
E_0 (j) =\frac{j^2}{8\Phi_r},
\eeq
and a continuum density of states above the gap given by
\beq
\label{eq:N3DOS}
\rho_j(E) = (2j) \frac{e^{S_0}\,\cosh\Big(2\pi\sqrt{2\Phi_r(E-E_0(j))}\Big)}{8\pi E \sqrt{2\Phi_r(E-E_0(j))}}.
\eeq
This result applies to both theories, although the range of $j$ is different, as emphasized above. We comment more on some aspects of this result below after discussing the commutation relations of conserved charges.

 \begin{figure}[h!]
\begin{center}
\includegraphics[scale=0.45]{"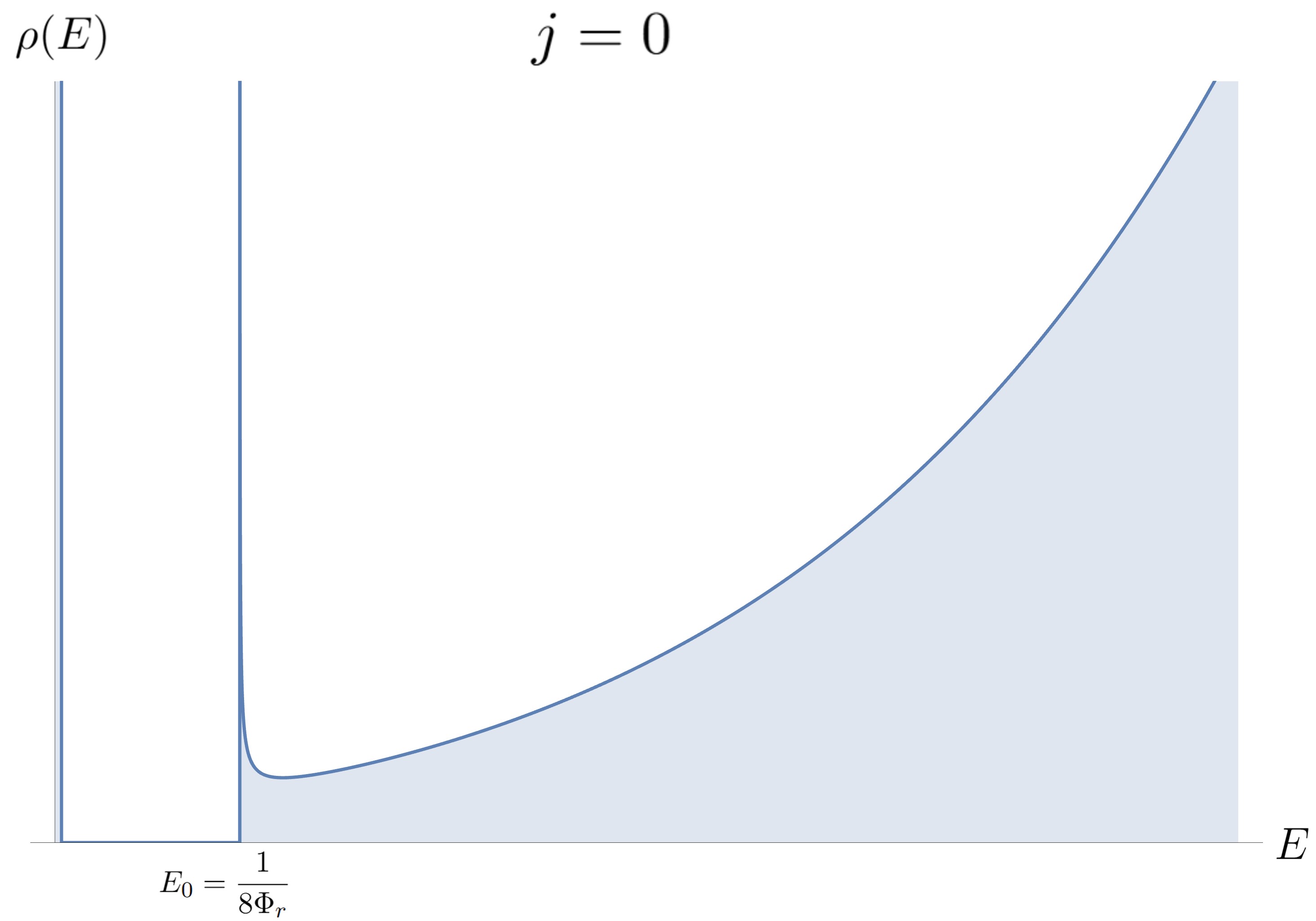"}
\caption{The microcanonical density of states for the non-anomalous $\mathcal{N}=3$ Schwarzian theory at fixed $j=0$. There is a degenerate set of BPS states at $E_{\text{BPS}}=0$ which are short multiplets. The apparent BPS bound at $E_0$ in this figure results from a $j=1$ long multiplet which has a $j=0$ state, but this bound is not actually saturated by any BPS state. Instead, we see the inverse square root edge behavior.} \label{fig:N3j0}
\end{center}
\end{figure}

We considered the above two theories, one with $\SO(3)$ symmetry or another with an anomaly. We can also define a third $\mathcal{N}=3$ Schwarzian theory with an $\SU(2)$ $R$-symmetry. No degree of freedom in the Schwarzian theory is in the spinor representation of $\SU(2)$, but other matter fields in the higher-dimensional theory could be. This affects the partition function of the Schwarzian theory only at the level of the sum over saddles, since we should sum over $n\in \mathbb{Z}$ given that fermions can distinguish $n$ and $n+1/2$.  Therefore, the partition function becomes precisely $\tilde{Z}$ above. The spectrum is the same as \eqref{eq:N311} but includes all integral and half-integral spins. Even though there are no half-integer fields in the Schwarzian sector, the theory has fermionic black hole microstates.  In a similar fashion, we can define any other $\OSp(n|2)$ generalization of the Schwarzian theory, with an $R$-symmetry that is $\text{Spin}(n)$ instead of $\SO(n)$, by simply changing the way we sum over saddles.

 \begin{figure}[h!]
\begin{center}
\includegraphics[scale=0.45]{"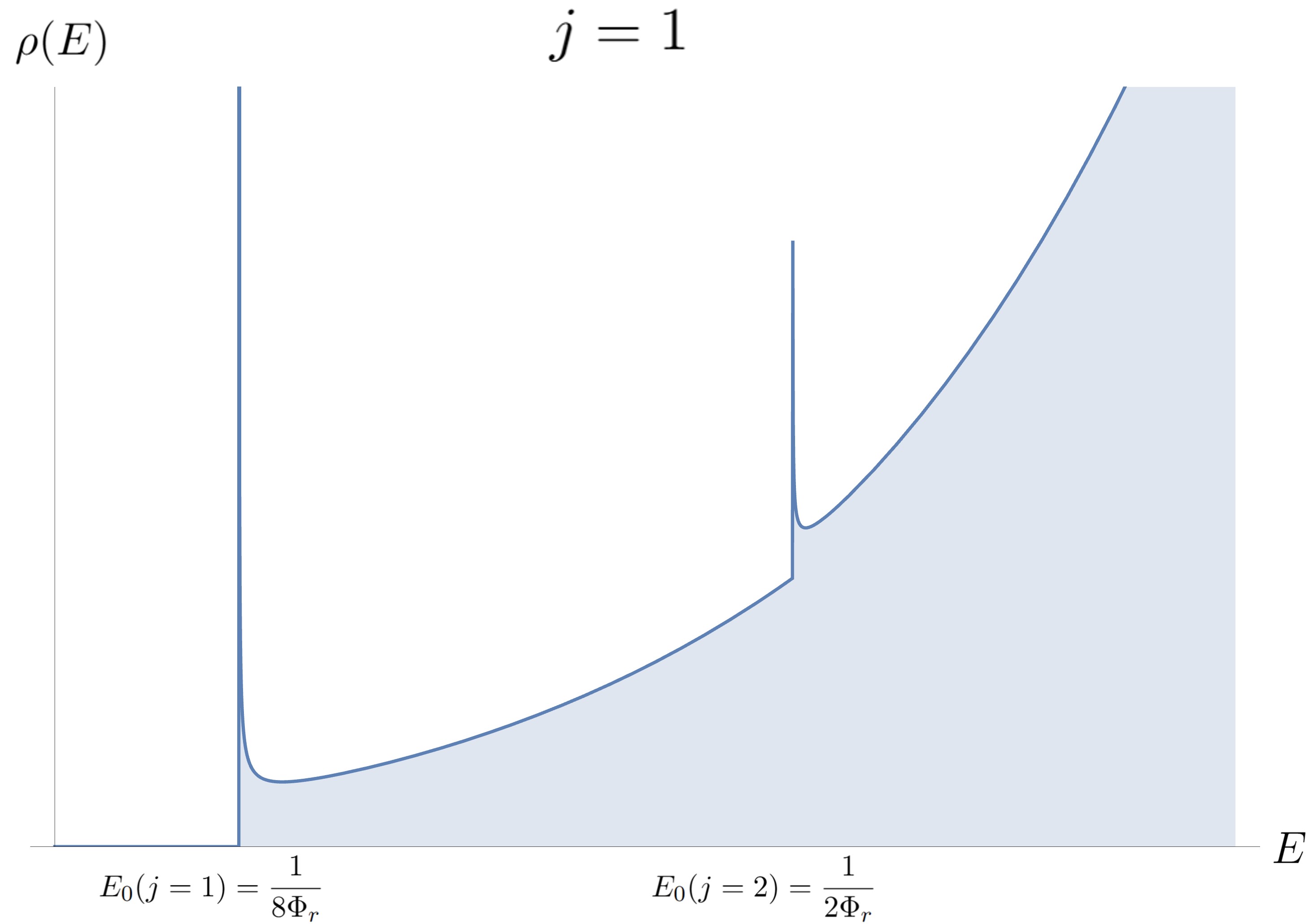"}
\caption{The microcanonical density of states for the non-anomalous $\mathcal{N}=3$ Schwarzian theory at fixed $j=1$. In contrast to the previous figure, there are no BPS states in this sector (they all have $j=0$). Instead, we see two apparent BPS bounds at $E_0(j=1,2)$ which comes from long multiplets containing $j=1$. Each apparent bound exhibits an inverse square root edge. To emphasize, there are no BPS states in this figure, only BPS bounds.} \label{fig:N3j1}
\end{center}
\end{figure}

\subsection{Commutation relations of conserved charges}
\label{ssec:nonlinearN3alg}

The conserved charges are the Hamiltonian $H$, three supercharges $Q^i$ with $i=1,2,3$, transforming in the adjoint of $\SO(3)$ generated by the remaining charges $J^i$. We could derive their commutation relations via an explicit construction of the charges from the super Schwarzian action, but instead of following this tedious route, the information we have so far uniquely determines the algebra. The $\SO(3)$ generators satisfy $[J^i,J^j]=\i \varepsilon_{ijk} \, J^k$, and since $Q^i$ are in the adjoint of $\SO(3)$ they satisfy $[J^i, Q^j] = \i \varepsilon_{ijk}\, Q^k$.
Both of these generators should commute with the Hamiltonian $[J^i,H]=[Q^i,H]=0$. We also define an operator $(-1)^{\sf F}$ such that it commutes with $H$ and $J^i$ and anticommutes with $Q^i$.

The only non-trivial relation is the anticommutator of the supercharges among themselves. Perhaps the most naive guess is 
\beq\label{eq:naive}
\{ Q^i , Q^j \} = \delta_{ij} H,~~~~~~~~\text{(Naive)}.
\eeq
This relation would imply that the Hamiltonian is a positive operator and that BPS states have zero energy and come in a representation with a unique spin $(j)$. This is certainly consistent with the spectrum we obtained; we found a large number of ground states with $H=0$ and $j=0$. 

This naive guess turns out to be wrong, which can be inferred by the following observation. Each supermultiplet $2(j)\oplus 2(j-1)$ has states only for energies above $E_0(j)$. If the BPS bound was simply $H=0$ as \eqref{eq:naive} suggests, we would expect a quantum chaotic system (with the Schwarzian spectrum as a leading approximation) to have a square-root edge everywhere except possibly at $H=0$. The spectrum we find above instead has an inverse square-root edge! Namely
\beq
\rho_j(E=E_0 + \epsilon) \sim e^{S_0} \frac{1}{\sqrt{\epsilon}},~~~\epsilon \gg E_0.
\eeq
This looks like a spectrum that is about to saturate a BPS bound at $E_{\text{BPS}}=E_0(j)$ based on expectations about random matrix theory with BPS bounds and BPS states~\cite{Stanford:2019vob,Turiaci:2023jfa}. Importantly, this behavior is inconsistent with the naive algebra \eqref{eq:naive}.


What are the possible modifications of \eqref{eq:naive}? One possibility is to allow the anticommutator of two supercharges to involve the $\SO(3)$ generators. The possible terms we can add are constrained by the $\SO(3)$ symmetry. The product of the vector representation decomposes as $3\otimes 3 = 5_S \oplus 3_A \oplus 1_S$, and since $\{Q_i,Q_j\}$ is symmetric in its two indeces, the only representations that can appear are either a singlet, or the symmetric traceless representation. This rules out any term linear in $J$ but at quadratic order we can add
\beq
\{ Q_i, Q_j\} = \delta_{ij} H + \upgamma_1 \, \delta_{ij} J^2 + \upgamma_2 \, \{J_{i},J_{j}\},
\eeq
where $\upgamma_1$ and $\upgamma_2$ are free parameters. We still need to check Jacobi identities. When evaluating $[\{Q_i,Q_j\},Q_k] + (\text{cyclic})$ we find it only vanishes when $\upgamma_1=0$. All values of $\upgamma_2$ are allowed so we found a one-parameter  \emph{nonlinear} correction of the algebra. The value of that parameter depends on the theory, and in our case it can depend on $\Phi_r$. We will see below that an analysis of the BPS spectrum leads to the following identification
\beq\label{eq:correct}
\{ Q_i, Q_j \} = \delta_{ij} H - \frac{1}{16\Phi_r} \{ J_i, J_j\}~~~~\text{(Correct)}.
\eeq
 The need for a non-linear correction is a novel feature that is not present in previously considered Schwarzian theories. The non-linear term is of order $\mathcal{O}(G_N)$ in terms of the Newton constant, since in physical models arising from higher-dimensions one can show $\Phi_r \sim \mathcal{O}(1/G_N)$. This does not necessarily imply the new term is small, since semiclassically the charges are of order $1/G_N$.


Let us derive more carefully the supermultiplets of such algebra and show that it indeed predicts BPS states at $E_0(j)$. We start by selecting a Cartan element $J^3$ and defining the rising and lowering generators $Q^{\pm}=(Q^1\pm \i Q^2)/\sqrt{2}$ and $J^\pm = (J^1 \pm \i J^2)/\sqrt{2}$ transforming with a definite $J^3$ charge. Representations of the algebra can be identified in terms of highest-weight states 
$$
| E, j\rangle.
$$
The labels corresponds to the following eigenvalue relations
\beq
H|E,j\rangle = E |E,j\rangle,~~~~J^3 |E,j\rangle = j |E,j\rangle, 
\eeq
while the highest-weight condition means that 
\beq
J^+ |E,j\rangle = Q^+  |E,j\rangle = 0.
\eeq
The action of $J^-$ generates an $\SO(3)$ multiplet with spin $j$. The action of $Q^3$ generates another spin $j$ multiplet of opposite fermion parity. A similar pair of spin $j-1$ multiplets and opposite fermion parity can be constructed by further acting with $Q^-$,\footnote{To ensure orthogonality within the supermultiplet, the bottom-left state in \eqref{fig:N3nonlinear} requires the linear combination of $Q^-|E,j\rangle$ and $J^- Q^3|E,j\rangle$, while the bottom-right state requires linear combination of $Q^- Q^{3} |E,j\rangle$ and $J^- |E,j\rangle$. See \cite{Schoutens:1988tg} for more details.} making a total multiplet $2(j) \oplus 2(j-1)$. This can be represented in a diagram
\bea\label{fig:N3nonlinear}
\begin{tikzpicture}[scale=1,baseline={([yshift=0cm]current bounding box.center)}]
\draw[very thick,->] (0,0.5) to (0,4);
\draw[left] (0,3) node {\footnotesize $j$};
\draw[left] (0,1) node {\footnotesize $j-1$};
\draw[thick,->] (1,3) to (2.95,3);
\draw[very thick] (-0.05,3) to (0.05,3);
\draw[very thick] (-0.05,1) to (0.05,1);
\draw[thick,->] (1,3) to (1,1.05);
\draw[above] (2,3) node {\footnotesize $Q^3$};
\draw[left] (1,2) node {\footnotesize $Q^-$};
\draw[above] (1,3) node {\footnotesize $|E,j\rangle$};
\draw[fill=black] (1,1) circle (0.05);
\draw[fill=black] (3,1) circle (0.05);
\draw[fill=black] (3,3) circle (0.05);
\draw[fill=black] (1,3) circle (0.05);
\end{tikzpicture}
\ea
This explains the structure found in \eqref{eq:N311}. We can bound the energy from \eqref{eq:correct} for $i=j=3$. Acting with this relation on the highest-weight state we obtain
\beq
\lVert Q^3 |E,j\rangle \rVert^2 = E - \frac{j^2}{8\Phi_r}~~~\Rightarrow~~~ E \geq E_0(j) =\frac{j^2}{8\Phi_r}.
\eeq
This gives us the correct ground state energy with the inverse square root edge found in \eqref{eq:N3DOS}. There is really no gap within the non-BPS multiplets, but there are also no BPS states except for the special case of $j=0$! This inverse edge implies the bound is not satisfied for $j\neq 0$ in gravity, but we can still contemplate what happens when it is. Interestingly, when the BPS bound is saturated, it means that the action of $Q^3$ becomes trivial, but not necessarily $Q^-$. The diagram is
\beq
\begin{tikzpicture}[scale=1,baseline={([yshift=0cm]current bounding box.center)}]
\draw[very thick,->] (0,0.5) to (0,4);
\draw[left] (0,3) node {\footnotesize $j$};
\draw[left] (0,1) node {\footnotesize $j-1$};
\draw[thick,->] (1,3) to (1,1.05);
\draw[left] (1,2) node {\footnotesize $Q^-$};
\draw[above] (1,3) node {\footnotesize $|E_0,j\rangle$};
\draw[fill=black] (1,1) circle (0.05);
\draw[fill=black] (1,3) circle (0.05);
\end{tikzpicture}
\eeq
Therefore the BPS multiplet is $(j)\oplus (j-1)$. 


Can we have BPS states preserving more than one supercharge? Let us first consider a supermultiplet constructed out of a highest-weight state with $j=0$. One can show that any such supermultiplet will be annihilated by all supercharges and have zero energy. Take the commutator
\beq
[Q^+, J^-]= Q^3.
\eeq
Since $j=0$ we have $J^-|E,j=0\rangle=0$ and being a highest-weight state $Q^+ |E,j=0\rangle=0$. This implies that $[Q^+,J^-] |E,j=0\rangle =0$ and therefore $Q^3|E,j=0\rangle=0$. Acting on $\SO(3)$ singlets we have the relation $\{Q^3,Q^3\} = H$ which implies that when $j=0$, $E=0$. Finally, for a state annihilated by $J^i$ and $Q^3$ the commutation relation $[J^\pm , Q^3] = \pm \i Q^\pm$ implies that $Q^-$ also annihilates the state, rendering it fully BPS. Could we have BPS states that transform nontrivially under $\SO(3)$ but also preserve all supercharges? The answer is no since
\beq
\{ Q^-, Q^-\} = - \frac{1}{8\Phi_r} J^-J^-,
\eeq
and similarly for $Q^+$. If the state preserves all supercharges then the LHS would vanish, but the RHS cannot vanish unless $j=0$\footnote{Another possibility is that $j=1/2$ since now $J^-J^-|+1/2\rangle = 0$. Only in this representation $\{J^i,J^j\}$ is proportional to $\delta_{ij}$. This case is special since even the non-BPS multiplet only includes two copies of $(1/2)$ while a BPS one (either fully or only preserving $Q^3$) only has one copy of $(1/2)$.}.


To summarize, the full spectrum associated to \eqref{eq:correct} has three types of representations with respective $\SO(3)$ characters
\bea
&&\text{Non-BPS:}~~~~~~~2\chi_j + 2\chi_{j-1},\nonumber\\
&&\text{Partial BPS:}~~~~~\chi_j + \chi_{j-1},\nonumber\\
&&\text{Full BPS:}~~~~~~~~~\chi_0,\label{eqn:Z3MGA}
\ea
This is precisely the structure we found as arising from the Schwarzian theory. Notice that within the $\mathcal{N}=3$ Schwarzian theories, the non-anomalous one contains both Full BPS and Non-BPS states, whereas the anomalous one contains only Non-BPS states. The partial BPS states did not appear in \eqref{eq:N311} and instead we found a continuum of Non-BPS multiplets up to the energy bound. This does not necessarily exclude them from a microscopic model, but in gravity with this amount of supersymmetry they do not seem to appear.

\subsection{The index for $\mathcal{N}=3$}

The Witten index of the model involves an insertion of $(-1)^{\sf F}$ in the partition function. The list above shows that this insertion not only removes non-BPS states but also removes partially BPS states. Therefore, the Witten index is sensitive only to fully BPS states. We can also use a different operator, $e^{\i \pi J^3}$, which also leads to vanishing contributions from both non- and partially-BPS states. Notice that $(-1)^{\sf F} \neq e^{\i \pi J^3}$ since $[Q^3,J^3]=0$ while $\{ Q^3, (-1)^{\sf F}\} =0$. The two possible indices we can define are
\beq
{\sf Index}=\text{Tr} \Big( e^{-\beta H}\,  (-1)^{\sf F} \Big),~~~~~~{\sf Index}'=\text{Tr} \Big( e^{-\beta H}\,  e^{\i \pi J^3} \Big).
\eeq
and for theories that satisfy the same commutation relations as the $\mathcal{N}=3$ Schwarzian theory, the two indices are equal. We can verify that $\tilde{Z}(\beta,1/4)=\frac{1}{2}e^{S_0}$ and therefore 
$$
{\sf Index}={\sf Index}' = e^{S_0} \delta_{\nu,0} \, ,
$$
reproducing the ground-state degeneracy. The advantage of ${\sf Index}'$ is that it leads to a clear bulk interpretation. Since this is equal to the grandcanonical partition function with a specific $\SO(3)$ fugacity, the bulk saddle is (in principle) a charged black hole with a smooth horizon. The bulk interpretation of ${\sf Index}$ is less clear, but since it takes the same value as ${\sf Index}'$ we will not worry about it.

\subsection{Linearization of the model, its solution, and the index}

It is possible to linearize the commutation relations between the charges by incorporating an additional field, a single free decoupled Majorana fermion. The explicit action of the full theory can be found in equation 5.20 of \cite{Kozyrev:2021agn}, but its explicit form will not be important. The larger linearized theory has an extra odd conserved charge which we denote $\tilde{Q}$ and is a singlet under $\SO(3)$. 

The algebra satisfied by the generators $H$, $J^i$, $Q^i$ and $\tilde{Q}$ can be determined as follows. Most commutators are fixed by $\SO(3)$ transformation properties. The only nontrivial ones to determine are the anticommutators of two supercharges. Assuming linearity, its most general form is
\beq
\{ Q^i, Q^j\} = \delta^{ij} H ,~~~~~~\{ \tilde{Q},\tilde{Q} \} = 8 \Phi_r,~~~~~~\{\tilde{Q}, Q^i\} =J^i.
\eeq
 The RHS of the middle equation is a free parameter which we relate to the Schwarzian coupling with some hindsight. The non-standard anticommutator between $Q^i$ and $\tilde{Q}$ effectively leads to the same non-linear relation between the BPS energy and the spin. We can see this by mapping this algebra to the previous non-linear relation via 
\beq
Q^i \to  Q_{NL}^i=Q^i - \frac{1}{8\Phi_r} \tilde{Q} J^i.
\eeq
This transforms $\{\tilde{Q}, Q^i\} = J^i $ into $\{ \tilde{Q}, Q_{NL}^i\}=0$ while the anticommutator between $Q^i$ with itself becomes $\{Q_{NL}^i,Q_{NL}^j\} =\delta_{ij} H  -\frac{1}{16 \Phi_r} \{ J^i, J^j\}$. This leads to the $\mathcal{N}=3$ algebra we studied above with a nonlinear relation, supplemented by a decoupled supercharge $\tilde{Q}$. Of course, the physical theory does not care how we choose to describe symmetry generators. 

The representations of the new linear algebra generated by $Q^i,\tilde{Q}$ and $J^i$ are easy to construct. The representations of $Q^i,J^i$ are the same as what we determined in the previous sections. The only modification is that now all states $|E,j\rangle$  are supplemented by their partner $\tilde{Q}|E,j\rangle$. Since $\{\tilde{Q},\tilde{Q}\} = 8 \Phi_r$, the operator $\tilde{Q}$ cannot annihilate any state and all states are two-fold degenerate. Even the fully BPS state we found earlier $(j=0)$ now has this two-fold degeneracy. This means that the index always vanishes since $\{(-1)^{\sf F},\tilde{Q}\}=0$, namely
\beq
{\rm Tr}\,(-1)^{\sf F} e^{-\beta H} = 0.
\eeq
This does not mean we cannot construct a protected partition function capturing the presence of BPS states (at least for the non-anomalous theory). Since the extra fermion in the linear theory decouples, its partition function is very simply related to the partition function we computed earlier 
\beq
Z_{\rm linear}(\beta,\alpha) = \sqrt{2} \cdot Z_{\rm non-linear}(\beta, \alpha).
\eeq
This implies that when $\alpha=1/2$ the partition function is again protected. The non-BPS characters are $2\cdot 2\cdot (\chi_j+\chi_{j-1})=0$ while the BPS characters are $2\cdot(\chi_j+\chi_{j-1}) = 0$ for partially BPS states and $2\cdot \chi_0= 2$ for fully BPS states. Therefore we conclude 
\beq
{\rm Tr} \, (-1)^{\sf F}\,  e^{-\beta H} = 0,~~~~~{\rm Tr}\,  e^{\i\pi J^3}\, e^{-\beta H} \neq 0.
\eeq
Therefore $Z_{\rm linear}(\beta,\alpha=1/2)$ is a protected quantity capturing fully-BPS states. (As far as we know we cannot define a protected quantity capturing partial BPS states.) The explanation is clear, since $[J^i,\tilde{Q}]=0$ while $\{ (-1)^{\sf F}, \tilde{Q}\}=0$, this again implies that $(-1)^{\sf F} \neq e^{\i \pi J}$. In the context of holographic 2d CFT this construction is lifted to the modified elliptic genus introduced in \cite{Eberhardt:2018sce}.

\section{Black holes with large $\mathcal{N}=4$}\label{sec:LN4SCH}

In this section we will consider black holes with large $\mathcal{N}=4$ supersymmetry, which is based on the superconformal algebra $\text{D}(2,1|\upalpha)$. This has 4 supercharges and the global symmetry $\SU(2)_+ \times \SU(2)_-$. Unlike the case for $\mathcal{N}=3$, there is a unique such theory (in the sense that there is no anomaly or discrete choices). However, as is well known, the choice of large $\mathcal{N}=4$ algebra does contain an additional continuous parameter called $\upalpha$ (which is quantized in string theory). 

A particularly interesting example of large $\mathcal{N}=4$ superconformal symmetry arises in string theory backgrounds such as $AdS_3 \times S^3 \times S^3 \times S^1$. These backgrounds (as well as their dual CFT's) are fairly special, and many properties are poorly understood. We will make specific contact with the precise string 
 theory constructions in upcoming work~\cite{WOP_US}, where we will analyze the supergravity background and gravitational index in detail, using our results for the large $\mathcal{N}=4$ Schwarzian theory from this section in an essential way.

\subsection{Solution of the model}
\label{sec:solutionD21}

According to section \ref{sec:GPF} the partition function is
\bea
Z &=& \sum_{n\in\mathbb{Z}}\sum_{m\in\mathbb{Z}} \frac{\Phi_r^{1/2}}{\beta^{1/2}}\frac{(\alpha_++n)}{\sin 2\pi \alpha_+ }\frac{(\alpha_-+m)}{\sin 2\pi \alpha_- }\, \cos \pi (\alpha_+ + \alpha_-) \cos \pi (\alpha_+ - \alpha_-)\nonumber\\
&&\times \frac{e^{{S_0+\frac{2\pi^2\Phi_r}{\beta}(1- \frac{4}{\gamma_-}(\alpha_+ +n)^2-\frac{4}{\gamma_+}(\alpha_- +m)^2)}}}{(1-4(\alpha_+ +\alpha_- + m + n)^2)(1-4(\alpha_+ -\alpha_- - m + n)^2)}.\label{eqn:LN422}
\ea
The power of temperature appearing in the one-loop determinant comes from the fact that $D(2,1|\upalpha)$ has $8$ fermionic and $9$ bosonic generators. The rest of the first line comes from the two $\SU(2)$ degrees of freedom and the fermions. The second line includes the classical action, and a denominator that removes fermion zero-modes from the one-loop determinants. Besides the Schwarzian coupling $\Phi_r$ the theory depends on one continuous parameter, since $\gamma_+ + \gamma_-=1$\footnote{In section \ref{ssec:exampleZsection}, we wrote this in terms of $\upalpha$ rather than introducing separate gauge couplings $\gamma_\pm$.}, which we can take to be their ratio $\upalpha=\gamma_-/\gamma_+$.


We first identify the conserved charges of the theory. Time translations are generated by the Hamiltonian $H$ and $\SU(2)_+ \times \SU(2)_-$ transformations are generated by
\beq
J_+{}^i,~~~J_-{}^i,~~~~i=1,2,3 \, ,
\eeq
which have the standard (independent) $SU(2)$ algebras. The presence of these bosonic generators implies, at the very least, that the partition function should be expanded into energy eigenstates where each comes in a specific representation of two copies of $\SU(2)$. The four fermionic generators can be labeled by 
\beq
Q_{A\dot{A}},~~~~A,\dot{A}=1,2 \, ,
\eeq
with the index $A$ and $\dot{A}$ transforming in the bifundamental of $\SU(2)_+ $ and $\SU(2)_-$, respectively. We can pick a basis such that $Q_{1\dot{B}}$ ($Q_{2\dot{B}}$) has eigenvalue $+1/2$ ($-1/2$) with respect to $J_+{}^3$, and similarly for the second index. We can also describe the $\SU(2)$ generators in the same language as $J_A{}^B = (\sigma^i)_{A}{}^B J_+^i$ and  $J_{\dot{A}}{}^{\dot{B}} = (\sigma^i)_{\dot{A}}{}^{\dot{B}} J_-^i$, where $\sigma^i$ are the Pauli matrices.

The information above fixes all the commutator and anticommutator relations except $\{ Q_{A\dot{A}}, Q_{B\dot{B}}\}$ which we will describe later. If the algebra was linear, we could fix $\{ Q_{A\dot{A}}, Q_{B\dot{B}}\} = \varepsilon_{AB} \varepsilon_{\dot{A}\dot{B}} H$, with $\varepsilon_{AB}$ and $\varepsilon_{\dot{A}\dot{B}}$ antisymmetric tensors with $\varepsilon_{12}=\varepsilon_{\dot{1}\dot{2}}=1$. This is the correct relation for the small $\mathcal{N}=4$ Schwarzian theory based on $\PSU(1,1|2)$, but we will see that the large $\mathcal{N}=4$ one requires a non-linear generalization similar to the one we introduced in \ref{ssec:nonlinearN3alg}.

A quite lengthy calculation which we leave for Appendix \ref{app:N4} shows that the partition function \eqref{eqn:LN422} can be expanded in the following way
\bea
Z &=& \sum_{j_+,j_-\geq 1/2} \chi^{\rm long}_{j_+j_-}(\alpha_+,\alpha_-)\, \int \d E\, e^{-\beta E} \rho_{j_+j_-}(E)\nonumber\\
&&~~~~+ \sum_{j_+,j_-\geq0} \chi^{\rm short}_{j_+ j_-}(\alpha_+,\alpha_-) e^{-\beta E_{\rm BPS}(j_+,j_-)} N_{j_+j_-}.\label{eq:ZN4EXP}
\ea
There are two terms corresponding to two types of supermultiplets. We will explain the origin of these multiplets later after specifying the algebra between the generators. The first line involves multiplets with a specific combination of $\SU(2)_+ \times \SU(2)_-$ representations 
\beq\label{eq:LN4LONGSMS}
\text{Long}_{j_+j_-}=\Big(j_- - 1, j_+ - \frac{1}{2}\Big) \oplus \Big(j_- , j_+ - \frac{1}{2}\Big)\oplus  \Big( j_- - \frac{1}{2},j_+ -1\Big)\oplus \Big(j_--\frac{1}{2}, j_+\Big) 
\eeq
This can be illustrated by the diagram
\beq
\begin{tikzpicture}[scale=1,baseline={([yshift=0cm]current bounding box.center)}]
\draw[very thick,->] (0,0) to (0,4);
\draw[very thick,->] (0,0) to (4,0);
\draw[left] (0,3) node {\footnotesize $j_-$};
\draw[left] (0,2) node {\footnotesize $j_--\frac{1}{2}$};
\draw[left] (0,1) node {\footnotesize $j_--1$};
\draw[below] (3,0) node {\footnotesize $j_+$};
\draw[below] (2.05,0) node {\footnotesize $j_+-\frac{1}{2}$};
\draw[below] (1,0) node {\footnotesize $j_+-1$};
\draw[very thick] (-0.05,3) to (0.05,3);
\draw[very thick] (-0.05,2) to (0.05,2);
\draw[very thick] (-0.05,1) to (0.05,1);
\draw[very thick] (3,-0.05) to (3,0.05);
\draw[very thick] (2,-0.05) to (2,0.05);
\draw[very thick] (1,-0.05) to (1,0.05);
\draw[thick,->] (3,2) to (2,3-0.05);
\draw[above] (3,2.5) node {\footnotesize $Q_{2\dot{1}}$};
\draw[fill=black] (2,3) circle (0.06);
\draw[fill=black] (2,1) circle (0.06);
\draw[fill=black] (1,2) circle (0.06);
\draw[fill=black] (3,2) circle (0.06);
\end{tikzpicture}
\eeq
 This is the naive structure of a supermultiplet generated by $Q_{A\dot{A}}$ which shift the spins along the diagonals (we show one example above). Notice that the labels $j_+$ and $j_-$ do not label the $\SU(2)_+ \times \SU(2)_-$ representations of any state in the multiplet. This happens because there is no state in the representation that is highest-weight with respect to all generators. The contribution to the partition function is
\bea
\chi^{\rm long}_{j_+ j_-}(\alpha_+,\alpha_-)&=& 4\cos \pi(\alpha_+ + \alpha_-) \cos \pi (\alpha_+-\alpha_-) \frac{\sin (2 j_+)2\pi \alpha_+}{\sin 2\pi \alpha_+}\frac{\sin (2 j_-)2\pi \alpha_-}{\sin 2\pi \alpha_-},\nonumber\\
&=& \chi_{j_--1}\chi_{j_+-\frac{1}{2}}+\chi_{j_-}\chi_{j_+-\frac{1}{2}}+\chi_{j_--\frac{1}{2}}\chi_{j_+-1}+\chi_{j_--\frac{1}{2}}\chi_{j_+},\label{eqn:N4LONGCHAR}
\ea
where we omit the $\alpha_\pm$ dependence to avoid cluttering. There are some special cases of non-BPS multiplets that are accidentally shorter without preserving any supercharge. These are
\bea
\text{Long}_{j_+=\frac{1}{2}\,j_-}&=&\Big(j_- - 1, 0\Big) \oplus \Big(j_- ,0\Big)\oplus  \Big(j_--\frac{1}{2},\frac12\Big) ,\\
\text{Long}_{j_+\,j_-=\frac{1}{2}}&=&  \Big(\frac12 , j_+ - \frac{1}{2}\Big)\oplus  \Big( 0,j_+ -1\Big)\oplus \Big(0, j_+\Big),\\
\text{Long}_{\frac12\,\frac{1}{2}}&=&  \Big(\frac12 , 0\Big)\oplus  \Big(0, \frac12\Big).
\ea
There are no multiplets labeled by either $j_+$ or $j_-$ equal to zero. This is similar to the $1/2 \oplus 0$ supermultiplet in the small $\mathcal{N}=4$ theory. For all other $j_-,j_+>1/2$ the expression \eqref{eq:LN4LONGSMS} is valid. The expressions we find below for the spectrum are valid for all $j_+,j_-$ as written, with the understanding that the cases above behave slightly differently.

The second line of \eqref{eq:ZN4EXP} involves shorter multiplets with spins
\beq
\text{Short}_{j_+j_-}=\Big(j_-,j_+\Big) \oplus \Big(j_- - \frac{1}{2}, j_+ - \frac{1}{2}\Big)
\eeq
represented by the diagram:
\beq
\begin{tikzpicture}[scale=1,baseline={([yshift=0cm]current bounding box.center)}]
\draw[very thick,->] (0,0) to (0,4);
\draw[very thick,->] (0,0) to (4,0);
\draw[left] (0,2.5) node {\footnotesize $j_-$};
\draw[left] (0,1.5) node {\footnotesize $j_--\frac{1}{2}$};
\draw[below] (2.5,0) node {\footnotesize $j_+$};
\draw[below] (1.5,0) node {\footnotesize $j_+-\frac{1}{2}$};
\draw[fill=black] (2.5,2.5) circle (0.06);
\draw[fill=black] (1.5,1.5) circle (0.06);
\end{tikzpicture}
\eeq
The diagram illustrates that these short multiplets preserve two supercharges and are annihilated by $Q_{1 \dot{2}}$ and $Q_{2 \dot{1}}$ making them partially BPS. The following case is special. When one of the $\SU(2)$ spin is zero the multiplet can become even shorter
\ie
\text{Short}_{j\,0}=(j,0),~~~\text{Short}_{0\,j}=(0,j),
\label{eqn:N4shorter}
\fe
and preserve four supercharges instead. The expressions we find below are valid for all $j_+,j_-$ including $j_+=0$ or $j_-=0$ as written, with the understanding that these cases are special. The presence of these fully BPS multiplets is important in order to recover the small $\mathcal{N}=4$ spectrum from the large one, since the BPS states of the former preserve all supercharges.

We find from the exact calculation \eqref{eq:largeN4disDOS} that the energy of these short multiplets is 
\beq
E_{\rm BPS}(j_-,j_+) = \frac{\upalpha}{2\Phi_r(1+\upalpha)^2}\left( j_+ + j_- + \frac{1}{2}\right)^2.\label{eq:N4LBPSEE}
\eeq 
A careful derivation requires knowledge of the full algebra of generators which we will present later. From the picture, it becomes clear that a long multiplet labeled by $(j_-,j_+)$ decomposes into two short multiplets with $(j_-,j_+-1/2)$ and $(j_--1/2,j_+)$. Their contribution to the partition function is a shorter character given by
\bea
\chi^{\rm short}_{j_+ j_-}(\alpha_+,\alpha_-)&=&\chi_{j_-}(\alpha_-) \chi_{j_+}(\alpha_+) +\chi_{j_--\frac{1}{2}}(\alpha_-) \chi_{j_+-\frac{1}{2}}(\alpha_+)
\ea
Before explaining the derivation of this structure, let us continue to present the result obtained from a direct evaluation using \eqref{eqn:LN422}.


The long non-BPS supermultiplet has a density of states
\bea
\rho_{j_+j_-}(E) &=&  e^{S_0}\frac{\upalpha^{3/2}\,j_+ j_-}{32\Phi_r\sqrt{2\pi}(1+\upalpha)^3}  \left(E-\frac{\upalpha (j_+- j_-)^2}{ 2\Phi_r(1+\upalpha)^2}\right)^{-1}\left(E-\frac{\upalpha (j_++ j_-)^2}{ 2\Phi_r(1+\upalpha)^2}\right)^{-1}\nonumber\\
&&\times \sinh \left(2\pi \sqrt{2\Phi_r \left(E-\frac{\upalpha j_+^2 + j_-^2}{2\Phi_r(1+\upalpha)}\right)}\right),\label{eq:NL4DOS}
\ea
and is nonvanishing only when the argument of the square root is positive
$$
E\geq E_0(j_+,j_-)= \frac{\upalpha j_+^2 + j_-^2}{2\Phi_r(1+\upalpha)}.
$$
This threshold energy is a consequence of the dynamics of the model and cannot be derived from the kinematics. Naively it seems that our expression for the supermultiplet density of states can become negative due to the denominators. Fortunately the theory avoids any potential non-unitarity since 
\beq
\frac{\upalpha j_+^2 + j_-^2}{(1+\upalpha)} - \frac{\upalpha(j_+ \pm j_-)^2}{(1+\upalpha)^2} = \frac{(\upalpha j_+ \pm j_-)^2}{(1+\upalpha)^2} \geq 0.
\eeq
The only potentially problematic case can be when the quantity above vanishes. For irrational values of $\upalpha$, given that $2j_\pm \in \mathbb{Z}$, this cannot happen. But given that string theory constructions such as $\text{AdS}_3 \times S^3 \times S^3\times S^1$ lead to a rational $\upalpha$, it is worth analyzing this case. It is easy to see from the expressions above that for supermultiplets with $\upalpha j_+ =j_-$ the typical square-root edge for arbitrary multiplets becomes an inverse square-root edge. We will explain this later as arising from the algebra; a BPS bound is about to be saturated.


Let us now analyze the consequences of \eqref{eqn:LN422} for the BPS spectrum. From our expression for the partition function, we find that short BPS multiplets with maximal spin $(j_-,j_+)$ have a fixed energy in terms of their spin $E_{\rm BPS}(j_-,j_+)$. The degeneracy of BPS states is given by 
\beq\label{eqn:NBPS}
N_{j_+j_-} =\begin{cases}
    e^{S_0} \frac{\sqrt{\pi\upalpha/2} }{32(1+\upalpha)} \left|\sin \left( \frac{\pi \upalpha(2j_++2j_-+1)}{1+\upalpha} \right)\right|,~~~~\text{if}   ~~~(j_+,j_-)\in R_{\rm BPS},\\
    \hspace{1cm} 0 \hspace{4.8cm}\text{if}   ~~~(j_+,j_-)\notin R_{\rm BPS},
    \end{cases}
\eeq
 $R_{\rm BPS}$ is defined by the equation above, as the set of short multiplets that present BPS states. We find
\beq
R_{\rm BPS} = \left\{ (j_+,j_-) \left| \, 0< \frac{2\upalpha j_+ - 2 j_- +\upalpha}{1+\upalpha}<1\right.\right\}.
\label{eq:BPSrangeD21}
\eeq
Let us emphasize that this result is inherently quantum mechanical. We can study the spectrum of BPS states in the classical approximation of the theory, which leads to $j_- \approx \upalpha j_+$ for large $j_\pm$.\footnote{A shortcut is to notice that the exponential term in the RHS of equation \eqref{eqn:LN422}, for $n=m=0$, is  the classical grand canonical potential. From this, we can extract the classical BPS mass and charge spectra by the usual rules of black hole thermodynamics.} Equation \eqref{eq:BPSrangeD21} gives the quantum BPS spectrum (the allowed set of R-charges), with order one accuracy in the spin, that has a degeneracy growing as $e^{S_0}$.

To describe the allowed spins of the BPS states more explicitly, consider multiplets with fixed total spin $j_+ + j_- = j$ and work with one value of $j$ at a time. For fixed $j$ we have $2j+1$ ways of distributing the total spin between the two $\SU(2)$ generators, namely 
\beq
\Big(j,0\Big),\,\Big(j-\frac{1}{2},\frac{1}{2}\Big),\,\ldots,\, \Big(\frac{1}{2},j-\frac{1}{2}\Big),\,\Big(0,j\Big).
\eeq
We find that for a given $j$ there is a unique choice of $(j_+,j_-)$ carrying all BPS states, but the precise way the total spin is distributed among the two $\SU(2)$ spins depends on the parameter $\upalpha$. For $\upalpha \sim 0$ our expression for $R_{\rm BPS}$ implies that the only multiplets with BPS states can be $(j_+=j,j_-=0)$. When we start raising $\upalpha$ there is a jump at $\upalpha=1/(2j)$ since the combination that appears in $R_{\rm BPS}$ approaches its upper bound. At $\upalpha=1/(2j)$ the representation with BPS states jumps from $(j,0)$ to $(j-1/2,1/2)$. As we continue to increase $\upalpha$ we encounter a finite number of jumps at
\beq
\upalpha_* = \frac{n}{2j_++2j_-+1-n},~~~n\in\mathbb{Z},
\eeq
where the spin of the representation with BPS states changes
\beq
\left( j-\frac{n}{2}, \frac{n}{2}\right) \to \left( j-\frac{n+1}{2}, \frac{n+1}{2}\right),
\eeq
This happens until we reach $\upalpha=2j$ and we transition from $(1/2,j-1/2)$ to $(0,j)$ at $\upalpha = 2j$. For any $\upalpha>2j$ the BPS spin is $(0,j)$. We emphasize that for a fixed value of $\upalpha$ and fixed $j=j_++j_-$ there is a unique BPS spin. Finally, we can check that the transition values $\upalpha_*$ take place precisely at the zeros of \eqref{eqn:NBPS}. 

 \begin{figure}[b]
\begin{center}
\includegraphics[scale=.45]{"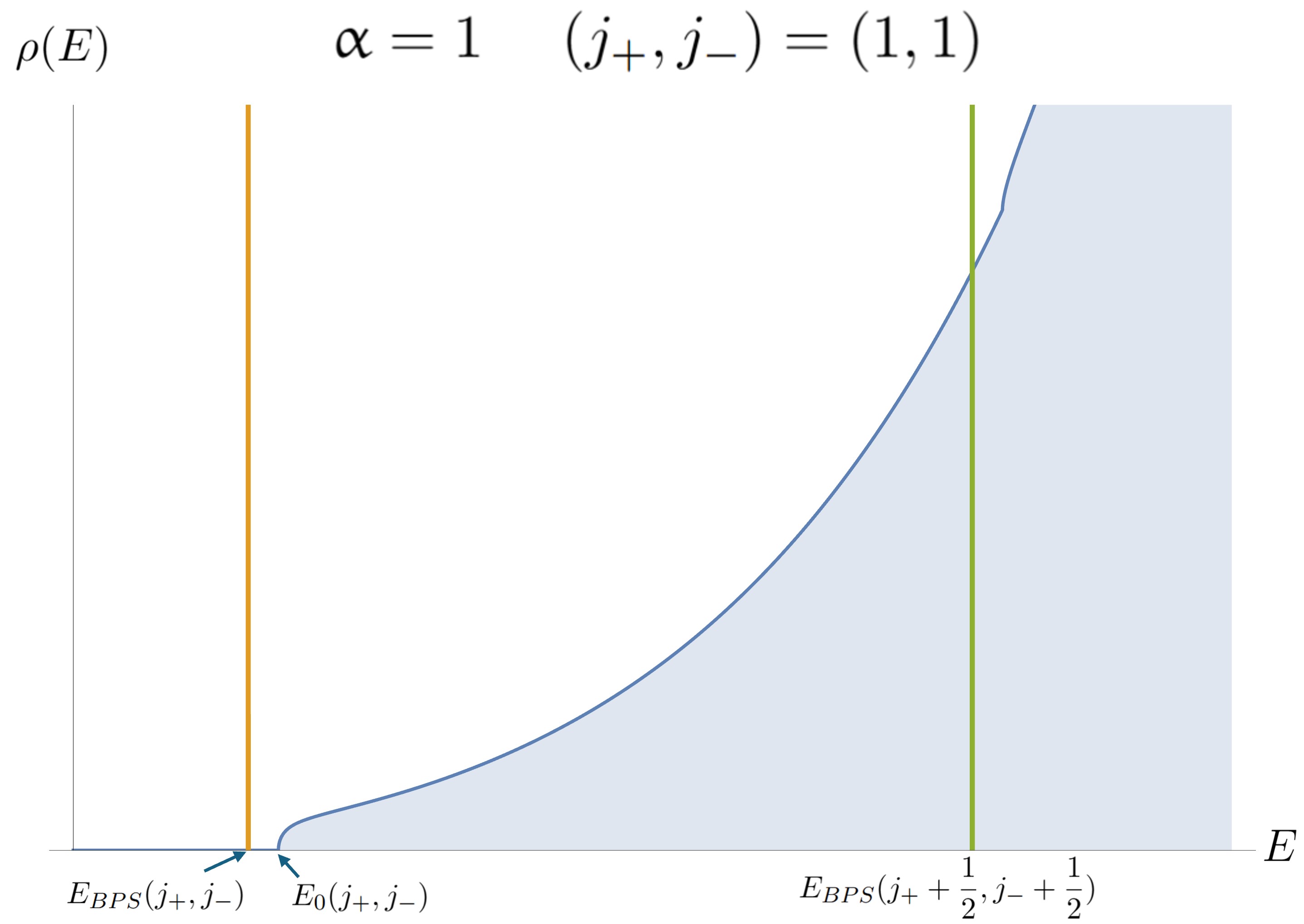"}
\caption{The microcanonical density of states for the large $\mathcal{N}=4$ Schwarzian theory at fixed $(j_+,j_-)=(1,1)$ and $\upalpha = 1$. The blue curve represents the contribution of all states (coming from four different long multiplets) which have these fixed R-charges. The edge $E_0$ is determined by the smallest edge of any multiplet containing this charge sector. The yellow and green delta functions come from two distinct short multiplets which each have a $(1,1)$ BPS state. At this value of $\upalpha$, both sets of BPS states are allowed and they have a gap to their respective non-BPS continuum.} \label{fig:N4a1}
\end{center}
\end{figure}

We can compare this phenomenon we found in large $\mathcal{N}=4$ to wall crossing.  $\upalpha$ is a continuous parameter and as we vary it, BPS states disappear from the spectrum and new BPS states appear with different $\SU(2)_- \times \SU(2)_+$ charges. In the context of the supergravity description of the $AdS_3 \times S^3 \times S^3 \times S^1$ background, this is a deformation that changes the relative sizes of the two 3-spheres. Of course, in the non-perturbative description the radii of the spheres in string units take discrete values and $\upalpha$ is not a continuous parameter since it takes rational values. Nevertheless, our analysis implies that as this parameter changes, the nature of the BPS states can jump.


Is there a gap between the BPS states and the non-BPS multiplets? To answer this question, it is important to compare the values of $E_{\rm BPS}(j_+,j_-)$ and $E_0 (j_+,j_-)$ for the correct multiplets. This can be clarified by analyzing how a long multiplet decomposes into a short multiplet 
\beq
\text{Long}_{j_+,j_-} = \text{Short}_{j_+,j_--\frac{1}{2}} + \text{Short}_{j_+-\frac{1}{2}, j_-}
\eeq
Therefore, to analyze whether there is a gap or not, one should compute
\beq
E_0(j_+,j_-)-E_{\rm BPS}(j_+-\frac{1}{2},j_-) = \frac{(\upalpha j_+ - j_-)^2}{2\Phi_r(1+\upalpha)^2} \geq0 
\eeq
and
\beq
E_0(j_+,j_-)-E_{\rm BPS}(j_+,j_--\frac{1}{2}) = \frac{(\upalpha j_+ - j_-)^2}{2\Phi_r(1+\upalpha)^2} \geq 0 
\eeq
The gap can only vanish when $\upalpha$ is rational and for spins such that $\upalpha j_+ = j_-$. As observed earlier, this is precisely the supermultiplet for which the density of states has an inverse-square-root edge. This is again consistent with expectations from random matrix behavior. Moreover, for those states with vanishing gaps, the number of BPS states automatically vanishes $N_{j_+-1/2,\upalpha j_+} \propto |\sin (2 \pi j_-) |= 0$ since $2j_-\in\mathbb{Z}$.

 \begin{figure}[h]
\begin{center}
\includegraphics[scale=.45]{"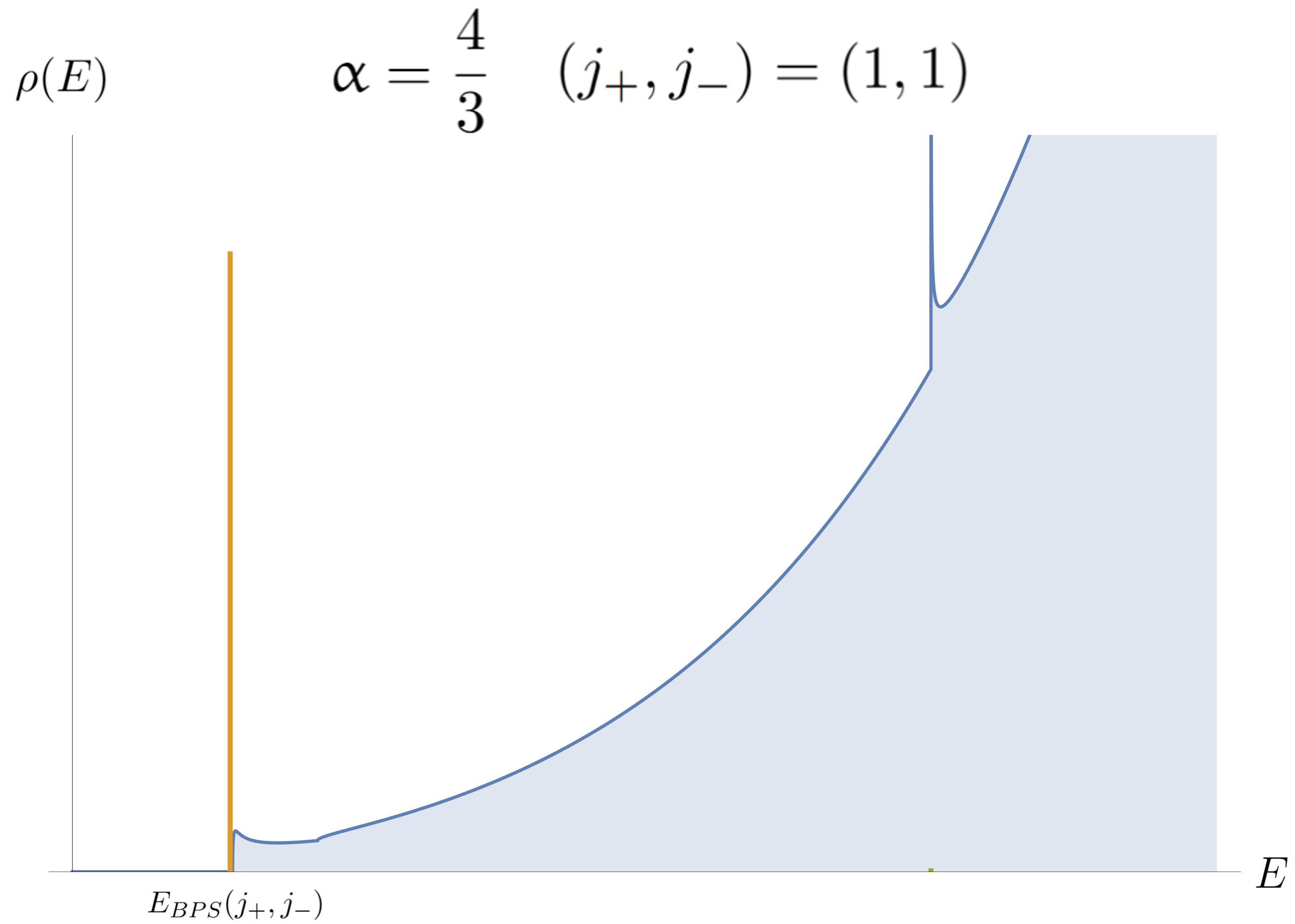"}
\caption{The microcanonical density of states for the large $\mathcal{N}=4$ Schwarzian theory at fixed $(j_+,j_-)=(1,1)$ and $\upalpha = \frac43$. As in the previous figure, the blue curve represents the contribution of different long multiplets which have these R-charges. In contrast, the edge $E_0$ is now colliding with a BPS bound (there remains a small gap which is not visible). The yellow delta function indicates that there is still a short multiplet with $(1,1)$ BPS states, but for this value of $\upalpha$, the second set of BPS states (green curve above) is absent. These would-be BPS states are instead replaced by long multipets approaching the BPS bound.} \label{fig:N4a43}
\end{center}
\end{figure}

 \begin{figure}[h]
\begin{center}
\includegraphics[scale=.45]{"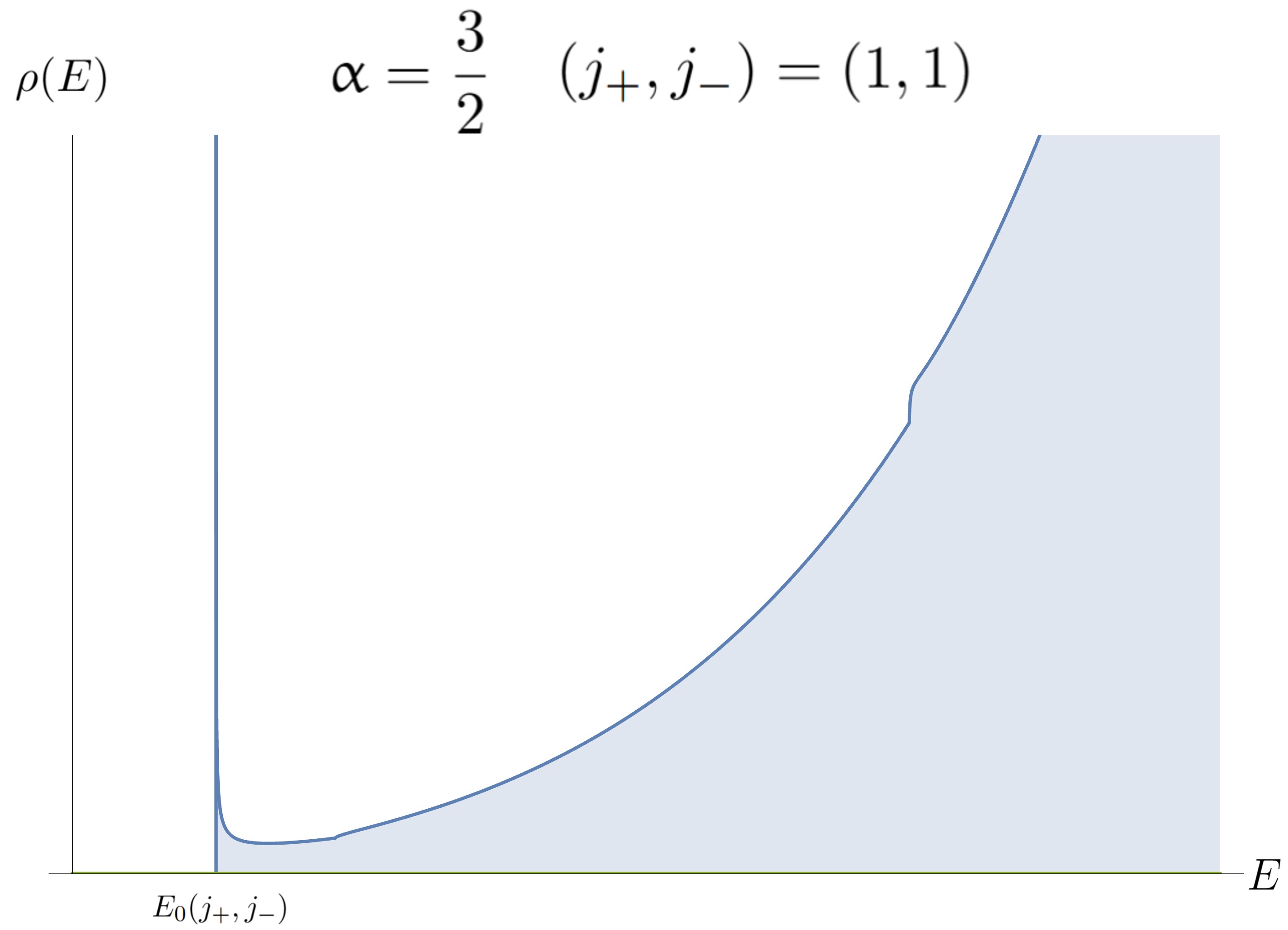"}
\caption{The microcanonical density of states for the large $\mathcal{N}=4$ Schwarzian theory at fixed $(j_+,j_-)=(1,1)$ and $\upalpha = \frac32$. Comparing with the previous figures, we see that all BPS states are now excluded for this value of $\upalpha$ and $(j_+,j_-)$. Instead we either have a shifted (gapped) set of long multiplets, or long multiplets colliding with the BPS bound. There are no BPS states in this figure, but the spike indicates many non-BPS states are close to, but do not saturate, the bound.} \label{fig:N4a32}
\end{center}
\end{figure}

The spectrum above simplifies in the case $\upalpha=1$ where the large $\mathcal{N}=4$ theory based on the superconformal group $\text{D}(2,1|\upalpha)$ reduces to $\OSp(2|4)$. In this case, all BPS states have $j_+= j_-$ and the BPS degeneracy is $N_{j_+,j_-}= e^{S_0} \sqrt{\pi}/4$. This is reasonable since for $\upalpha=1$ the two factors of the $R$-symmetry are completely symmetric under permutation $\SU(2)_+ \leftrightarrow \SU(2)_-$. The multiplets $\text{Long}_{j,j}$ have vanishing gap. These multiplets decompose as $\text{Short}_{j\pm\frac{1}{2},j}$ and therefore have no BPS states. Nevertheless let us emphasize that the Schwarzian based on $\text{D}(2,1|\upalpha)$ and $\OSp(4|2)$ are different since the former involves the $R$-symmetry $\text{Spin}(4)$ (and therefore includes half-integer spins) while the latter involves $\SO(4)$ and only has bosonic states.

\subsection{Commutation relations of conserved charges}
\label{sec:commutatorofcharges}
Next we look at the conserved charges and its supermultiplet structure. We could determine the algebra via Poisson bracket of the symmetry generators. Instead, similar to the $\mathcal{N}=3$ case, we will assume a minimal modification by allowing quadratic terms appearing in the anticommutator $\{Q_{A\dot{A}},Q_{B\dot{B}}\}$, imposing Jacobi identities and comparing with the spectrum we calculated. 


Let us first use the $\SU(2)$ invariance of the model. The right hand side of $\{Q_{A\dot{A}},Q_{B\dot{B}}\}$ should involve tensors with the right indices that moreover have to be symmetric under simultaneous $A\leftrightarrow B$ and $\dot{A}\leftrightarrow\dot{B}$. This allows for the structure $\varepsilon_{AB}\varepsilon_{\dot{A}\dot{B}}$, multiplied by any operator that is a singlet of $\SU(2)_+ \otimes \SU(2)_-$. If we restrict to at most quadratic deformations of the algebra, there are three types of possibilities. First, we can add
\beq
\{ Q_{A\dot{A}},Q_{B\dot{B}}\}  \supset \varepsilon_{AB}\varepsilon_{\dot{A}\dot{B}} H,~\varepsilon_{AB}\varepsilon_{\dot{A}\dot{B}} J_+^2,~\text{or}~\varepsilon_{AB}\varepsilon_{\dot{A}\dot{B}} J_-^2,
\eeq
 where $J_+^2 = \frac{1}{2}J_{A}^{B} J_{B}^{A}$ and $J_-^2=\frac{1}{2}J_{\dot{A}}^{\dot{B}} J_{\dot{B}}^{\dot{A}}$ are the $\SU(2)$ quadratic Casimir operators. The second possible structure is 
 \beq
\varepsilon_{AB}J_{\dot{A}\dot{B}},~~\text{or}~~\varepsilon_{\dot{A}\dot{B}}J_{AB}.
 \eeq
These two structures are ruled out by the fact that the anticommutator has to be symmetric under $A\leftrightarrow B$ and $\dot{A}\leftrightarrow \dot{B}$. The tensor $\varepsilon$ is obviously antisymmetric while $J_{AB}$ and $J_{\dot{A}\dot{B}}$ are symmetric.  The third possible structure is $J_{AB} J_{\dot{A}\dot{B}}$ which transforms in the same representation of $\SU(2)_+ \times \SU(2)_-$ as $\{ Q_{A\dot{A}},Q_{B\dot{B}}\}$ and is symmetric. The most general \emph{non-linear} extension of the algebra is 
\beq
\{ Q_{A\dot{A}},Q_{B\dot{B}}\}  \supset \updelta \varepsilon_{AB}\varepsilon_{\dot{A}\dot{B}} J_+^2 +\upbeta \varepsilon_{AB}\varepsilon_{\dot{A}\dot{B}} J_-^2 + \upgamma 2 J_{AB} J_{\dot{A}\dot{B}},
\eeq
with $\updelta$, $\upbeta$ and $\upgamma$ arbitrary coefficients. Imposing Jacobi identities between the supercharges $[\{Q_{A\dot{A}},Q_{B\dot{B}}\},Q_{C\dot{C}}]+\text{(cyclic)}=0 $ leaves only one free parameter since it constrains $\updelta=\upbeta=\upgamma$. The final version of the algebra is
\beq
\{ Q_{A\dot{A}},Q_{B\dot{B}}\} = -\varepsilon_{AB}\varepsilon_{\dot{A}\dot{B}} H  + \upgamma \left\{\varepsilon_{AB}\varepsilon_{\dot{A}\dot{B}} \left[J_+^2 + J_-^2+ \frac{1}{4} \right] - 2 J_{AB}J_{\dot{A}\dot{B}} \right\},
\eeq
where we also shifted the Hamiltonian by $H \to H-\upgamma/4$ for later convenience. 


The last step is to determine $\upgamma$ for the specific case of the Schwarzian theory, by comparing the BPS spectrum. Imposing that $\{Q_{1\dot{2}},Q_{2\dot{1}}\}$ is a positive operator acting on a highest-weight state with $J_{12}=J_+^3=j_+$ and $J_{\dot{2}\dot{1}}=J_-^3=j_-$ leads to 
\beq
E - \upgamma ( j_+(j_++1)+j_-(j_-+1) +2 j_+ j_- + 1/4) = E - \upgamma (j_+ + j_- + 1/2)^2 \geq 0.
\eeq
This bound is saturated by partially BPS states that are annihilated by $Q_{1\dot{2}}$ and $Q_{2\dot{1}}$, and a comparison with \eqref{eq:N4LBPSEE} implies that 
\beq
\upgamma = \frac{\upalpha}{2\Phi_r(1+\upalpha)^2}.
\eeq
We have therefore constructed the minimal modification of the algebra that accommodates the spectrum found by directly quantizing our theory. As for the case with three supercharges, this result can be verified by relating our theory to large $\mathcal{N}=4$ Virasoro CFT, similarly to \cite{Mertens:2017mtv}. Because of the close relationship of this calculation to $AdS_3$ gravity, we will perform this in a companion paper~\cite{WOP_US} which focuses on the string background and predictions for the dual CFT$_2$.

\subsection{The index for large $\mathcal{N}=4$}\label{sec:LN4INDEX}

We shall analyze the protected supersymmetric index one can define for theories with large $\mathcal{N}=4$ supersymmetry. Its properties are very different than those of less supersymmetric cousins. By index, we mean a partition function which receives contributions solely from BPS states. Therefore, we need to determine a choice of boundary conditions such that the contribution from all supermultiplets $\text{Long}_{j_+j_-}$ vanishes. 

From equation \eqref{eqn:N4LONGCHAR} it is evident that the contributions from non-BPS states will cancel as long as 
\beq
\cos \pi (\alpha_++\alpha_-) \cos \pi (\alpha_+ - \alpha_-) = 0.
\eeq
This will happen as long as
\beq
\alpha_- = \alpha_+ + 1/2.
\eeq
This choice is unique up to the trivial symmetries of the partition function under $\alpha_+ \to - \alpha_+$ and $\alpha_+ \to \alpha_++1$, and similarly for the $\SU(2)_-$ chemical potential. The second property we should check is that the contribution from BPS multiplets $\text{Short}_{j_+j_-}$ are non-vanishing. We obtain
\beq
\chi_j^{\rm short} (\alpha_+,\alpha_++1/2) = \frac{\sin 4\pi \alpha_+(j_-+j_++\frac{1}{2})}{\sin 2\pi \alpha_+}(-1)^{2j_-}. \label{eq:cal}
\eeq


The supersymmetric partition function implied by the previous analysis is
\bea
{\sf Index}(\sigma) &=& Z(\sigma,\sigma+1/2),\nonumber\\
&=&\text{Tr}\,\Big[e^{-
\beta H} \, (-1)^{\sf F}\,  e^{4\pi \i \sigma (J_- ^3+J_+^3)}\Big],
\ea
where the fermion parity operator is $(-1)^{\sf F}= e^{2\pi \i J_+^3}$. We will now evaluate it directly from \eqref{eqn:LN422}. Since for $\alpha_-=\alpha_++1/2$ the numerator arising from the fermion determinant vanishes, the only saddles that can potentially contribute correspond to the values of $(m,n)$ where the denominator also vanishes. We find two choices
\beq
m=n,~~~\text{or}~~~m=n-1.
\eeq
The sum over saddles collapses to the following 
$$
{\sf Index}(\sigma)= \frac{\pi}{\sin 2\pi \sigma} \sum_{n\in\mathbb{Z}}\sum_{\epsilon=\pm} \frac{\epsilon}{64\sqrt{\beta/(\Phi_r)}}\, e^{S_0+\frac{2\pi^2\Phi_r}{\beta}\left(1-4(1+\upalpha)(\sigma +n)^2-4(1+\upalpha^{-1})(\sigma+n-\frac{\epsilon}{2})^2\right)}.
$$
We can apply Poisson resummation to this expression and obtain an expansion as a sum over spins. We obtain the index.
$$
{\sf Index}(\sigma) = \sum_{j \geq 0} \chi_{j}(\sigma) \, e^{S_0}\frac{\sqrt{\pi \upalpha/2}}{32(1+\upalpha)}\,\sin \Big(\frac{\pi (2j+1)}{1+\upalpha}\Big)\, e^{-\beta \frac{\upalpha (j+1/2)^2}{2\Phi_r(1+\upalpha)^2}}.$$
The first obvious observation is the fact that the index depends on the temperature. BPS states have a fixed energy that depends \emph{non-linearly} on $R$-charges, and therefore we cannot get rid of temperature dependence by a judicious choice of parametrization of the potentials as e.g. in \cite{Cabo-Bizet:2018ehj}. A partition function where temperature couples to the anticommutator between supercharges would be temperature independent, but it is not easy to implement in the gravity path integral due to the non-linearity. Second, it is clear that $j$ plays the role of $j_++j_-$ by comparing the $\sigma$ dependence with \eqref{eq:cal}. Given this identification, the index does not involve visible cancellation; the only remaining unprotected information on the BPS spectrum is how the total spin $j$ is divided between $(j_+,j_-)$, which we determined earlier.

\subsection{Linearization of the model, its solution, and the index}\label{sec:XLN4SCH}

It is possible to define a theory of large $\mathcal{N}=4$ near-BPS black holes where the supercharges satisfy a linear algebra. In order to achieve this, it is necessary to add a free multiplet consisting of four fermions transforming in the bifundamental of $\SU(2)_+ \otimes \SU(2)_-$ and a bosonic $\U(1)$ mode. The extra fields are free and do not interact with the fields of the theory analyzed so far, but they generate new charges that can be used to linearize the commutation relations. An explicit expression for the action of the enlarged theory can be found in equation 4.33 of \cite{Kozyrev:2021icm}.

We have four supercharges $Q_{A\dot{B}}$, a Hamiltonian $H$ and two $\SU(2)$ generators $J_{AB}$ and $J_{\dot{A}\dot{B}}$. In the ``extra large'' model that linearized the commutation relations, we supplement these generators with four fermions $\tilde{Q}_{A\dot{B}}$ in the same representation as the supercharges and a generator of a $\U(1)$ symmetry conventionally denoted by $U$. The algebra is as follows 
\beq
\label{eq:extralargeRR}
\{ Q_{A\dot{A}}, Q_{B\dot{B}}\} = -\varepsilon_{AB} \varepsilon_{\dot{A}\dot{B}} H,~~~~~\{ \tilde{Q}_{A\dot{A}}, \tilde{Q}_{B\dot{B}}\} = \varepsilon_{AB} \varepsilon_{\dot{A}\dot{B}} \frac{4(1+\upalpha)^2\Phi_r}{\upalpha}
\eeq
and 
\beq
\{ \tilde{Q}_{A\dot{A}}, Q_{B\dot{B}}\} = \varepsilon_{AB} \varepsilon_{\dot{A}\dot{B}} \i\, U + \varepsilon_{\dot{A}\dot{B}} J_{AB}-\varepsilon_{AB} J_{\dot{A} \dot{B}} 
\eeq
All the fermionic generators commute with $H$ and $U$, while their commutators with $J_{AB}$ or $J_{\dot{A} \dot{B}}$ are fixed by their representation. In our convention $(Q_{1\dot 1})^\dagger=-Q_{2\dot 2}$, $(Q_{1\dot 2})^\dagger=Q_{2\dot 1}$, $(\tilde Q_{1\dot 1})^\dagger=\tilde Q_{2\dot 2}$, $(\tilde Q_{1\dot 2})^\dagger=-\tilde Q_{2\dot 1}$.

How does the presence of extra fermions and $\mathrm{U}(1)$ degrees of freedom affect the representations? Essentially, each state in the large algebra gets replaced by a four-dimensional multiplet generated by $\tilde{Q}_{A\dot{B}}$, such that non-BPS multiplets have 16 states while BPS multiplets have 8 states.  We can start from the supermultiplets studied earlier, and each $\SU(2)_+ \otimes \SU(2)_-$ representation is split into four representations by the $\tilde{Q}_{A\dot{B}}$. The resulting long and short multiplets are 
\beq
\label{eq:XLN4SM}
\hspace{-0.3cm}
\begin{tikzpicture}[scale=1,baseline={([yshift=0cm]current bounding box.center)}]
\draw[very thick,->] (0,0) to (0,6);
\draw[very thick,->] (0,0) to (6,0);
\draw[left] (0,5) node {\footnotesize $j_-+\frac{1}{2}$};
\draw[left] (0,4) node {\footnotesize $j_-$};
\draw[left] (0,3) node {\footnotesize $j_- - \frac{1}{2}$};
\draw[left] (0,2) node {\footnotesize $j_--1$};
\draw[left] (0,1) node {\footnotesize $j_--\frac{3}{2}$};
\draw[below] (3,0) node {\footnotesize $j_+-\frac{1}{2}$};
\draw[below] (2.05,0) node {\footnotesize $j_+-1$};
\draw[below] (1,0) node {\footnotesize $j_+-\frac{3}{2}$};
\draw[below] (4,0) node {\footnotesize $j_+$};
\draw[below] (5,0) node {\footnotesize $j_++\frac{1}{2}$};
\draw[very thick] (-0.05,5) to (0.05,5);
\draw[very thick] (-0.05,4) to (0.05,4);
\draw[very thick] (-0.05,3) to (0.05,3);
\draw[very thick] (-0.05,2) to (0.05,2);
\draw[very thick] (-0.05,1) to (0.05,1);
\draw[very thick] (3,-0.05) to (3,0.05);
\draw[very thick] (2,-0.05) to (2,0.05);
\draw[very thick] (1,-0.05) to (1,0.05);
\draw[very thick] (4,-0.05) to (4,0.05);
\draw[very thick] (5,-0.05) to (5,0.05);
\draw[fill=black] (4,4) circle (0.06);
\draw[fill=black] (4,2) circle (0.06);
\draw[fill=black] (3,3) circle (0.06);
\draw[fill=black] (5,3) circle (0.06);
\draw[fill=black] (2,2) circle (0.06);
\draw[fill=black] (3,1) circle (0.06);
\draw[fill=black] (1,3) circle (0.06);
\draw[fill=black] (2,4) circle (0.06);
\draw[fill=black] (3,5) circle (0.06);
\draw[thick] (2,3) circle (0.06);
\draw[thick] (3,2) circle (0.06);
\draw[thick] (4,3) circle (0.06);
\draw[thick] (3,4) circle (0.06);
\end{tikzpicture}
~~
\begin{tikzpicture}[scale=1,baseline={([yshift=0cm]current bounding box.center)}]
\draw[very thick,->] (0,0) to (0,6);
\draw[very thick,->] (0,0) to (5,0);
\draw[left] (0,4) node {\footnotesize $j_-+\frac{1}{2}$};
\draw[left] (0,3) node {\footnotesize $j_-$};
\draw[left] (0,2) node {\footnotesize $j_--\frac{1}{2}$};
\draw[left] (0,1) node {\footnotesize $j_--1$};
\draw[below] (3,0) node {\footnotesize $j_+$};
\draw[below] (2.05,0) node {\footnotesize $j_+-\frac{1}{2}$};
\draw[below] (1,0) node {\footnotesize $j_+-1$};
\draw[below] (4,0) node {\footnotesize $j_++\frac{1}{2}$};
\draw[very thick] (-0.05,4) to (0.05,4);
\draw[very thick] (-0.05,3) to (0.05,3);
\draw[very thick] (-0.05,2) to (0.05,2);
\draw[very thick] (-0.05,1) to (0.05,1);
\draw[very thick] (3,-0.05) to (3,0.05);
\draw[very thick] (2,-0.05) to (2,0.05);
\draw[very thick] (1,-0.05) to (1,0.05);
\draw[very thick] (4,-0.05) to (4,0.05);
\draw[fill=black] (3,4) circle (0.06);
\draw[fill=black] (3,2) circle (0.06);
\draw[fill=black] (2,3) circle (0.06);
\draw[fill=black] (4,3) circle (0.06);
\draw[fill=black] (1,2) circle (0.06);
\draw[fill=black] (2,1) circle (0.06);
\draw[thick] (2,2) circle (0.06);
\draw[thick] (3,3) circle (0.06);
\end{tikzpicture}
\eeq
where the white dots denote the states before the additional free fermions are incorporated. All states have the same $\U(1)$ charge which we denote $u$. Another new aspect is that $\SU(2)_+\otimes \SU(2)_-$ representations have degeneracies. For example, for the long multiplet depicted in the left panel of the diagram above, there are $4$ representations with spins $(j_+-1/2,j_--1/2)$, $2$ representations with spins $(j_+-1,j_--1)$,$(j_+-1,j_-)$, $(j_+,j_--1)$ and $(j_+,j_-)$, and the rest have only one. For the short multiplets all representations appear once except $(j_+,j_--1/2)$ and $(j_+-1/2,j_-)$ which appear twice.

To derive the new BPS bound we consider the $2\times 2$ matrix made of the anticommutators of $Q_{1\dot{2}}$ and $\tilde{Q}_{1\dot{2}}$ and demand its determinant vanishes (which implies some linear combination annihilates the state). We find
\beq
E_{\rm BPS}(j_+,j_-,u) = \frac{\upalpha}{2\Phi_r(1+\upalpha)^2} \left\{ \left( j_+ + j_- + \frac{1}{2} \right)^2 + u^2\right\}.
\eeq
This is the same as in the earlier discussion in this section with a simple additive contribution from the $\U(1)$ mode. Next, we will reproduce all these features from the partition function.

How is the partition function of the ``extra large'' $\mathcal{N}=4$ theory related to the one we computed earlier? The answer is very simple since in the action the new fermions and $\U(1)$ modes are completely decoupled from the large $\mathcal{N}=4$ Schwarzian modes studied before. This implies that the partition function
\beq
Z_{linear}(\beta,\alpha_+,\alpha_-,\mu) = {\rm Tr} \left[ e^{-\beta H} e^{4\pi \i \alpha_+ J^3_+} e^{ 4\pi \i \alpha_- J^3_-} e^{2\pi \i \mu U} \right]
\eeq
is given by
\bea
Z_{linear}(\beta,\alpha_+,\alpha_-,\mu) &=& Z_{\text{ $\mathcal{N}=4$}}(\beta,\alpha_+,\alpha_-) \cdot  \underbrace{4\prod_{\pm} \cos \pi (\alpha_+ \pm \alpha_-)}_{\text{Contribution from new fermions}} \nonumber\\
&&\cdot \underbrace{\sqrt{\frac{2\pi}{\beta}\frac{\Phi_r(1+\upalpha)^2}{\upalpha}} \sum_{n\in\mathbb{Z} \cdot u_0^{-1}} e^{-\frac{2\pi^2 \Phi_r(1+\upalpha)^2}{\upalpha}\frac{(\mu+n)^2}{\beta}}}_{\text{Contribution from $\U(1)$ mode}}
\ea
In the second line we introduced a free parameter $u_0$ such that the $\U(1)$ charge quantization condition reads $u \in \mathbb{Z}\cdot u_0$. At the level of the Schwarzian theory, this is a free parameter which should be determined via a matching with a microscopic model.

In the RHS of the equation above, replace the large $\mathcal{N}=4$ partition function with the expansion \eqref{eq:ZN4EXP}. The partition function of the linear theory can then be written as
\bea
Z &=& \sum_{u\in u_0 \cdot \mathbb{Z}}\sum_{j_+,j_-} \chi^{\rm long}_{j_+j_-u}(\alpha_+,\alpha_-,\mu)\, \int \d E\, e^{-\beta E} \rho_{j_+j_-u}(E)\nonumber\\
&&~~~~+ \sum_{u\in u_0 \cdot \mathbb{Z}}\sum_{j_+,j_-} \chi^{\rm short}_{j_+ j_-u}(\alpha_+,\alpha_-,\mu) e^{-\beta E_{\rm BPS}(j_+,j_-,u)} N_{j_+j_-u}.\label{eq:ZN4EXP2}
\ea
We leave implicit the fact that these quantities correspond to the larger linear theory with the $\U(1)$ multiplet. This should be clear by the appearance of a $u$ dependence. The effect of the fermion contribution is to enlarge the long and short multiplet character as implied by the diagrams \eqref{eq:XLN4SM}. For example, it simply generates the long multiplet of the linear algebra
\bea
\chi^{\text{long}}_{j_+j_-u} &=& e^{2\pi \i \mu u} \cdot 4 \cos \pi(\alpha_+ +\alpha_-) \cos \pi (\alpha_+ - \alpha_-) \cdot \chi^{\rm long}_{j_+,j_-} \nonumber\\
&=& e^{2\pi \i \mu u} \cdot\Big(\chi_{j_+-\frac{3}{2}}\chi_{j_- -\frac{1}{2}} + 2 \chi_{j_+-1}\chi_{j_-}+2\chi_{j_+-1}\chi_{j_--1}+\chi_{j_+-\frac{1}{2}}\chi_{j_-+\frac{1}{2}} + 4 \chi_{j_+-\frac{1}{2}}\chi_{j_--\frac{1}{2}}\nonumber\\
&&~~~+\chi_{j_+-\frac{1}{2}}\chi_{j_--\frac{3}{2}}+2\chi_{j_+}\chi_{j_-}+2\chi_{j_+}\chi_{j_--1}+\chi_{j_++\frac{1}{2}}\chi_{j_--\frac{1}{2}}\Big)
\ea
and the short multiplet of the linear algebra
\bea
\chi^{\text{short}}_{j_+,j_-,u} &=& e^{2\pi \i \mu u} \cdot 4 \cos \pi(\alpha_+ + \alpha_-) \cos \pi (\alpha_+ - \alpha_-)\cdot  \chi^{\text{short}}_{j_+,j_-},\nonumber\\
&=& e^{2\pi \i \mu u} \cdot\Big(\chi_{j_+} \chi_{j_-+\frac{1}{2}} +2\chi_{j_+-\frac{1}{2}}\chi_{j_-} + \chi_{j_++\frac{1}{2}}\chi_{j_-}\nonumber\\
&&~~+2\chi_{j_+}\chi_{j_--\frac{1}{2}}+\chi_{j_+-1}\chi_{j_--\frac{1}{2}}+\chi_{j_+-\frac{1}{2}}\chi_{j_--1}\Big).
\ea
At this level, the density of states of supermultiplets is unchanged, only the supermultiplet structure is adjusted. Moreover, all supermultiplets of the previous theory become families of supermultiplets of different $\U(1)$ charges. This can be seen by a simple Poisson resummation of the expression above with respect to the chemical potential $\mu$. This also has the simple effect of adding an extra contribution to the energy $E \to E + \frac{\upalpha}{2\Phi_r(1+\upalpha)^2} u^2$. For short supermultiplets, this has the effect of producing the correct BPS energy when comparing with the algebra. The density of non-BPS supermultiplets is simply related to $\rho_{j_+,j_-}(E)$ by this shift. To summarize
\beq
N_{j_+ j_- u} = N_{j_+ j_-},~~~~\rho_{j_+j_- u}(E) = \rho_{j_+, j_-}\left( E - \frac{\upalpha}{2\Phi_r(1+\upalpha)^2} u^2\right)
\eeq
The quantities on the right-hand side are the earlier BPS degeneracy \eqref{eqn:NBPS} and density of states \eqref{eq:NL4DOS}. Notice that the BPS degeneracy is completely independent of $u$, while the dependence of the density of states is only through a simple charge-dependent shift of energies.

Finally, we discuss the index of the ``extra large'' $\mathcal{N}=4$ theory. Since all fermions (including the ones that lead to the original supercharges as well as the ones that are part of the $\U(1)$ multiplet) are charged under the $\SU(2)_+\otimes \SU(2)_-$ symmetry, we do not have the problem that arises for $\mathcal{N}=3$. We can define a fermion-parity operator as $(-1)^{\sf F}=e^{2\pi \i J_+^3}$ and insert it into the partition function by choosing a chemical potential, namely $\alpha_-=\alpha_++1/2=\sigma$. It is easy to see that $\chi^{\text{long}}_{j_+j_-u}(\sigma,\sigma+1/2)=0$ and that the index is indeed protected. In this case a problem arises since even though $\chi_{j_+j_-}^{\text{short}}\neq 0$, the contribution of $\tilde{Q}_{A\dot{A}}$ to $\chi^{\text{short}}_{j_+j_-u}$ vanishes. 

A solution to this problem was proposed in \cite{Gukov:2004ym,Gukov:2004fh} in the context of 2d CFT and we can implement it here as well. The idea is based on the observation that 
$$\chi^{\rm long}_{j_+j_-u}(\sigma+\varepsilon,\sigma+1/2) \sim O(\varepsilon^2),~~~
\chi^{\rm short}_{j_+j_-u}(\sigma+\varepsilon,\sigma+1/2) \sim O(\varepsilon),
$$
A solution is to take a derivative with respect to, for example, $\alpha_+$ such that the long character still vanishes but not the short one. Concretely,
\bea
\frac{\d \chi^{\rm long}_{j_+j_-u}}{\d \alpha_+}(\sigma,\sigma+1/2,\mu)  &=& 0,\\
\frac{\d \chi^{\rm short}_{j_+j_-u}}{\d \alpha_+}(\sigma,\sigma+1/2,\mu)  &=& e^{2\pi \i \mu u}4\pi \sin 2 \pi \sigma \underbrace{\frac{\sin 4\pi \sigma(j_++j_-+\frac{1}{2})}{\sin 2\pi \sigma} (-1)^{2 j_+}}_{= \chi_{j_+j_-}^{\rm short}(\sigma,\sigma+1/2)}
\ea
Using the result above for the non-linear theory we can easily derive the index as
\bea
\frac{1}{4\pi}\frac{\d Z}{\d \alpha_+} (\sigma,\sigma+1/2,\mu) &=& \sum_{j \geq 0}\sum_{u \in u_0 \cdot \mathbb{Z}}e^{2\pi \i \mu u}\, \sin 2 \pi \sigma \, \chi_{j}(\sigma) \nonumber\\
&&~e^{S_0}\frac{2\sqrt{\pi \upalpha}}{(1+\upalpha)}\,\sin \Big(\frac{\pi (2j+1)}{1+\upalpha}\Big)\, e^{-\beta \frac{\upalpha }{2\Phi_r(1+\upalpha)^2}((j+1/2)^2+u^2)}.\label{eq:blabla}
\ea
This object correctly captures the protected information regarding the BPS states. Therefore the useful index to define in the linearized large $\mathcal{N}=4$ algebra is given by
\beq
\text{Index}(\sigma) = \text{Tr}\left[e^{-\beta H}(-1)^{\sf F}\, J_+^3\, e^{4\pi \i \sigma(J_-^3+J_+^3)}  \right].
\eeq
The insertion of $J_+^3$ is equivalent to taking an $\alpha_+$ derivative, and therefore, this partition function is equal to \eqref{eq:blabla} up to an overall prefactor. This expression is useful since it allows us to formulate a computation of the index without breaking supersymmetry (which a derivative with respect to the chemical potential would require). The role of the insertion of $J_+^3$ is to soak up integrals over fermion zero modes that appear in the calculation. In that sense, this index is precisely the helicity supertrace introduced for flat space black holes \cite{Kiritsis:1997hj}.

\section{Cases with non-linearly realized symmetries}
\label{sec:N>4}

In this section, we analyze Schwarzian theories with more than four supercharges. We refer to Table \ref{tab:scalgebras} for a comprehensive list of simple superconformal groups that are used to construct such theories. Using group theory information reviewed in Appendix \ref{app:roots}, we construct the exact partition function using the method outlined in section \ref{sec:GPF}. 
Our analysis reveals that the $\OSp(4^*|2n)$ and $\SU(1,1|n)$ series do not result in well-behaved Schwarzian theories and therefore cannot possibly arise in the near-BPS limit of black holes described by unitary quantum systems. 
For the nonpathological cases, we explicitly calculate the spectrum for each theory, highlighting particularly the specific multiplet structure and the associated BPS conditions.

For the $\cN\le 4$ cases studied in the previous sections, the (anti)commutation relations of conserved (super)charges can involve non-linear combinations of the generators. In those cases, we explicitly showed how the (anti)commutation relations can be linearized by adding new matter fields that result in an enlarged symmetry group. In this section, it will become clear that for $\cN>4$ cases, only the non-linear formulation is available as shown by Schoutens \footnote{From the viewpoint of extended superconformal algebras, the linearized $\OSp(n|2)$ superconformal algebra is explicitly ruled out for $n>4$ in \cite{Schoutens:1988ig}. Although a general proof of non‑linearizability is lacking, no successful linearization of the superconformal algebra based on $\text{F}(4)$ and $\text{G}(3)$ has been reported.}. This parallels the construction of superconformal generalizations of the Virasoro group developed by Knizhnik \cite{Knizhnik:1986wc}, see also \cite{Bershadsky:1986ms}. 

When the Schwarzian effective theories arise from the near-BPS limit of higher dimensional black holes, a linear algebra of conserved charges in higher dimensions implies a linear algebra of conserved charges in the Schwarzian theory (being a subset of their higher-dimensional counterparts). This fact severely restricts the scenarios in which the theories analyzed in this section can arise, but they are interesting theories nonetheless.

\subsection{Ruling out $\OSp(4^*|2n)$ and $\SU(1,1|n)$}
\label{sec:rulingout}

For the supergroups $\SU(1,1|n),~n > 2$ and $\OSp(4^*|2n)$, we will argue that their corresponding Schwarzian theories have divergent partition functions and are thus ill defined. Let us first recall how to derive the Schwarzian action in the BF formulation.
According to Section \ref{sec:GPF}, the boundary action is given by a supertrace over the connection
\ie
\oint \,{\rm Str}\,A_t^2 \, .
\fe
In the case of $\OSp(4^*|2n)$ and $\SU(1,1|n)$, the supergroup admits a supermatrix representation, which offers a easier way to compute the relative normalization $q$.
A general supermatrix takes the block form $a=\left(\begin{array}{c|c}
\alpha & \beta \\
\hline \gamma & \delta
\end{array}\right) $, with ordinary matrices $\alpha$ and $\delta$, as well as fermionic matrices $\beta$ and $\gamma$. The supertrace is defined as
\ie
{\rm Str}(a)={\rm Tr}\, \alpha-{\rm Tr}\,\delta \, .
\fe
Usually, the $\SL(2,\mathbb{R})$ part of the bosonic subgroup lies in the upper-left block, while the $R$-symmetry in the lower-right. Then the minus sign in the supertrace explains the different signature between Schwarzian mode and gauge field in the action.  However, for the case of $\OSp(4^*|2n)$, the supermatrix is given by
\begin{equation}
\left(\begin{array}{c|c}
\SO^*(4) & \text { fermionic } \\
\hline \text { fermionic } & \text{Sp}(2n)
\end{array}\right) .
\end{equation}
The $\SO^*(4)$ in the upper-left block is a complexification of $\SO(4)$, $\SO^*(4)\cong \SL(2,\mathbb{R})\times \SU(2)$. The $R$-symmetry is given by the product $\SU(2)\times \Sp(2n)$. After taking the supertrace, the two simple components of the $R$-symmetry get a relative negative sign from the supertrace:
\ie
I_{\OSp(4^*|2n)}=-\Phi_r \int \d\tau \,\left\{\text{Sch}(f,\tau)  - {\rm Tr} [ (g_{\SU(2)}^{-1} \partial_\tau g_{\SU(2)})^2] + {\rm Tr} [ (g_{\Sp(2n)}^{-1} \partial_\tau g_{\Sp(2n)})^2] +\text{fermions} \right\}
\fe
It's easier to see the trouble of this difference of signature in the on-shell action. Consider the simplest $\nu=0$ saddle and using the fact that $g$ is in the $(2,2n)$ representation, we obtain the on-shell action:
\ie
I_{\text{on-shell}}=-\frac{2\pi^2\Phi_r}{\beta}-\frac{2\pi^2\Phi_r}{\beta}16\,\alpha_{\SU(2)}^2 + \frac{2\pi^2\Phi_r}{\beta}\sum_{i=1}^n 8\,\alpha_{\text{Sp(2n)},i}^2
\fe
where we have defined $g_{\SU(2)} = {\rm exp}({ 4\pi \i \,\alpha_{\SU(2)} H_{\SU(2)}})$ and $g_{\Sp(2n)} = {\rm exp}({4\pi \i \sum_{i} \alpha_{{\rm Sp}(2n),i} H_i })$. The $\alpha$'s are the chemical potentials associated to the Cartan subalgebra of $\Sp(2n)$, generated by $H_i$ with $i=1,\ldots, n$, and $\SU(2)$, generated by $H_{\SU(2)}$. The full path integral involves summing over shifts $\alpha_{\SU(2)}\to \alpha_{\SU(2)}+ m$, which are just classical Euclidean solutions. However, for very large values of $\alpha_{\SU(2)}+m$, the action can become arbitrarily large and negative. Summing over such configuration leads to a divergence in the partition function within the grand canonical ensemble. Furthermore, quantum corrections will not eliminate these pathological saddles. This suggests that the Schwarzian theory is not well defined for the supergroup $\OSp(4^*|2n)$.
The same pathology exists at the level of superconformal algebra \cite{Knizhnik:1986wc}: The Jacobi identity imposes a relation between the level of $\widehat{\SU(2)}$ and $\widehat{\Sp(2n)}$ current algebras and forces them to have opposite signs at large levels, leading to non-unitary representations.

The second case is the superconformal group with global subgroup $\SU(1,1|n)$. Now the $R$-symmetry group is $\SU(n)\times\U(1)$. The $\U(1)$ generator is, by definition, $\SL(2,\mathbb{R})$ and $\SU(n)$ invariant and has zero supertrace. We denote it as $X$ and when we represent the superconformal group as a supermatrix it takes the form
\ie
X=\text{diag}(n,n,\underbrace{2,\dots,2}_{\text{$n$ entries}}) \, ,
\fe
with the coefficients chosen such that $\text{Str}\,X=0$. Notice that again we have an $R$-symmetry generator that leaks into the upper-left corner of the supermatrix. As we evaluate the Schwarzian action $\text{Str}\,A_t^2$ corresponding to this $\U(1)$ mode, we need to square this generator. The coefficient of the $\U(1)$ action will be given by the following supertrace
\ie
\text{Str}\, X^2=2n^2-4n=2n(n-2)>0,~\text{for} ~ n>2
\fe
which is again a wrong sign for the contribution of the gauge group! A sum over saddles with non-trivial winding of the $\U(1)$ phase mode will lead to a Schwarzian partition function that diverges. Although this was unavoidable for non-Abelian gauge groups such as $\SU(2)$, for $\U(1)$ we could choose not to sum over these saddles. In this case, the partition function might be finite (after a rotation of the contour of integration), but the spectrum of charges would certainly be continuous, which is unphysical. The case $n=2$ is ruled out, since the generator $\U(1)$ becomes proportional to the identity matrix, allowing us to eliminate that generator and instead consider the group ${\rm PSU}(1,1|2)$. The case $n=1$ is also well-behaved since ${\rm Str}\, X^2 = -2$ has the correct sign, which should be expected since $\SU(1,1|1) \sim \OSp(2|2)$. 

Therefore, the Schwarzian theories associated to both $\SU(1,1|n)$ with $n>2$ and $\OSp(4^*|2n)$ are ruled out. The next set of examples do not present this explicit pathology, but still have some usual features.

\subsection{$\OSp(n|2)$}

\subsubsection*{Odd $n$ case: Spectrum}

Now we turn to the supergroup $\OSp(n|2)$. The $R$-symmetry is $\SO(n)$, whose characters for irreducible representations are collected in Appendix \ref{app:roots}. Throughout this section, we label the irreducible representations of the $R$-symmetry groups by their weights $\lambda=\sum_{i=1}^{\text{rank}\,R}\,m_i\mu_i$, where $\mu_i$'s are the fundamental weights, or explicitly by these non-negative integers $m_1,\dots,m_{\text{rank}\,R}$. 
We begin by analyzing the case where $n$ is odd. Denote $n=2l+1$ with $l$ the rank of the $R$-symmetry group.  To extract the density of states, we need to perform the inverse Legendre transform with respect to $\beta$ and the Fourier transform with respect to $\alpha_i$, in the partition function \eqref{eqn:ptospodd}. The derivation is outlined in Appendix \ref{app:Osp}. We find that the partition function can be written in terms of density of states as:
\ie\substack{}
Z^{\text{Osp}(2l+1|2)}&= \sum_{\substack{m_1,\dots,m_{l-1}\in\mathbb{Z}_{\ge 0} \\m_l\in 2\mathbb{Z}_{\ge 0}}}\chi^{\text{long}}_{m_1,\dots,m_l}(\alpha_1,\dots,\alpha_l)\,\int \d E \,e^{-\beta E}\rho_\lambda(E)\\
&~~~~+ \sum_{m_1,\dots,m_{l-1}\in\mathbb{Z}_{\ge 0}}\chi^{\text{short}}_{m_1,\dots,m_{l-1},0}(\alpha_1,\dots,\alpha_l) e^{-\beta E_{\rm BPS}}N_{m_1,\dots,m_{l-1},0}
\fe
Here $m_l$ takes the value only in even integers to guarantee all states in the multiplets have only integer charges. In the first line, it will be convenient to express the density of states directly in terms of $\lambda$, instead of the $m$'s. The spectrum is comprised of two types of states, continuous and discrete ones:

\begin{itemize}
\item The long non-BPS supermultiplets within the continuous spectrum consist of a combination of $2\times 2^{l}$  $\SO(2l+1)$ states.
\ie
\chi^{\text{long}}_{m_1,\dots,m_l}(\alpha_1,\dots,\alpha_l)=2\sum_{e_1,\dots,e_l\in\{0,1\}}\chi_{m_1-e_1+e_2,m_2-e_2+e_3,\dots,m_{l-1}-e_{l-1}+e_{l},m_l-2e_{l}}(\alpha_1,\dots,\alpha_l)
\label{eqn:SOoddlong}
\fe
See \eqref{eqn:SOoddweight} for explicit forms of the $\SO(2l+1)$ group characters. The origin of the multiplet structure from the symmetry algebra will be explained soon. There exists a spectral gap separating the ground state from the continuum of higher-energy non-BPS multiplets:
\ie
E_0(\lambda)=\frac{1}{8\Phi_r}(\lambda+w-1/2)^2
\fe
Here $w$ is the Weyl vector, the half-sum of the positive roots. The spectral density of the multiplets in the continuous spectrum is given by\footnote{The normalization between the density of states and the partition function is off by a numerical factor we will not track down but can be absorbed in $S_0$ anyways.}
\ie
\rho_\lambda(E)=\frac{e^{S_0}}{2\pi}\frac{\prod_{r\in R_+} r\cdot (\lambda+w-1/2)}{\prod_{i=1}^l \left(E-\frac{\sum_{j\neq i}(\lambda+w-1/2)_j^2}{8\Phi_r}\right)}\frac{\cosh\left(2\pi\sqrt{2\Phi_r(E-E_0)}\right)}{\sqrt{2\Phi_r(E-E_0)}}
\fe
It is easy to check that the density of multiplets is always positive at energies above the spectral gap. In other words, when $E>E_0$ one can check that the terms in the denominator, as well as the product of roots that appears in the numerator, are all positive.

\item We also find BPS states appearing in short multiplets, each containing $2^{l-1}$ states
\ie
\chi^{\text{short}}_{m_1,\dots,m_{l-1},0}(\alpha_1,\dots,\alpha_l)=\sum_{e_1,\dots,e_{l-1}\in\{0,1\}}\chi_{m_1-e_1+e_2,m_2-e_2+e_3,\dots,m_{l-1}-e_{l-1},0}(\alpha_1,\dots,\alpha_l)
\label{eqn:SOoddshort}
\fe
Notice that all the BPS states in the short multiplets have $m_l=0$, indicating they have zero weight in the $l$-direction. These short multiplets exist at the BPS energy, which is given by (denote $\lambda_{\rm BPS}=m_1\mu_1+\dots+m_{l-1}\mu_{l-1}+0$)
\ie
E_{\rm BPS}=\frac{1}{8\Phi_r}\sum_{i=1}^{l-1}(\lambda_{\rm BPS}+w-1/2)_i^2
\label{eqn:SOoddgap}
\fe
The BPS degeneracy is
\ie
N_{\lambda_{\rm BPS}}=e^{S_0}\prod_{r\in R^{\prime l}_+} r\cdot (\lambda_{\rm BPS}+w-1/2).
\fe
where $R^{\prime l}_+$ is defined as the subset of positive roots that have zero component in the $l$-th direction.  We observe that: (1) $\,$The BPS energy at $\lambda_{\rm BPS}$ takes the same form as the spectral gap in the continuum spectrum, except that $m_l$ is forced to be zero in the BPS spectrum; (2) $\,$The factor  $\prod_{r\in R^{\prime l}_+} r\cdot (\lambda_{\rm BPS}+w-1/2)$ is always a positive integer for any non-negative $m_1,\dots,m_{l-1}$, so that the BPS degeneracy is always positive.

\end{itemize} 

Above, we have discussed the spectrum in terms of long and short multiplets. It is also worth examining the spectrum within specific charge sectors. For a given charge sector $(m_1,\dots,m_l)$, there exist only continuous states when $m_l>0$, or there exist both BPS states and continuous states when $m_l=0$. For the latter case, since $\rho_{\lambda_{\rm BPS}}=0$, the continuous states with $(m_1,\dots,m_{l-1},m_l=0)$ can originate only from supermultiplets with a different highest weight $\lambda_{\rm BPS}+2\mu_l-\mu_{l-1}$, or explicitly $(m_1,\dots,m_{l-1}-1,m_l=2)$. We can verify that there is always a gap $E_{\rm gap}=E_0(\lambda_{\rm BPS}+2\mu_l-\mu_{l-1})-E_{\rm BPS}(\lambda_{\rm BPS})=\frac{1}{8\Phi_r}$ between the degenerate BPS states and the continuum, consistent with the general expectation of \cite{Turiaci:2023jfa,Johnson:2023ofr,Johnson:2024tgg} based on random matrix theory.

\subsubsection*{Odd $n$ case: Conserved charges}

To understand the multiplet structure appearing in the spectrum, we turn our attention to the algebra of symmetry generators of $\text{OSp}(2l+1|2)$. They are time translation generators $H$, $\SO(2l+1)$ generators $J$'s, and the global supersymmetry $Q$'s. In the Cartan-Weyl basis $J^{i},~J^{\alpha}~,J^{-\alpha}~~~i=1,\dots,l, ~\alpha\in R_+$, the $\SO(2l+1)$ generators have the following commutation relations
\ie
&[J^i,J^j]=0,~~~~~~~~~~~~~~~~~[J^i,J^{\pm\alpha}]=\pm\alpha^i J^{\pm\alpha},
\\
&[J^{\alpha},J^{-\alpha}]=\sum_{i}\alpha^i J^i ,~~~~~[J^\alpha,J^\beta]=N_{\alpha\beta}J^{\alpha+\beta}
\fe
The second equation of the second line does not include the case $\alpha = - \beta$ and involves numerical factors $N_{\alpha\beta}$ fixed by Jacobi identities. The $2l+1$ supercharges lie in the fundamental representation of $\SO(2l+1)$. We can choose the complex basis in which $Q^{\bar i}$ has the weight vector $e^i$, $Q^i$ has $-e^i$, and $Q^0$ has a vanishing weight vector; namely
\ie
[J^i,Q^{\bar j}]&=\delta_{ij}Q^{\bar j},\\
[J^i,Q^{ j}]&=-\delta_{ij}Q^{j},
\\
[J^i,Q^0]&=0
\fe
The remaining non-trivial algebras are the anti-commutators between supercharges. The most general form is given by
\ie
\{Q^k,Q^{l}\}&=\delta_{kl}(H-c)+Kt^a_{kl}J^a+\gamma \Pi^{ab}_{kl}J^a J^b
\fe
Here, $k,l$ run over $1,\dots 2l+1$ while $J^a,J^b$ span all the generators in $\SO(2l+1)$. Each generator $J^a$ is represented by $t^a_{ij}$ in the fundamental representation. The tensor $\Pi^{ab}_{kl}$ is defined as $\Pi_{i j}^{a b}=t_{i m}^a t_{m j}^b+t_{i m}^b t_{m j}^a+2 \delta^{a b} \delta_{i j}$. The constants $c,K,\gamma$ can be determined through embedding in the $\SO(n)$-extended Virasoro algebra \cite{Knizhnik:1986wc,Bershadsky:1986ms}. However, we leave them free here to be matched later from our spectrum.
Among these anticommutators, we are particularly interested in the one involving zero weight supercharges:
\ie
\{Q^0,Q^0\}&=H-c-\upgamma C_2
\fe
where $C_2$ is the Casimir operator of $\SO(2l+1)$.
A highest weight state $\ket{E,\lambda}$ is defined to be annihilated by all the `raising' operators
\ie
Q^{\bar i}\ket{E,\lambda}=&0, ~~~~~ ~~~~~~J^{\alpha}\ket{E,\lambda}=0
\\
H\ket{E,\lambda}=&E\ket{E,\lambda}, ~~ ~J^{i}\ket{E,\lambda}=\lambda^i \ket{E,\lambda}
\fe
The long multiplet is generated from $\ket{E,\lambda}$ by applying supercharges $Q^i$, which modify the weight as $\lambda^i\to\lambda^i-1$, and $Q^0$. 
Or equivalently, expressed in the basis of the fundamental weights:
\ie
Q^i\ket{E,(\dots,m_{i-1},m_i,\dots)}&\sim \,\ket{E,(\dots,m_{i-1}+1,m_i-1,\dots)} ~~~{\rm for}~i=1,\dots,l-1
\\
Q^l\ket{E,(\dots,m_{l-1},m_l)}&\sim \,\ket{E,(\dots,m_{l-1}+1,m_l-2)}
\\
Q^0\ket{E,(\dots,m_{i},\dots)}&\sim \,\ket{E,(\dots,m_{i},\dots)}
\fe
Here we use $\sim$ because these states require linear combinations with $J^\alpha\ket{E,\lambda}$ to ensure orthogonality, a detail we will not work out. $l$ different $Q^i$ operators generate a multiplet of $2^l$ states. The action of $Q^0$ on each state within this multiplet generates another state with identical charges but opposite fermion parity, effectively doubling the number of states in the multiplet while having the same quantum numbers. This is exactly the long multiplet structure that we found in the spectrum \eqref{eqn:SOoddlong}.

We can derive a bound on the energy by considering the unitary bound for $Q^0$ \cite{Schoutens:1988tg},
\ie\label{eq:SOOBPS}
|Q^0\ket{E,\lambda}|^2\ge 0~~\Rightarrow~~E\ge \upgamma\lambda(\lambda+2w)+c.
\fe
Compared with the BPS bound in equation \eqref{eqn:SOoddgap}, we find $\upgamma=1/8\Phi_r$, $c=(w^2+1/4)/8\Phi_r$, and a shift $\lambda\to\lambda-1/2$. This shift origins from a mismatch between the highest weight in a supermultiplet and its average weight. When the BPS bound is saturated, $Q^0$ annihilates the highest weight state, halving the number of states in the long multiplet. The BPS states identified in the Schwarzian spectrum are of this type. Furthermore, these states further satisfy the condition $m_l=0$, which indicates that they are also annihilated by $Q^l$. To see this, we consider the following commutator
\ie
\label{eqn:Ql}
[J^{\alpha=-e^l},Q^{0}]=-\frac{1}{\sqrt{2}}Q^l
\fe
The states with $m_l=0$ have zero weight in the $l$-th direction of the root space, meaning they are in the trivial representation under the corresponding $\SU(2)$ subalgebra. As a result, the lowering operator $J^{\alpha=-e^l}$, which acts as an annihilation operator in that $\SU(2)$ subspace, must annihilate these states. Combined with the fact that $Q^0\ket{E_{\rm BPS},m_l=0}=0$, we can conclude that $Q^l$ annihilates these BPS state. In total, the short multiplets in \eqref{eqn:SOoddshort} preserve three supersymmetries $Q^0,\, Q^l,\,Q^{\bar l}$. 

Can these states preserve more supersymmetries for certain values of $m_1,\dots,m_{l-1}\in\mathbb{Z}_{\ge 0}$? We can verify this by examining similar commutators $[J^{\alpha=-e^i},Q^{0}]=-\frac{1}{\sqrt{2}}Q^i$. In the nonzero directions of the weight vector, when $J^{\alpha=-e^i}$ acts on BPS states, it can generate states with weights $\lambda_i^\prime=\lambda_i-1$ in the $i$-th direction, making them deviate from the BPS energy. Therefore, from the algebra $\frac{1}{\sqrt{2}}Q^i\ket{E_{\rm BPS},\lambda}=Q^0 J^{\alpha=-e^i}\ket{E_{\rm BPS},\lambda}\neq 0$ we can conclude that for these $i$, $Q^i$ cannot annihilate the BPS states. On the other hand, if $m_l,m_{l-1},\dots,m_{l-k}$ are all zero, we can deduce from the same argument that these BPS states preserve $2(k+1)+1$ supersymmetries. The energy of these BPS states is still given by \eqref{eq:SOOBPS}. As an example, when all $m$'s are zero and $k=l-1$, the states are fully BPS and in a singlet with energy $E_{\rm BPS}=\frac{(l-1)l(2l-1)}{48\Phi_r}$.

To summarize, most BPS short multiplets in this $\OSp(2l+1|2)$ theory preserve three supercharges. However, setting certain charges to zero can make these multiplets even shorter and preserve additional supercharges ($2l+1$ in total at most), as discussed above. This is analogous to the distinction between standard BPS multiplets and shorter BPS multiplets \eqref{eqn:N4shorter} in the large $\mathcal{N}=4$ theory. This structure holds universally for all the following cases, but we will only describe the supersymmetry that general BPS states can preserve there.

\subsubsection*{Even $n$ case}

When $n$ is even, the structure is similar. We refer again to Appendix \ref{app:roots} for our group theory conventions. In the canonical ensemble, the partition function is written as
\ie
Z^{\text{Osp}(2l|2)}&= \sum_{\substack{m_1,\dots,m_{l-1}\in\mathbb{Z}_{\ge 0} \\m_l+m_{l-1}\in 2\mathbb{Z}_{\ge 0}}}\chi^{\text{long}}_{m_1,\dots,m_l}(\alpha_1,\dots,\alpha_l)\,\int \d E \,e^{-\beta E}\rho_\lambda(E)\\
&~~~~+ \sum_{m_1,\dots,m_{l-1}\in\mathbb{Z}_{\ge 0}}\chi^{\text{short}}_{m_1,\dots,m_{l-1},0}(\alpha_1,\dots,\alpha_l) \,e^{-\beta E_{\rm BPS}}\,N_{m_1,\dots,m_{l-1},0}
\fe
Similarly we find both continuous and discrete states in the spectrum. In the continuous spectrum, the generic non-BPS long multiplets are given by
\ie
\chi^{\text{long}}_{m_1,\dots,m_l}(\alpha_1,\dots,\alpha_l)=\sum_{e_1,\dots,e_l\in\{0,1\}}\chi_{m_1-e_1+e_2,m_2-e_2+e_3,\dots,m_{l-1}-e_{l-1}+e_{l},m_l-e_{l-1}-e_{l}}(\alpha_1,\dots,\alpha_l)
\label{eqn:SOevenlong}
\fe
These long multiplets are similarly constructed from a highest weight state by the action of $l$ $Q$'s with positive weight, resulting in $2^l$ states. The key distinction from the odd $n$ case is that there is no zero weight state in the fundamental representation of $\SO(2l)$, so that the previous `second copy' does not appear in the long multiplets.

The non-BPS states are placed above a supermultiplet-dependent spectral gap
\ie
E_0(\lambda)=\frac{1}{8\Phi_r}(\lambda+w-1/2)^2
\fe
with their density of states given by
\ie
\rho_\lambda(E)=\frac{e^{S_0}}{2\pi}\frac{\prod_{r\in R_+} r\cdot (\lambda+w-1/2)}{\prod_{i=1}^l \left(E-\frac{\sum_{j\neq i}(\lambda+w-1/2)_j^2}{8\Phi_r}\right)}\sinh\left(2\pi\sqrt{2\Phi_r(E-E_0)}\right).
\fe
which is always positive for all supermultiplets. 

The BPS short multiplets contain $2^{l-1}$ states
\ie
\chi^{\text{short}}_{m_1,\dots,m_{l-1},m_{l-1}}(\alpha_1,\dots,\alpha_l)=\sum_{e_1,\dots,e_{l-1}\in\{0,1\}}\chi_{m_1-e_1+e_2,m_2-e_2+e_3,\dots,m_{l-1}-e_{l-1},m_{l-1}-e_{l-1}}(\alpha_1,\dots,\alpha_l)
\label{eqn:SOevenshort}
\fe
The shortening condition $m_{l}=m_{l-1}$ indicates that the weight of the states in the short multiplets is zero in the $l$-th direction(see \eqref{eq:charactersoeven}). Therefore, these BPS states are annihilated by two supercharges $Q^l$ and $Q^{\bar l}$ since they are singlets in the $\SU(2)$ subspace spanned by $J^l$.

Similarly, the BPS energy and degeneracy is given by
\ie
E_{\rm BPS}=\frac{1}{8\Phi_r}\sum_{i=1}^{l-1}(\lambda_{\rm BPS}+w-1/2)_i^2
\label{eqn:SOevengap}
\fe
\ie
N_{\lambda_{\rm BPS}}=e^{S_0}\prod_{r\in R^{\prime l}_+} r\cdot (\lambda_{\rm BPS}+w-1/2).
\fe
but this time $\lambda_{\rm BPS}=m_1\mu_1+\dots+m_{l-1}\mu_{l-1}+m_{l-1}\mu_{l}$. We can again check the existence of the gap between BPS degeneracy and continuous states: continuous states with charge $\lambda_{\rm BPS}$ come from long multiplets with the highest weight $\lambda_{\rm BPS}+\mu_l-\mu_{l-1}$, so there exists a gap $\frac{1}{8\Phi_r}$ between them and the BPS states with the same charge. 

As a concrete example, the spectrum for the $\OSp(4|2)$ Schwarzian theory is given by
\ie
Z^{\OSp(4|2)}=&\int \d E\,e^{-\beta E}\,\sum_{m_1+m_2\in 2\mathbb{Z}_{\ge 0}}\chi^{\rm long}_{m_1,m_2}\frac{e^{S_0}}{2\pi}\frac{m_2(1+m_1)\sinh\left(2\pi\sqrt{2\Phi_r(E-\frac{(1+m_1)^2+m_2^2}{16\Phi_r})}\right)}{(E-\frac{(m_1+m_2+1)^2}{32\Phi_r})(E-\frac{(m_1-m_2+1)^2}{32\Phi_r})}
\\
&+\sum_{m\in \mathbb{Z}_{\ge 0}}\chi^{\rm short}_{m,m} \,e^{S_0}\,\delta\left(E-\frac{(2m+1)^2}{32\Phi_r}\right)
\fe
From the superalgebra isomorphism $D(2,1|1)\cong \OSp(4|2)$, we expect the spectrum of these two theories should be identical. Actually, the BPS condition \eqref{eq:BPSrangeD21} for $D(2,1|\upalpha)$ becomes $j_+=j_-$ for $\upalpha=1$. Then we can match the spectrums of the two theories by identifying $\alpha_-=\alpha_1-\alpha_2,\alpha_+=\alpha_1+\alpha_2$; $j_+,j_-=\frac{m_1+1}{2},\frac{m_2}{2}$ in the continuous spectrum, and $j=\frac{m}{2}$ in the discrete spectrum. However, the restriction to integer-spin representations, $m_1+m_2\in 2\mathbb{Z}_{\ge 0}$, does not apply to $D(2,1|\upalpha)$, since its $R$-symmetry $\SU(2)\times\SU(2)$ is a double cover of $\SO(4)$.

\subsection{Anomalies and extensions of $\OSp(n|2)$}
\label{sec:anomalyosp}
Here we want to emphasize that for each $\OSp(n|2)$ there are three possible Schwarzian theories one can define. The one-loop determinants are equal in all cases but the way the sum over saddles is performed changes:

\begin{itemize}
    \item The first has a $R$ -symmetry $\text{Spin}(n)$. This assumes the presence of fields in spinorial representations and, therefore, we need to sum over even shifts of the chemical potentials $n_i\in 2\mathbb{Z}$ in the partition function \eqref{eqn:ptospeven} and \eqref{eqn:ptospodd}, instead of $n_i\in\mathbb{Z}$. The Laplace transform has multiplets with all representations, with no constraint on $m_l$ ($m_l+m_{l-1}$) being even for $n=2l +1$ odd ($n=2l$ even). Other than this, the expressions we found the density of states above still hold.
    \item The second theory has non-anomalous $\SO(n)$. This is the theory we consider above in this section, summing over $n_i\in\mathbb{Z}$ shifts, and the spectrum has only nonspinorial representations of $\SO(n)$.
    \item Finally, we can consider an anomalous $\SO(n)$ theory. This can be achieved by including a phase 
    \beq
     I \to I + \i \pi \int_{\mathcal{M}_{\text{bulk}}} w_2(\SO(n)),
    \eeq
    where $\mathcal{M}_{\text{bulk}}$ is any choice of 2d space with the time circle as boundary. This is the same Stiefel-Whitney class but now involving an $\SO(n)$ bundle. See \cite{Cordova:2017vab} for some relevant facts about $\SO(n)$ and its anomalies. It is easy to derive that the spectrum of such theories have solely spinorial representations. Other than changing the range of weights of the representations, the formulas for the density of states derived above are still valid. This is entirely analogous to what we observed for the $\OSp(3|2)$ case with anomalous $\SO(3)$.
\end{itemize}

\subsection{Exceptional superconformal groups}
\label{sec:exceptionalgroup}
Finally, for completeness, we repeat some of the analysis for the Schwarzian theories based on G(3) or F(4).
\subsubsection*{$\text{G}(3)$}
In the case of the exceptional supergroup $\text{G}(3)$, we can determine $q$ to be $\frac{2}{3}$ \cite{Bowcock:1992bm}.
By dimension matching, the fermions in this $\cN=7$ super-Schwarzian theory belong to the $7$-dimensional fundamental representation of $\text{G}_2$.  The partition function is then given by
\ie
Z_0(\beta,\alpha_1,\alpha_2)=\frac{\Phi_r^{3/2}}{\beta^{3/2}}
\frac{\cos\pi(\alpha_1+\sqrt{3}\alpha_2)\cos\pi(\alpha_1-\sqrt{3}\alpha_2)\cos2\pi\alpha_1}{\left(1-4(\alpha_1+\sqrt{3}\alpha_2)^2\right)\left(1-4(\alpha_1-\sqrt{3}\alpha_2)^2\right)\left(1-16\alpha_1^2\right)}\frac{\alpha_1\sqrt{3}\alpha_2}{\sin 2\pi\alpha_1\sin2\pi\sqrt{3}\alpha_2}
\\
\frac{(\alpha_1-\sqrt{3}\alpha_2)(\alpha_1+\sqrt{3}\alpha_2)(3\alpha_1-\sqrt{3}\alpha_2)(3\alpha_1+\sqrt{3}\alpha_2)}{\sin\pi(\alpha_1-\sqrt{3}\alpha_2)\sin\pi(\alpha_1+\sqrt{3}\alpha_2)\sin\pi(3\alpha_1-\sqrt{3}\alpha_2)\sin\pi(3\alpha_1+\sqrt{3}\alpha_2)}
e^{S_0+\frac{2\pi^2\Phi_r}{\beta}\left(1-16\alpha_1^2-16\alpha
_2^2\right)}
\fe
\ie
Z^{\text{G}(3)}(\beta,\alpha_1,\alpha_2)=\sum_{n_1,n_2\in\mathbb{Z}}Z_0(\beta,\alpha_1+\frac{n_1+n_2}{2},\alpha_2+\frac{n_1-n_2}{2\sqrt{3}})
\fe
where the irrational shift in $\alpha_2$ is essential to maintain the periodicity of gauge fields. Solving the spectrum, we similarly find both BPS and non-BPS states,
\ie
Z^{\text{G}(3)}&= \sum_{m_1,m_2\in\mathbb{Z}_{\ge 0}}\chi^{\text{long}}_{m_1,m_2}(\alpha_1,\alpha_2)\,\int \d E \,e^{-\beta E}\rho(E)\\
&~~~~+ \sum_{m_2\in\mathbb{Z}_{\ge 0}}\chi^{\text{short}}_{0,m_2}(\alpha_1,\alpha_2) \,e^{-\beta E_{\rm BPS}}\,N_{\rm BPS}
\fe
The long multiplet is given by
\ie
\chi^{\text{long}}_{m_1,m_2}(\alpha_1,\alpha_2)=&2\big(\chi_{m_1,m_2}
\\
&+\chi_{m_1-2,m_2+1}+\chi_{m_1-1,m_2}+\chi_{m_1+1,m_2-1}
\\
&+\chi_{m_1-1,m_2}+\chi_{m_1-3,m_2+1}+\chi_{m_1,m_2-1}
\\
&+\chi_{m_1-2,m_2}\big)
\fe
The long multiplet structure is analogous to the $\OSp(2l+1|2)$ case: the overall factor of 2 comes from the action of a supercharge that is neutral under the $R$-symmetry; the action of three `lowering' supercharge generates the multiplet of $2^3=8$ states, making the total dimension $2^4=8$ The explicit form of each character can be found in \eqref{eqn:G2character}. The density of long multiplets is given by
\ie
\rho(E)=\frac{e^{S_0}}{2\pi}\frac{m_1(m_1+2m_2+2)}{E-\frac{3}{32\Phi_r}(m_1+2m_2+2)^2}\,\frac{\cosh2\pi\sqrt{2\Phi_r(E-E_0)}}{\sqrt{2\Phi_r(E-E_0)}}
\fe
where $E_0$ is the spectral gap 
\ie
E_0=\frac{1}{8\Phi_r} \left(m_1^2 + 3 m_1 (1 + m_2) + 3 (1 + m_2)^2\right).
\fe
The BPS states manifest as short multiplets
\ie
\chi^{\text{short}}_{0,m_2}(\alpha_1,\alpha_2)=\chi_{0,m_2}+\chi_{1,m_2-1}+\chi_{0,m_2-1}
\fe
We can see that the shortening condition is $Q^0\ket{{\rm hws}}=0$, as well as $m_1=0$. Three supercharges are therefore preserved
\ie
Q_{(0,0)},\, Q_{(1,0)},\, \bar Q_{(-1,0)}.
\fe
which are labeled by their weights. The energy of these BPS states is given by
\ie
E_{\rm BPS}=\frac{3 (1 + m_2)^2}{8\Phi_r} 
\fe
with degeneracy
\ie
N_{\rm BPS}=e^{S_0}(m_2+1).
\fe
Focusing on the sector that contains only states in $(0,m_2>0)$ representation, we find two degenerate BPS states from short multiplets $(0,m_2)$ and $(0,m_2+1)$, as well as four non-BPS continuum of states from long multiplets $(1,m_2)$, $(2,m_2)$, $(2,m_2-1)$ and $(3,m_2-1)$.

\subsubsection*{$\text{F}(4)$}
In this subsection, we explore another example with $\cN=8$ supersymmetry, the $\text{F}(4)$ Schwarzian theory. The eight fermions transform under the spinor representation of the $R$-symmetry $\text{Spin}(7)$, therefore have weights $(\pm\frac{1}{2},\pm\frac{1}{2},\pm\frac{1}{2})$. Given this information, along with the fact that $q=\frac{3}{4}$ for $\text{F}(4)$, we can write down the partition function as
\ie
Z_0(\beta,\alpha_1,\alpha_2,\alpha_3)=\frac{\Phi_r^4}{\beta^4}\frac{\cos\pi(\alpha_1+\alpha_2+\alpha_3)\cos\pi(\alpha_1-\alpha_2+\alpha_3)\cos\pi(\alpha_1+\alpha_2-\alpha_3)}{(1-4(\alpha_1+\alpha_2+\alpha_3)^2)(1-4(\alpha_1-\alpha_2+\alpha_3)^2)(1-4(\alpha_1+\alpha_2-\alpha_3)^2)}
\\
\frac{\cos\pi(\alpha_1-\alpha_2-\alpha_3)}{(1-4(\alpha_1-\alpha_2-\alpha_3)^2)}\times \prod_{r\in R_+}\frac{r\cdot \alpha}{\sin 2\pi r\cdot \alpha}\,e^{\frac{2\pi^2\Phi_r}{\beta}(1-12\alpha_1^2-12\alpha_2^2-12\alpha_3^2)}
\fe
\ie
Z^{\text{F}(4)}(\beta,\alpha_1,\alpha_2,\alpha_3)=\sum_{n_{1,2,3}\in\mathbb{Z}}Z_0(\beta,\alpha_1+\frac{n_2+n_3}{2},\alpha_2+\frac{n_1-n_2}{2},\alpha_3+\frac{n_1-n_3}{2})
\fe 
Similarly, we can solve its spectrum, which is given by
\ie
Z^{\text{F}(4)}&= \sum_{m_1,m_2,m_3\in\mathbb{Z}_{\ge 0}}\chi^{\text{long}}_{m_1,m_2,m_3}(\alpha_1,\alpha_2,\alpha_3)\,\int \d E \,e^{-\beta E}\rho(E)\\
&~~~~+ \sum_{\substack{m_1,m_2\in\mathbb{Z}_{\ge 0}\\ m_3\in\{2m_1,2m_1+1\}}}\chi^{\text{short}}_{m_1,m_2,m_3}(\alpha_1,\alpha_2,\alpha_3) \,e^{-\beta E_{\rm BPS}}\,N_{\rm BPS}
\fe
The character formula for $\text{Spin}(7)$ has the same form as that for $\SO(7)$ given in \eqref{eqn:SOoddweight}. However, the key difference is that now $m_3$ is allowed to include half-integer spin representations, beside the integer ones. The non-BPS long multiplet is given by
\ie
&\chi^{\text{long}}_{m_1,m_2,m_3}(\alpha_1,\alpha_2,\alpha_3)
\\
&=\chi_{m_1,m_2,m_3}
\\
&+\chi_{m_1,m_2,m_3-1}+\chi_{m_1,m_2-1,m_3+1}+\chi_{m_1-1,m_2+1,m_3-1}+\chi_{m_1+1,m_2,m_3-1}
\\
&+\chi_{m_1,m_2-1,m_3}+\chi_{m_1-1,m_2+1,m_3-2}+\chi_{m_1+1,m_2,m_3-2}+\chi_{m_1-1,m_2,m_3}+\chi_{m_1+1,m_2-1,m_3}+\chi_{m_1,m_2+1,m_3-2}
\\
&+\chi_{m_1,m_2,m_3-1}+\chi_{m_1,m_2+1,m_3-3}+\chi_{m_1+1,m_2-1,m_3-1}+\chi_{m_1-1,m_2,m_3-1}
\\
&+\chi_{m_1,m_2,m_3-2}
\fe
The long multiplet has dimension $16=2^4$, where $4$ is half the number of total supercharges. It is generated from a highest weight state $(m_1, m_2, m_3)$, which is annihilated by four out of eight supercharges, with each subsequent line of states obtained by acting with one additional remaining supercharge. For example, the second line involves the action one supercharge, the third line involves two, and the last line involves all four supercharges. The density of long multiplets is given by
\ie
\rho(E)=\frac{e^{S_0}}{2\pi}(4+2m_1+2m_2+m_3)(2+2m_2+m_3)m_3(1+m_1)(3+m_1+2m_2+m_3)
\\
(2+m_1+m_2)(2+m_1+m_2+m_3)(1+m_2)(1+m_2+m_3)\frac{\sinh\,2\pi\sqrt{2\Phi_r(E-E_0)}}{E-E_0}
\\
\times\frac{1}{E-\frac{1}{18\Phi_r}\left((1+m_1)^2+(2+m_1+m_2)^2+(1+m_2)^2\right)}
\\
\times\frac{1}{E-\frac{1}{18\Phi_r}\left((1+m_1)^2+(2+m_1+m_2+m_3)^2+(1+m_2+m_3)^2\right)}
\\
\times\frac{1}{E-\frac{1}{18\Phi_r}\left((3+m_1+2m_2+m_3)^2+(2+m_1+m_2)^2+(1+m_2+m_3)^2\right)}
\\
\times\frac{1}{E-\frac{1}{18\Phi_r}\left((3+m_1+2m_2+m_3)^2+(2+m_1+m_2+m_3)^2+(1+m_2)^2\right)}
\fe
where $E_0$ is the spectral gap
\ie
E_0=\frac{1}{24\Phi_r}\left((4+2m_1+2m_2+m_3)^2+(2+2m_2+m_3)^2+m_3^2\right)
\fe
The continuous spectrum begins at $E_0$ and maintains a positive density for all energies above this value. We also find BPS short multiplets in the spectrum, which preserve two out of eight supercharges $Q_{(-\frac{1}{2},\frac{1}{2},\frac{1}{2})}$, $\bar Q_{(\frac{1}{2},-\frac{1}{2},-\frac{1}{2})}$. The corresponding short multiplet is composed of the following states
\ie
&\chi^{\text{short}}_{m_1,m_2,m_3}(\alpha_1,\alpha_2,\alpha_3)
\\
=&\chi_{m_1,m_2,m_3}
\\
+&\chi_{m_1,m_2-1,m_3+1}+\chi_{m_1-1,m_2+1,m_3-1}+\chi_{m_1,m_2,m_3-1}
\\
+&\chi_{m_1-1,m_2,m_3}+\chi_{m_1,m_2-1,m_3}+\chi_{m_1-1,m_2+1,m_3-2}
\\
+&\chi_{m_1-1,m_2,m_3-1}
\fe
These BPS short multiplets have energy
\ie
E_{\rm BPS}=\frac{1}{18\Phi_r}\left((3+m_1+2m_2+m_3)^2+(2+m_1+m_2+m_3)^2+(1+m_2)^2\right)
\fe
and have degeneracy
\ie
N_{\rm BPS}=e^{S_0}(3+m_1+2m_2+m_3)(2+m_1+m_2+m_3)(1+m_2)
\fe

There is a further BPS condition imposed by the Schwarzian theory: $m_3=2m_1+1$ or $m_3=2m_1$. As a result, the BPS states can be either bosonic or fermionic.

\section{Conclusions and future directions}

Based on the known classification of superconformal groups, in this work we have classified and solved all the effective theories of near-BPS black holes. The 1-loop exactness of the Schwarzian theory allowed us to extract the partition function, spectrum, and conserved charges as a function of a few effective field theory parameters, using only the supergroup theoretic data. Based on the algebra of charges, if we restrict the commutation relations to be linear, we found that BPS black holes cannot preserve more than four supercharges, explaining a feature observed in string theory constructions of such black holes. To justify this assumption, we assume that those generators are a subset of the symmetry generators in the UV and are not emergent symmetries in the IR. 

\smallskip

We described the quantization of the new near-BPS black hole effective theories we constructed, which was essentially the disc partition function in the near-horizon region. Following \cite{Turiaci:2023jfa}, one could attempt to define a notion of random matrix ensembles for these theories (introducing higher topologies and a genus expansion) and therefore produce a precise definition of chaos both in their BPS and non-BPS sectors. To make progress in this direction, it would be interesting to define an extended supersymmetric pure JT gravity model and show whether they can be described as  random matrix theories. This is currently underway \cite{WOP_JK}.

\smallskip

An application of our analysis is to small black holes. According to \cite{Chen:2021dsw}, a transition to a string gas should happen before we reach any singularity, ruling out small string-sized black holes. It would be interesting to find more examples along these lines where a potential small black hole that preserves too many supercharges is replaced by another phase. We mentioned other examples involving $AdS_3$ in the introduction, where previous results suggested the black holes and hypothetical dual SCFT would maintain a symmetry group we have seemingly ruled out. 

\smallskip

The specific example of $AdS_3$ is a potential loophole for our argument. This is because there are proposed string theory backgrounds involving an $AdS_3$ factor that supports more than four supercharges. These examples involve superconformal groups with unitary affine extensions, so they could potentially exist \cite{Dibitetto:2018ftj, Legramandi:2020txf, Macpherson:2023cbl, Hong:2019wyi,Conti:2025djz,Lozano:2025ief}. However, we believe that the fact that the algebras are nonlinear implies that these throats cannot be embedded in higher-dimensional flat space. It would be interesting to see how a nonlinear algebra can arise from the higher-dimensional (but not asymptotically flat) perspective.

One final future direction concerns whether there could exist quantum field theories or quantum mechanical systems which flow in the infrared to the $\mathcal{N}$-extended Schwarzian theories we have discussed here. Of course, for $\mathcal{N} \leq 4$, we have known brane systems in string theory dual to supersymmetric black holes and AdS spacetimes, but it is currently unknown how to derive the Schwarzian theory as the exact infrared description of these at low energies.\footnote{The only exception are holographic 2d CFT for which one can derive the low-energy description using modular invariance \cite{Ghosh:2019rcj}, but a direct derivation is still lacking.} One class of quantum mechanical models which provably has an $\mathcal{N}=2$ super-Schwarzian mode in the infrared are $\mathcal{N}=2$ SYK models~\cite{Fu:2016vas,Peng:2017spg,Berkooz:2020xne,Heydeman:2022lse,Peng:2020euz,Benini:2024cpf,Heydeman:2024ohc}. However, it seems difficult to find an interacting SYK-like model with even $\mathcal{N}=4$ supersymmetry which has both a stable superconformal 2-point function at $T=0$ as well as an $\mathcal{N}=4$ Schwarzian mode (see the arguments in the conclusion of \cite{Heydeman:2020hhw} for some details.) This might imply $\mathcal{N}>4$ SYK models are even more difficult to construct. If one \emph{could} construct these theories with the correct infrared behavior, it would be interesting to see how the nonlinearities and instabilities discovered in this paper are confirmed or resolved in a microscopic model.

\paragraph{Acknowledgements} It is a pleasure to thank Jan Boruch, Luca Iliesiu, Daniel Jafferis, Clifford Johnson, Maciej Kolanowski, Ho Tat Lam, Juan Maldacena, Alexey Milekhin, Mukund Rangamani, Ashoke Sen, Marc Spradlin, Andy Strominger, Cumrun Vafa, Anastasia Volovich, and Edward Witten for valuable discussions. We thank Clifford Johnson and Mukund Rangamani for bringing the $\mathcal{N}=3$ case to our attention. MTH is supported by Harvard University and the Black Hole Initiative, funded in part by the Gordon and Betty Moore Foundation (Grant 8273.01) and the John Templeton Foundation (Grant 62286). XS and GJT are supported by the University of Washington and the DOE award DE-SC0011637.

\appendix
\section{Technical details on the $\mathcal{N}=3$ and $\mathcal{N}=4$ theories}
In this appendix, we collect some intermediate steps in the derivation of the density of states for the specific cases of the $\mathcal{N}=3$ and large $\mathcal{N}=4$ Schwarzian theories. 

\subsection*{Derivation for $\mathcal{N}=3$}\label{app:N3}

We focus on the partition function $\tilde{Z}(\beta,\alpha)$. As explained in the main text, it is straightforward to extract from it the full spectrum.  The partition function is then
\beq
\tilde{Z}(\beta,\alpha) = \sum_{n\in\mathbb{Z}} \frac{8\cos(2\pi \alpha)}{\pi(1-16 (\alpha+n)^2)} \frac{\alpha+n}{\sin(2\pi \alpha)} \, e^{S_0+\frac{2\pi^2\Phi_r}{\beta}(1-16 (\alpha+n)^2)}
\eeq
Following \cite{Turiaci:2023jfa}, to derive the spectrum we take a derivative of the partition function with respect to $\beta$. This gives
\beq
-\partial_\beta \tilde{Z}= e^{S_0}2\cos 2\pi \alpha \sum_{n\in\mathbb{Z}} \frac{\alpha+n}{\sin 2\pi \alpha} \frac{8\pi\Phi_r}{\beta^2}  \, e^{\frac{2\pi^2\Phi_r}{ \beta}(1-16(\alpha+n)^2)}.
\eeq
We can Fourier transform in the $\SO(3)$ chemical potential by using Poisson resummation, which leads to a relation of the form 
\beq
\sum_{n\in\mathbb{Z}} \frac{\alpha+n}{\sin 2\pi \alpha} \frac{32(2\pi\Phi_r)^{3/2}}{\beta^{3/2}}  \, e^{-\frac{32\Phi_r\pi^2(\alpha+n)^2}{\beta}} = \sum_{j=\frac{1}{2},1,\ldots} (2j) \,\chi_{j-\frac{1}{2}}(\alpha)\, e^{- \beta \frac{j^2}{8\Phi_r}}
\eeq
Using this identity, and introducing the energy $E_0(j)=\frac{j^2}{8\Phi_r}$, we get
\bea
-\partial_\beta \tilde{Z} &=& 2\cos 2\pi \alpha \sum_{j} \chi_{j-\frac{1}{2}}(\alpha)\, (2j)\, e^{- \beta E_0(j)} \frac{1}{\beta^{1/2}8\sqrt{2\pi\Phi_r}} e^{S_0+\frac{2\pi^2\Phi_r}{ \beta}},\\
&=& 2\cos 2\pi \alpha \sum_{j} \chi_{j-\frac{1}{2}}(\alpha) \, \int_{E_0(j)}^\infty \d E \, e^{-\beta E} \, \frac{e^{S_0}(2j)\,\cosh(2\pi\sqrt{2\Phi_r(E-E_0(j))})}{8\pi\sqrt{2\Phi_r(E-E_0(j))}}.
\ea
Integrating over $\beta$, and using the fact that 
$$
2\cos (2 \pi \alpha) \, \chi_{j}(\alpha) = \chi_{j+1/2}(\alpha) + \chi_{j-1/2}(\alpha),
$$
gives the result
\bea
\tilde{Z} &=& \sum_{j=\frac{1}{2},1,\ldots} \int_{E_0(j)}^\infty \d E ( \chi_{j}(\alpha) + \chi_{j-1}(\alpha)) \,\rho_j(E) e^{-\beta E}+ \text{constant.}
\ea
where the density of states is essentially the same as in $-\partial_\beta \tilde{Z}$ up to an overall factor of energy in the denominator, reproducing the formula quoted in the main text.

Finally we need to evaluate the temperature independent (but possibly $\alpha$ dependent) constant. If we take $\beta \to \infty$ the continuous part above vanishes since $E_0(j) >0$ for all $j$. Therefore the constant is simply equal to $\tilde{Z}(\beta\to\infty,\alpha)$, namely
\beq
\sum_{n\in\mathbb{Z}} \frac{8\cos(2\pi \alpha)}{\pi(1-16 (\alpha+n)^2)} \frac{\alpha+n}{\sin(2\pi \alpha)} \, e^{S_0} = \frac{e^{S_0}}{2}.
\eeq
This completes the derivation. 

\subsection*{Derivation for $\mathcal{N}=4$}\label{app:N4}
In this case the derivation is more complex since the $R$-symmetry group of $D(2,1|\upalpha)$, $\SU(2)\times \SU(2)$, has rank 2. The partition function can be written as
\ie
Z\left(\beta, \alpha_{+}, \alpha_{-}\right)=&e^{S_0} \sum_{m, n} \frac{\Phi_r^{1/2}}{\beta^{1/2}} \frac{\cos \pi\left(\alpha_{+}+\alpha_{-}\right) \cos \pi\left(\alpha_{+}-\alpha_{-}\right)(\alpha_++n)(\alpha_-+m)}{ \sin 2 \pi \alpha_{+} \sin 2 \pi \alpha_{-}} 
\\
&\quad\times\frac{e^{\frac{2 \pi^2 \Phi_{r}}{\beta}\left[1-\frac{4}{\gamma_{-}}\left(n+ \alpha_{+}\right)^2-\frac{4}{\gamma_{+}}\left(m+ \alpha_{-}\right)^2\right]}}{\left(1-4(\alpha_++n-\alpha_--m)^2\right)\left(1-4(\alpha_++n+\alpha_-+m)^2\right)}
\\
=&e^{S_0} \sum_{m, n} \sum_{\epsilon_{ \pm}} Z_0(\beta,\epsilon_+(\alpha_++n),\epsilon_-(\alpha_-+m))
\fe
Here we introduce another sum over $\epsilon_\pm=\pm 1$ to emphasize $\SU(2)\times \SU(2)$ Weyl invariance, which acts by $\alpha_i\to -\alpha_i$. The individual saddle is given by
\ie
Z_0\left(\beta,\epsilon_+(\alpha_++n),\epsilon_-(\alpha_-+m)\right)=&\frac{\Phi_r^{1/2}}{\beta^{1/2}}\frac{\cos \pi\left(\epsilon_+(\alpha_{+}+n)+\epsilon_-(\alpha_{-}+m)\right) \cos \pi\left(\epsilon_+(\alpha_{+}+n)-\epsilon_-(\alpha_{-}+m)\right)}{64 \epsilon_+\epsilon_-\sin 2 \pi \alpha_{+} \sin 2 \pi \alpha_{-}}
\\
&\times \frac{e^{\frac{2 \pi^2 \Phi_{r}}{\beta}\left[1-\frac{4}{\gamma_{-}}(\alpha_{+}+n)^2-\frac{4}{\gamma_{+}} (\alpha_{-}+m)^2\right]}}{\left(\frac{1}{2}+\epsilon_+(\alpha_{+}+n)+\epsilon_-(\alpha_{-}+m)\right)}
\fe
To extract the spectrum, we first perform a Fourier transform in the chemical potentials on individual saddles,
\ie
\hat Z(\beta,\epsilon_+\alpha_+,\epsilon_-\alpha_-|k_+,k_-)=&\int \d n\d m \, e^{2\pi \i nk_++2\pi \i mk_-}Z_0\left(\beta,\epsilon_+(\alpha_++n),\epsilon_-(\alpha_-+m)\right)
\\
=&\frac{\Phi_r^{1/2}}{\beta^{1/2}}\frac{e^{-2\pi \i\epsilon_+\alpha_+k_+-2\pi \i\epsilon_-\alpha_-k_-}}{64 \epsilon_+ \epsilon_- \sin 2 \pi \alpha_{+} \sin 2 \pi \alpha_{-}}\int \d n\d m\, e^{2\pi \i nk_++2\pi \i mk_-}
\\
&\quad\times\cos\pi(n+m)\cos\pi(n-m)\frac{e^{\frac{2\pi^2\Phi_r}{\beta}\left(1-\frac{4}{\gamma_-}n^2-\frac{4}{\gamma_+}m^2\right)}}{\frac{1}{2}+ n+m}
\fe
The above integral can be simplified by introducing new variables $t_1=n+m$, $t_2=n-m$,
\ie
\int &\d n\d m\, e^{2\pi \i n k_++2\pi \i m k_-} \cos \pi \left(n+m\right) \cos\pi(n-m)\frac{e^{\frac{2\pi^2\Phi_r}{\beta}\left(1-\frac{4}{\gamma_-}n^2-\frac{4}{\gamma_+}m^2\right)}}{\frac{1}{2}+n+m}
\\
=&\int \frac{-1}{2}\d t_1 \d t_2 e^{\pi \i (k_++k_-)t_1+\pi \i (k_+-k_-)t_2}\frac{e^{\frac{ 2\pi^2\Phi_r }{\beta}\left[1-(1+\upalpha^{-1})(t_1+t_2)^2-(1+\upalpha)(t_1-t_2)^2\right]}}{\frac{1}{2}+t_1}\cos\pi t_1\cos\pi t_2
\\
=&\sqrt{\frac{\beta}{8 \pi\Phi_r (2+\upalpha+\upalpha^{-1})}}\Bigg(e^{\frac{2 \pi^2 \Phi_r}{\beta}-\frac{\beta(k_+-k_-+1)^2}{8\Phi_r(2+\upalpha+\upalpha^{-1})}}\int \frac{-\d t_1}{1+2t_1}e^{-\frac{2 \pi^2 \Phi_r}{\beta}4t_1^2+2\pi \i\frac{\upalpha (k_++\frac{1}{2})+(k_--\frac{1}{2})}{\upalpha+1}t_1}\cos\pi t_1
\\
&\quad +e^{\frac{2 \pi^2 \Phi_r}{\beta}-\frac{\beta(k_+-k_--1)^2}{8\Phi_r(2+\upalpha+\upalpha^{-1})}}\int \frac{-\d t_1}{1+2t_1}e^{-\frac{2 \pi^2 \Phi_r}{\beta}4t_1^2+2\pi \i\frac{\upalpha (k_+-\frac{1}{2})+(k_-+\frac{1}{2})}{\upalpha+1}t_1}\cos\pi t_1\Bigg)
\fe
We have performed the Gaussian integral over $t_2$, leaving with an integral over $t_1$. We can formally apply the Poisson summation formula to obtain
\ie
Z\left(\beta, \alpha_{+}, \alpha_{-}\right)=&e^{S_0} \sum_{\epsilon_{ \pm}}\sum_{m, n}  Z_0(\beta,\epsilon_+\alpha_++n,\epsilon_-\alpha_-+m)
\\
=&e^{S_0} \sum_{\epsilon_{ \pm}}\sum_{k_+, k_-}\hat Z(\beta,\epsilon_+\alpha_+,\epsilon_-\alpha_-|k_+,k_-)
\\
=&e^{S_0} \sum_{\epsilon_{ \pm}}\sum_{k_+, k_-}\frac{e^{-2\pi \i\epsilon_+\alpha_+(k_++\frac{1}{2})-2\pi \i\epsilon_-\alpha_-(k_--\frac{1}{2})}}{ \epsilon_+\epsilon_-\sin 2 \pi \alpha_{+} \sin 2 \pi \alpha_{-}}\frac{\sqrt{\upalpha}}{64\sqrt{2\pi}(1+\upalpha)}\cos\pi (\epsilon_+\alpha_+-\epsilon_-\alpha_-)
\\
&\times e^{\frac{2 \pi^2 \Phi_r}{\beta}-\frac{\beta(k_+-k_-+1)^2}{8\Phi_r(2+\upalpha+\upalpha^{-1})}}\int \frac{-\d  t_1}{1+2t_1}e^{-\frac{2 \pi^2 \Phi_r}{\beta}4t_1^2+2\pi \i\frac{\upalpha (k_++\frac{1}{2})+(k_--\frac{1}{2})}{\upalpha+1}t_1}\cos\pi t_1
\fe
To derive the spectrum $\rho(E)$, the next step involves performing the integral and applying the inverse Laplace transform. However, before proceeding with the integral, we can pause to examine the dependence on $\beta$ in the above integrand,
\ie
\,&e^{\frac{2 \pi^2 \Phi_r}{\beta}(1-4t_1^2)-\frac{\beta( k_+-k_-+1)^2}{8\Phi_r(2+\upalpha+\upalpha^{-1})}}
\\
=&\int_{E_0} \d E e^{-\beta E}\left(\sqrt{2 \pi^2 \Phi_r(1-4t_1^2)}\frac{I_1(2\sqrt{2 \pi^2 \Phi_r(1-4t_1^2)(E-E_0)})}{\sqrt{E-E_0}}+\delta(E-E_0)\right)
\\
=&\left(e^{\frac{2 \pi^2 \Phi_r}{\beta}(1-4t_1^2)}-1\right)e^{-\frac{\beta( k_+-k_-+1)^2}{8\Phi_r(2+\upalpha+\upalpha^{-1})}}+\int \d E e^{-\beta E}\delta(E-E_0)
\label{eqn:betadependence}
\fe
where $E_0=\frac{( k_+-k_-+1)^2}{8\Phi_r(2+\upalpha+\upalpha^{-1})}$. The appearance of a Dirac delta function suggests that the entire spectrum is divided into continuous states and discrete states with energy $E_0(k_+,k_-)$. It can also be seen by taking the large $\beta$ limit. The next step is to consider the two terms that represent the continuous and discrete spectra separately. The integral for discrete states is simplified after inserting the delta function
\ie
Z_{\text{discrete}}=&e^{S_0} \sum_{\epsilon_{ \pm}}\sum_{k_+, k_-}\frac{e^{-2\pi \i\epsilon_+\alpha_+(k_++\frac{1}{2})-2\pi \i\epsilon_-\alpha_-(k_--\frac{1}{2})}}{ \epsilon_+\epsilon_-\sin 2 \pi \alpha_{+} \sin 2 \pi \alpha_{-}}\frac{\sqrt{\upalpha}}{64\sqrt{2\pi}(1+\upalpha)}\cos\pi (\epsilon_+\alpha_+-\epsilon_-\alpha_-)
\\
&\times \int \d E e^{-\beta E}\delta(E-\frac{( k_+-k_-+1)^2}{8\Phi_r(2+\upalpha+\upalpha^{-1})})\int \frac{-\d t_1}{1+2t_1}e^{2\pi \i\frac{\upalpha (k_++\frac{1}{2})+(k_--\frac{1}{2})}{\upalpha+1}t_1}\cos\pi t_1
\\
=&e^{S_0} \sum_{k_+, k_-}\frac{-2\sin2\pi \alpha_+(k_++1)\sin 2\pi\alpha_-(k_--1)-2\sin2\pi\alpha_+ k_+\sin2\pi\alpha_- k_-}{ \sin 2 \pi \alpha_{+} \sin 2 \pi \alpha_{-}}\frac{\sqrt{\upalpha}}{64\sqrt{2\pi}(1+\upalpha)}
\\
&\times \int \d E e^{-\beta E}\delta(E-\frac{( k_+-k_-+1)^2}{8\Phi_r(2+\upalpha+\upalpha^{-1})})\frac{\pi}{2}e^{-\pi \i\frac{\upalpha (k_++\frac{1}{2})+(k_--\frac{1}{2})}{\upalpha+1}}\Theta(\frac{\upalpha  (k_++\frac{1}{2})+ (k_--\frac{1}{2})}{\upalpha+1})
\label{eqn:thetafunction}
\fe
where $\Theta(x)=1$ if $|x|<\frac{1}{2}$, $\Theta(x)=\frac{1}{2}$ if $|x|=\frac{1}{2}$ and 0 otherwise. 
This imposes a constraint on the allowed range of charges for BPS states. To highlight the appearance of irreducible representations, we further replace $k_+,k_-\in \mathbb{Z}$ with the SU(2) spins $j_+,j_-\in\frac{\mathbb{Z}^{\geq 0}}{2}$ which are more natural for $SU(2)$ characters. After making this substitution, the contribution to the partition function from discrete states becomes:
\ie
Z_{\text{discrete}}=&e^{S_0}\sum_{j_+,j_-\in R_{\text{BPS}}}(\chi_{j_+}(\alpha_+)\chi_{j_-}(\alpha_-)+\chi_{j_+-\frac{1}{2}}(\alpha_+)\chi_{j_--\frac{1}{2}}(\alpha_-))
\\
\quad&\times\int \d E e^{-\beta E}\delta\left(E-\frac{\upalpha(j_++j_-+\frac{1}{2})^2}{2\Phi_r(1+\upalpha)^2}\right)
  \frac{\sqrt{\upalpha\pi/2}}{32(1+\upalpha)}\left|\sin\left(\frac{2\pi\upalpha(j_++j_-+\frac{1}{2})}{1+\upalpha}\right)\right| \label{eq:largeN4disDOS}
\fe
Here $R_{\text{BPS}}=\{j_+,j_-|0< \frac{\upalpha}{\upalpha+1}(2j_++2j_-+1)-2j_-< 1\}$.

Next, we start to extract the continuous spectrum from the partition function,
\ie
\,&Z_{\text{continuous}}\left(\beta, \alpha_{+}, \alpha_{-}\right)
\\
=&e^{S_0} \sum_{\epsilon_{ \pm}}\sum_{k_+, k_-}\frac{e^{-2\pi \i\epsilon_+\alpha_+(k_++\frac{1}{2})-2\pi \i\epsilon_-\alpha_-(k_--\frac{1}{2})}}{ \epsilon_+\epsilon_-\sin 2 \pi \alpha_{+} \sin 2 \pi \alpha_{-}}\frac{\sqrt{\upalpha}}{64\sqrt{2\pi}(1+\upalpha)}\cos\pi (\epsilon_+\alpha_+-\epsilon_-\alpha_-)
\\
&\times e^{-\frac{\beta(k_+-k_-+1)^2}{8\Phi_r(2+\upalpha+\upalpha^{-1})}}\int \frac{-\d t_1}{1+2t_1}(e^{\frac{2 \pi^2 \Phi_r}{\beta}(1-4t_1^2)}-1)e^{2\pi \i\frac{\upalpha (k_++\frac{1}{2})+(k_--\frac{1}{2})}{\upalpha+1}t_1}\cos\pi t_1
\\
=&e^{S_0} \sum_{\epsilon_{ \pm}}\sum_{k_+, k_-}\frac{e^{-2\pi \i\epsilon_+\alpha_+k_+-2\pi \i\epsilon_-\alpha_-k_-}}{ \epsilon_+\epsilon_-\sin 2 \pi \alpha_{+} \sin 2 \pi \alpha_{-}}\frac{\sqrt{\upalpha}}{64\sqrt{2\pi}(1+\upalpha)}\cos\pi (\alpha_+-\alpha_-)\cos\pi (\alpha_++\alpha_-)
\\
&\times e^{-\frac{\beta(k_+-k_-)^2}{8\Phi_r(2+\upalpha+\upalpha^{-1})}}\int \frac{-\d t_1}{1+2t_1}(e^{\frac{2 \pi^2 \Phi_r}{\beta}(1-4t_1^2)}-1)e^{2\pi \i\frac{\upalpha k_++k_-}{\upalpha+1}t_1}\cos\pi t_1
\fe
We denote the last line as $f(\beta)$. By an `integration by parts' procedure, we can eliminate the pole and do the Gaussian integral
\ie
\frac{\d f}{\d\beta}=&-\frac{( k_+-k_-)^2}{8\Phi_r(2+\upalpha+\upalpha^{-1})}f(\beta)
\\
&+e^{-\frac{\beta( k_+-k_-)^2}{8\Phi_r(2+\upalpha+\upalpha^{-1})}}\int d t_1\frac{2\pi^2\Phi_r(1-2t_1)}{\beta^2}e^{\frac{2 \pi^2 \Phi_r}{\beta}(1-4t_1^2)+2\pi i\frac{\upalpha k_++k_-}{\upalpha+1}t_1}\cos\pi t_1
\\
=&-\frac{( k_+-k_-)^2}{8\Phi_r(2+\upalpha+\upalpha^{-1})}f(\beta)
\\
&+e^{\frac{2\pi^2 \Phi_r}{\beta}}e^{-\frac{\beta(\upalpha k_+^2+k_-^2)}{8\Phi_r(\upalpha+1)}}\left( \beta^{-3/2}\sqrt{\frac{\pi^3\Phi_r}{2}}-\beta^{-1/2}\frac{\i\frac{\upalpha k_++k_-}{\upalpha+1}}{4\sqrt{2\Phi_r/\pi}}
 \right)
 \label{eqn:bypart1}
\fe
We can drop the second term in the last parenthesis since it is odd under $k_\pm\to-k_\pm$ and will vanish after summing over them. Performing inverse Laplace transform on the first part, we obtain the density of states
\ie
-E\tilde f(E)=-\frac{( k_+-k_-)^2}{8\Phi_r(2+\upalpha+\upalpha^{-1})}\tilde f(E)+\frac{1}{2}\sinh\sqrt{8\pi^2\Phi_r \left(E-\frac{(\upalpha k_+^2+k_-^2)}{8\Phi_r(\upalpha+1)}\right)}
\\
\Rightarrow\quad \tilde f(E)=\frac{1}{\frac{( k_+-k_-)^2}{8\Phi_r(2+\upalpha+\upalpha^{-1})}-E}\frac{1}{2}\sinh\sqrt{8\pi^2\Phi_r \left(E-\frac{(\upalpha k_+^2+k_-^2)}{8\Phi_r(\upalpha+1)}\right)}
\label{eqn:bypart2}
\fe
Similarly, we rewrite the sum over $k_+$ and $k_-$ in terms of spins $j_+, j_-$, which makes the character for the long multiplets manifest.  We finally arrive at
\ie
Z\left(\beta, \alpha_{+}, \alpha_{-}\right)&=e^{S_0}\frac{\sqrt{\upalpha}}{64\sqrt{2\pi}(1+\upalpha)}\sum_{j_+,j_-}\chi^{\rm long}_{j_+,j_-}(\alpha_+,\alpha_-)\int_{E_0}^\infty \d E e^{-\beta E}
\\
&\frac{\frac{2\upalpha}{\Phi_r(1+\upalpha)^2}j_+j_-}{\left(E-\frac{\upalpha( j_++j_-)^2}{2\Phi_r(1+\upalpha)^2}\right)\left(E-\frac{\upalpha( j_+-j_-)^2}{2\Phi_r(1+\upalpha)^2}\right)}\sinh\sqrt{8\pi^2\Phi_r \left(E-\frac{\upalpha j_+^2+j_-^2}{2\Phi_r(\upalpha+1)}\right)}
\fe
The structure of the group character for the long multiplets is explained around \eqref{eqn:N4LONGCHAR}.

\section{Technical details on the theories 
 with non-linearly realized symmetries}
\label{app:roots}
The derivation of the spectrum for the $\cN>4$ theories in Section \ref{sec:N>4} closely follows the procedure used in the large $\cN=4$ case. As the steps are analogous, we omit the detailed and lengthy derivation and only summarize the group characters appearing in the multiplets for clarity.

\subsection*{$\OSp(n|2)$}
\label{app:Osp}
In the following table, we collect information on the group theory for $\SO(n)$, the $R$-symmetry of $\OSp(n|2)$.
\begin{figure}[h!]
    \centering
\begin{tabular}{c|c|c}
     & {$\SO(2l+1)$} & {$\SO(2l)$}  \\
     \hline
    Positive roots $r$ & {$e^j\pm e^k$ for $j<k$, $e^j$} & {$e^j\pm e^k$ for $j<k$} \\
    \hline
    Fundamental weights $\mu_i$   & {\makecell{$e^1+\dots+e^{j}$ for $j\le l-1$, \\$\frac{1}{2}(e^1+\dots+e^l)$}} & {\makecell{$e^1+\dots+e^{j}$ for $j\le l-2$, \\$\frac{1}{2}(e^1+\dots-e^l)$, $\frac{1}{2}(e^1+\dots+e^l)$}}  \\
    \hline
        Weyl vector $w$  & { $\sum_{j=1}^l (l-j+\frac{1}{2})e^j$} & { $\sum_{j=1}^l (l-j)e^j$} \\
        \hline
    Weyl group & {$\mathbb{Z}_2^l \rtimes S_n$} & {$\mathbb{Z}_2 \rtimes S_n$}\\
\end{tabular}
\end{figure}

By Weyl character formula, the character for irreducible representations in $\SO(2l+1)$ with highest weight $\lambda=\sum_{i=1}^l m_i \mu_i$ is given by
\ie
\chi_\lambda(\alpha_1\,\dots,\alpha_l)=\frac{\sum_{\sigma\in S_l}\left(\text{sign}(\sigma)\prod_{j=1}^l \sin 4\pi \alpha_j \left(\lambda+w\right)_{\sigma(j)}\right) }{(2\i)^{l(l-1)}\prod_{r\in R_+}\sin 2\pi r\cdot \alpha}
\\
\text{with}~~~\lambda=(m_1+\dots+m_{l-1}+\frac{m_l}{2},\dots,m_{l-1}+\frac{m_l}{2},\frac{m_l}{2})
\label{eqn:SOoddweight}
\fe
while for $\SO(2l)$ the character is given by
\ie
\chi_\lambda(\alpha_1\,\dots,\alpha_l)=\frac{\sum_{\sigma\in S_l}\left(2\,\text{sign}(\sigma)\,\cos \sum_{j=1}^l 4\pi\alpha_j \left(\lambda+w\right)_{\sigma(j)}\right) }{(2\i)^{l(l-1)}\prod_{r\in R_+}\sin 2\pi r\cdot \alpha}
\\
\text{with}~~~\lambda=(m_1+\dots+\frac{m_{l-1}}{2}+\frac{m_{l}}{2},\dots,\frac{m_{l-1}}{2}+\frac{m_{l}}{2},-\frac{m_{l-1}}{2}+\frac{m_{l}}{2})
\label{eq:charactersoeven}
\fe
Note that $m_l$ in the case of $\SO(2l+1)$ and $m_{l-1}+m_l$ in the case of $\SO(2l)$ should take values only in even positive integers to ensure all representations get integer spin. The possibility of taking other values is discussed in Section \ref{sec:anomalyosp}.

The partition functions of the $\OSp(n|2)$ Schwarzian theories are given in \eqref{eqn:ptospeven} and \eqref{eqn:ptospodd}. Since the derivation of the spectrum closely parallels that of the previous $D(2,1|\upalpha)$ case, we provide only a brief outline of the procedure here.

To change from a sum over saddles to a sum over charges, we again perform a Fourier transform on the $n_i$'s,
\ie
~&\prod_i \int \d n_i e^{2\pi\i n_i k_i}Z^{\OSp(n|2)}(\beta,\alpha_i+n_i/2)
\\
=&\sum_i (\dots) \int\prod_{k\neq i} \d n_k e^{\sum_{p\neq i} 2\pi \i k_p n_p}\left(\prod_{r\in R^{\prime i}_+} r\cdot n\right)e^{-\frac{2\pi^2\Phi_r}{\beta}\sum_{j\neq i}4n_j^2}
\\
&\times \int \d n_i e^{2\pi \i (k_i-1/2) n_i}\frac{\cos 2\pi n_i}{1-4n_i^2}e^{S_0+\frac{2\pi^2\Phi_r}{\beta}\left(1-4n_i^2\right)}
\fe
where we isolate the integral over $n$'s and absorb all other $n$-independent terms in the ellipses. Except for $n_i$ that appears in the denominator and produces a pole, all other integrals are of standard Gaussian type and give $\left(\prod_{r\in R^{\prime i}_+} r\cdot k\right)  e^{-\frac{\beta}{8\Phi_r} \sum_{j\neq i}k_j^2}\times\left(\frac{\beta}{8\Phi_r \pi}\right)^{\frac{(n-2)(n-3)}{4}}$.
One can check that the factor of $\beta^{\frac{(n-2)(n-3)}{4}}$ arising from the Fourier transform precisely cancels the $\beta$ dependence from the one-loop quantum correction, where the exponent number is given by the difference between the bosonic and fermionic generators of the superalgebra. This cancellation is a general feature that also reflects the presence of BPS states in these theories. 

The remaining dependence on $\beta$ is then $e^{\frac{2\pi^2\Phi_r}{\beta}(1-4n_i^2)-\frac{\beta}{8\Phi_r} \sum_{j\neq i}k_j^2}$. We still need to perform an inverse Laplace transform to transition to the microcanonical ensemble. The subtlety lies in the $n_i$ integral, which contains a pole, a manifestation of the fermion zero mode associated with supersymmetry breaking. In principle, one can evaluate it analytically by carefully altering the contour in the complex plane. However, this complexity can be avoided by separating the BPS and non-BPS contributions prior to integration. The low-temperature asymptotics $e^{-\frac{\beta}{8\Phi_r} \sum_{j\neq i}k_j^2}$ indicate the existence of BPS states with energy $\frac{1}{8\Phi_r} \sum_{j\neq i}k_j^2$. We then perform the inverse Laplace transform separately on BPS and non-BPS parts, yielding a Dirac delta function at BPS energy plus a modified spherical Bessel function of the continuous energy above the BPS energy, analogous to \eqref{eqn:betadependence}.

The next step is to evaluate the integrals for the two parts separately. For the BPS spectrum, all the $\beta$ dependence is contained in the Dirac delta function at the BPS energy. We can further apply partial fraction to obtain integrals of the form, $$\int \d n_i e^{2\pi\i k_i' n_i}\frac{\cos2\pi n_i}{1\pm 2n_i}$$The result is again given by a special function $\Theta(k_i')$, defined around \eqref{eqn:thetafunction}, which restricts the allowed values of charges $k_i$ that a BPS state can carry. This feature appears universally for the large $\cN=4$ theory and all the $\cN\ge4$ theories studied in this appendix: after a suitable partial fraction, we arrive at the same type of the  integral, whose outcome is the $\Theta$ function constraining the BPS charges. For the non-BPS spectrum, the integral involving the spherical modified Bessel function can be equivalently done by the "integration by parts" step performed in \eqref{eqn:bypart1} and \eqref{eqn:bypart2}, which is a more convenient method for carrying out the computation.

\subsection*{$\text{G}(3)$}
The $R$-symmetry of $\text{G}(3)$ is the exceptional group $\text{G}_2$
\begin{figure}[h!]
    \centering
\begin{tabular}{c|c}
     &  {$\text{G}_2$}  \\
     \hline
    Positive roots  & {$(1,0),(0,\sqrt{3}),(\pm\frac{1}{2},\frac{\sqrt{3}}{2}),(\pm\frac{3}{2},\frac{\sqrt{3}}{2})$} \\
    \hline
    Fundamental weights    & {$(\frac{1}{2},\frac{\sqrt{3}}{2}),(0,\sqrt{3})$} \\
    \hline
        Weyl vector   & {$(\frac{1}{2},\frac{3\sqrt{3}}{2})$} \\
\end{tabular}
\end{figure}

Using the Weyl character formula, the character for the irreducible representations of $\text{G}_2$ with highest weight $\lambda=m_1\mu_1+m_2\mu_2=(\frac{m_1}{2},\frac{\sqrt{3}m_1+2\sqrt{3}m_2}{2})$ is given by
\ie
\label{eqn:G2character}
\,&\chi_{m_1,m_2}(\alpha_1,\alpha_2)=\frac{1}{16\sin2\pi\alpha_1\sin2\pi\sqrt{3}\alpha_2\sin\pi(\alpha_1-\sqrt{3}\alpha_2)\sin\pi(3\alpha_1-\sqrt{3}\alpha_2)}
\\
&\frac{1}{\sin\pi(\alpha_1+\sqrt{3}\alpha_2)\sin\pi(3\alpha_1+\sqrt{3}\alpha_2)}\bigg(\sin\alpha_1\frac{m_1+1}{2}\sin\alpha_2\frac{\sqrt{3}m_1+2\sqrt{3}m_2+3\sqrt{3}}{2}
\\
&-\sin\alpha_1\frac{m_1+3m_2+4}{2}\sin\alpha_2\frac{\sqrt{3}m_1+\sqrt{3}m_2+2\sqrt{3}}{2}+\sin\alpha_1\frac{2m_1+3m_2+5}{2}\sin\alpha_2\frac{\sqrt{3}m_2+\sqrt{3}}{2}\bigg)
\fe

\subsection*{$\text{F}(4)$}
The $R$-symmetry of $\text{F}(4)$ is $\text{Spin}(7)$. Its character formula has the same form as that of $\SO(7)$ given in \eqref{eqn:SOoddweight}, except that $m_3$ can take any positive integer value to accommodate the inclusion of half-integer spin representations.

\bibliographystyle{utphys2}
{\small \bibliography{Biblio}{}}

\end{document}